\documentclass[twoside,twocolumn,9pt]{article}
\usepackage{extsizes}
\usepackage[super,sort&compress,comma]{natbib} 
\usepackage[version=3]{mhchem}
\usepackage[left=1.5cm, right=1.5cm, top=1.785cm, bottom=2.0cm]{geometry}
\usepackage{balance}
\usepackage{mathptmx}
\usepackage{sectsty}
\usepackage{graphicx} 
\usepackage{lastpage}
\usepackage[format=plain,justification=justified,singlelinecheck=false,font={stretch=1.125,small,sf},labelfont=bf,labelsep=space]{caption}
\usepackage{float}
\usepackage{fancyhdr}
\usepackage{fnpos}
\usepackage[english]{babel}
\addto{\captionsenglish}{%
  
}
\usepackage{array}
\usepackage{droidsans}
\usepackage{charter}
\usepackage[T1]{fontenc}
\usepackage[usenames,dvipsnames]{xcolor}
\usepackage{setspace}
\usepackage[compact]{titlesec}
\usepackage{hyperref}
\usepackage[caption=false]{subfig}
\usepackage{bm}
\usepackage{gensymb}

\usepackage{epstopdf}

\definecolor{cream}{RGB}{222,217,201}

\begin{document}

\pagestyle{fancy}
\thispagestyle{plain}
\fancypagestyle{plain}{
\renewcommand{\headrulewidth}{0pt}
}

\makeFNbottom
\makeatletter
\renewcommand\LARGE{\@setfontsize\LARGE{15pt}{17}}
\renewcommand\Large{\@setfontsize\Large{12pt}{14}}
\renewcommand\large{\@setfontsize\large{10pt}{12}}
\renewcommand\footnotesize{\@setfontsize\footnotesize{7pt}{10}}
\makeatother

\renewcommand{\thefootnote}{\fnsymbol{footnote}}
\renewcommand\footnoterule{\vspace*{1pt}%
\color{cream}\hrule width 3.5in height 0.4pt \color{black}\vspace*{5pt}} 
\setcounter{secnumdepth}{5}

\makeatletter 
\renewcommand\@biblabel[1]{#1}            
\renewcommand\@makefntext[1]%
{\noindent\makebox[0pt][r]{\@thefnmark\,}#1}
\makeatother 
\renewcommand{\figurename}{\small{Fig.}~}
\sectionfont{\sffamily\Large}
\subsectionfont{\normalsize}
\subsubsectionfont{\bf}
\setstretch{1.125} 
\setlength{\skip\footins}{0.8cm}
\setlength{\footnotesep}{0.25cm}
\setlength{\jot}{10pt}
\titlespacing*{\section}{0pt}{4pt}{4pt}
\titlespacing*{\subsection}{0pt}{15pt}{1pt}

\fancyfoot{}
\fancyfoot[LO,RE]{\vspace{-7.1pt}\includegraphics[height=9pt]{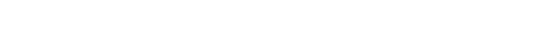}}
\fancyfoot[CO]{\vspace{-7.1pt}\hspace{13.2cm}\includegraphics{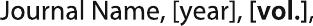}}
\fancyfoot[CE]{\vspace{-7.2pt}\hspace{-14.2cm}\includegraphics{head_foot/RF}}
\fancyfoot[RO]{\footnotesize{\sffamily{1--\pageref{LastPage} ~\textbar  \hspace{2pt}\thepage}}}
\fancyfoot[LE]{\footnotesize{\sffamily{\thepage~\textbar\hspace{3.45cm} 1--\pageref{LastPage}}}}
\fancyhead{}
\renewcommand{\headrulewidth}{0pt} 
\renewcommand{\footrulewidth}{0pt}
\setlength{\arrayrulewidth}{1pt}
\setlength{\columnsep}{6.5mm}
\setlength\bibsep{1pt}

\makeatletter 
\newlength{\figrulesep} 
\setlength{\figrulesep}{0.5\textfloatsep} 

\newcommand{\topfigrule}{\vspace*{-1pt}%
\noindent{\color{cream}\rule[-\figrulesep]{\columnwidth}{1.5pt}} }

\newcommand{\botfigrule}{\vspace*{-2pt}%
\noindent{\color{cream}\rule[\figrulesep]{\columnwidth}{1.5pt}} }

\newcommand{\dblfigrule}{\vspace*{-1pt}%
\noindent{\color{cream}\rule[-\figrulesep]{\textwidth}{1.5pt}} }

\makeatother

\twocolumn[
  \begin{@twocolumnfalse}
{
\includegraphics[width=18.5cm]{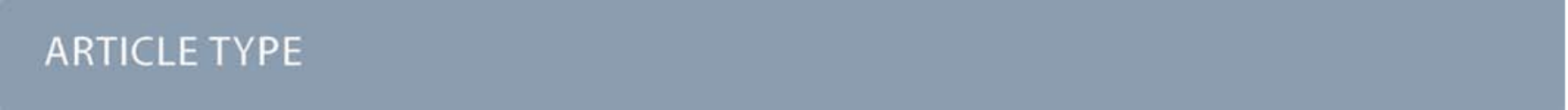}}\par
\vspace{1em}
\sffamily
\begin{tabular}{m{4.5cm} p{13.5cm} }

\includegraphics{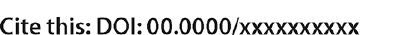} & \noindent\LARGE{\textbf{Granular mixtures discharging through a silo with eccentric orifice location$^\dag$}} \\
\vspace{0.3cm} & \vspace{0.3cm} \\

 & \noindent\large{A. Vamsi Krishna Reddy \textit{$^{a}$} and K. Anki Reddy $^{\ast}$\textit{$^{a}$}} \\

\includegraphics{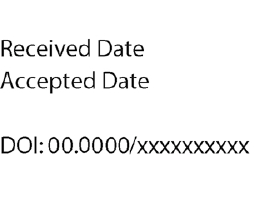} & \noindent\normalsize{We used the discrete element method (DEM) to study the flow dynamics  of a mixture of dumbbells and discs for two silo cases: 1. orifice situated on the $lateral$ wall and 2. $multiple$ orifices placed on the base of a two-dimensional silo. The time-averaged flow fields of various parameters like velocity, area fraction, pressure, shear stress etc, obtained from coarse-graining technique are presented for both the above-mentioned cases. A modified Beverloo scaling is reported for the flow discharging through the lateral orifice. The dynamic friction is found to increase with an increase in the addition of dumbbells. This resulted in a decrease in the flow rate through a $lateral$ orifice with the addition of dumbbells. Self-similar velocity profiles are observed for mixtures for the whole range of lateral orifice widths studied. The flow rate decreases with an increase in the inter-orifice distance $L/d$ for the multiple orifice case. The velocity profile and flow fields revealed an interaction zone between the two orifices at small $L/d$ which is almost absent when the orifices are wide apart. This interaction zone or flow through an orifice influenced by the presence of an adjacent orifice is the reason for a higher velocity above each of the orifices thus resulting in a higher flow rate at small $L/d$. The stagnant zone between the orifices is found to expand with an increase in the inter-orifice distance which yields in an increase in the shear stress between each of the orifices and the stagnant zone.} \\

\end{tabular}

 \end{@twocolumnfalse} \vspace{0.6cm}

  ]

\renewcommand*\rmdefault{bch}\normalfont\upshape
\rmfamily
\vspace{-1cm}


\footnotetext{\textit{$^{a}$Department of Chemical Engineering, Indian Institute of Technology Guwahati, Guwahati 781039, India. Fax : + 91 361 2582291; Tel: + 91 361 2583532; E-mail: anki.reddy@iitg.ac.in}}

\footnotetext{\dag~Electronic Supplementary Information (ESI) available: See DOI: 10.1039/cXsm00000x/}



\section{Introduction.}

The flow of an assembly of solid particles reminiscent of fluid flow can be witnessed in the form of landslides or during unloading of sand from trucks in a construction site or the flow of sand in an hourglass. In industries, handling of solid raw materials is involved mainly during production or storage. Understanding the flow of a collection of solid particles or the rheology of granular particles is of great importance in industries for effective plant operations and for producing desired products. The common scenarios of granular particulate flow are the flow of particles on an inclined surface \cite{Bagnoldscaling} or particles discharging through an orifice of a silo \cite{Mankoc2007} or hopper \cite{rodsphere}. 


The flow of particles through an orifice entails interesting phenomena like ratholing, clogging, pressure saturation etc, based on the material properties, the shape of the particles and the silo geometry. During a silo discharge, two types of flow patterns \cite{mass} can occur namely mass flow and funnel flow depending upon the inter-particle friction and the shape of the particles. In mass flow, there is little difference in the velocities of the particles that are flowing in the centre and those near the walls. Whereas, in a funnel type flow, the flow is mainly concentrated in the central part of a silo with a significant difference in the velocities of particles in the centre and those near the wall. An extreme situation of funnel type flow is rat-holing \cite{rathole} where the flow occurs only in a small section at the centre of the silo and in the other regions, the particles are completely stationary. This type of flow has been witnessed in the systems of elongated particles with high aspect ratios. Clogging phenomena \cite{clog1,clog2} or a sudden stoppage of the flow due to the formation of a stable structure of particles that blocks the orifice is common in silo flows involving small orifices. The average major dimension of the elongated particles was found to align almost along the streamlines of the flow due to shear-induced orientation of the particles \cite{elongalign}. The pressure is found to get saturated at depths greater than the width of the silos \cite{janseen} as the load of the particles is partially taken by the side walls through force chains and this is named as Janssen effect. The above-mentioned are some of the phenomena observed during the flow of particles through a silo. Beverloo \textit{et al}. \cite{Beverloo_1961} proposed a model to compute flow rate for a system of spherical particles discharging through a silo as $Q=C\rho_b\sqrt{g}(W-kD)^{n-\frac{1}{2}}$. Here, $Q$ is the flow rate, $C$ is a constant whose value depends on the material properties, $\rho_b$, $g$ and $W$ are bulk density, acceleration due to gravity and orifice width. Moreover, $k$, $D$ are dimensionless coefficient, particle diameter and $n=2,3$ corresponds to a two-dimensional and three-dimensional silo. In the recent decades, many models have been proposed for computing flow rate  \cite{Mankoc2007,Artega_1990,Chevoir_2007,Benyamine} of particles discharging through an orifice on the silo base depending on the particle shapes, mixtures of particles or the range of orifice widths considered. 

In silos, the orifice is usually placed at the centre of the base due to the practical applications in industries. One of the unconventional orifice positionings in a silo is on the sidewall which can be witnessed during an accidental leakage. In the silos with lateral exits, wall thickness plays a significant role which is not the case in silos having bottom apertures. With an increase in the wall thickness outflow capacity was found to decrease in a cylindrical tube \cite{Bagrintsev1977}.  The flow rate dependence on the diameter/hydraulic diameter of the orifice and the wall thickness has been experimentally investigated \cite{lateral1} for an assembly of particles discharging through circular, rectangular and triangular orifices on the vertical walls. Recently, Serrano \textit{et al.} \cite{lateral2} proposed a correlation for computing flow rate $Q$ as a function of $D$ and $w$ of dry cohesionless particles discharging through circular orifices of diameter $D$ on a vertical wall of thickness $w$. Further, the authors showed that Hagen-law (Q $\alpha$ D$^{5/2}$) can be used for computing the flow rate of particles discharging through orifices placed on very thin vertical walls. Zhou \textit{et al.} \cite{Zhou2017} performed experiments, discrete and continuous simulations and proposed an empirical relation for computing flow rate based on the dimensions of the lateral orifice of a silo with thin walls. Moreover, clogging phenomena at lateral orifice for a variety of granular materials was studied and the minimum orifice width where the continuous flow can be expected was proposed by Davies and Desai \cite{DAVIES2008436}. 

Granular particles discharging through $multiple$ orifices placed on the base of the silo is another eccentricity in the silo flows apart from a $lateral$ orifice. The usage of multiple orifices is one of the practical industrial solutions for mitigating the clogging of particles discharging through narrow orifices. Before the system gets clogged, the average number of particles discharged from each of the two small orifices as compared with a single orifice silo was found to increase by an order of magnitude \cite{multi1} just by varying the inter-orifice distance. The fluctuations due to an intermittent flow from an orifice resulted in the resumption of the flow in another jammed orifice thus increasing the time before the system gets clogged. In another study \cite{multi3}, the effect of inter-particle friction has been investigated on the flow and jamming behaviour of the particles exiting through multiple orifices. Correlations for the flow rate of the particles discharging out of two orifices placed on the base considering various outlet sizes and inter-orifice distances has been proposed by performing simulations \cite{multiori} as well as experiments \cite{multi2}. Fullard \textit{et al.}\cite{multi4} noticed a non-monotonic dependence of flow rate on the inter-orifice distance. The authors reported that inter-particle friction is the reason for this kind of behaviour. Maiti \textit{et al.} \cite{multi6} studied the influence of inter-orifice distance between symmetrically as well as asymmetrically placed orifices on the base. Further, they reported the existence of a neutral axis between the two orifices which bifurcates the flow fields due to each orifice inside a silo when the orifices are wide apart. Kamath \textit{et al.}\cite{multi5} studied mixing characteristics in a silo having multiple orifices where it is noticed that an intermittent flow through a narrow orifice due to its small size influences the mixing of particles discharged.

In the silo problems involving either lateral orifice or multiple orifices, mostly the systems were involving spherical particles or disc particles. However, understanding the dynamics of mixtures involving non-spherical particles is more useful for practical applications since in reality, the systems involve mixtures of particles varying in size and shape. In this work, we studied the flow of a mixture of dumbbells and discs in two eccentric silo flow situations. In the first case, we analyzed how the fraction of dumbbells influence the dynamics of the particles flowing through a lateral orifice. Whereas in the second case, we studied how the distance between the two orifices placed on the silo base affects the flow dynamics. Moreover, we have presented time-averaged flow fields of various parameters in both the cases which can be verified with the results obtained from the continuum models using $\mu(I)$ rheology. The paper is organised in the following way: in the next section, geometries of both the silos and the simulation technique are explained. In Section \ref{sec:lateral}, the results pertaining to the $lateral$ orifice case are discussed and in section \ref{sec:multi}, the results obtained for the $multiple$ orifice case are elucidated. Finally, in section \ref{sec:conclusion}, our important findings are summarized.

\begin{figure}
 \centering
 \includegraphics[width=0.4\linewidth]{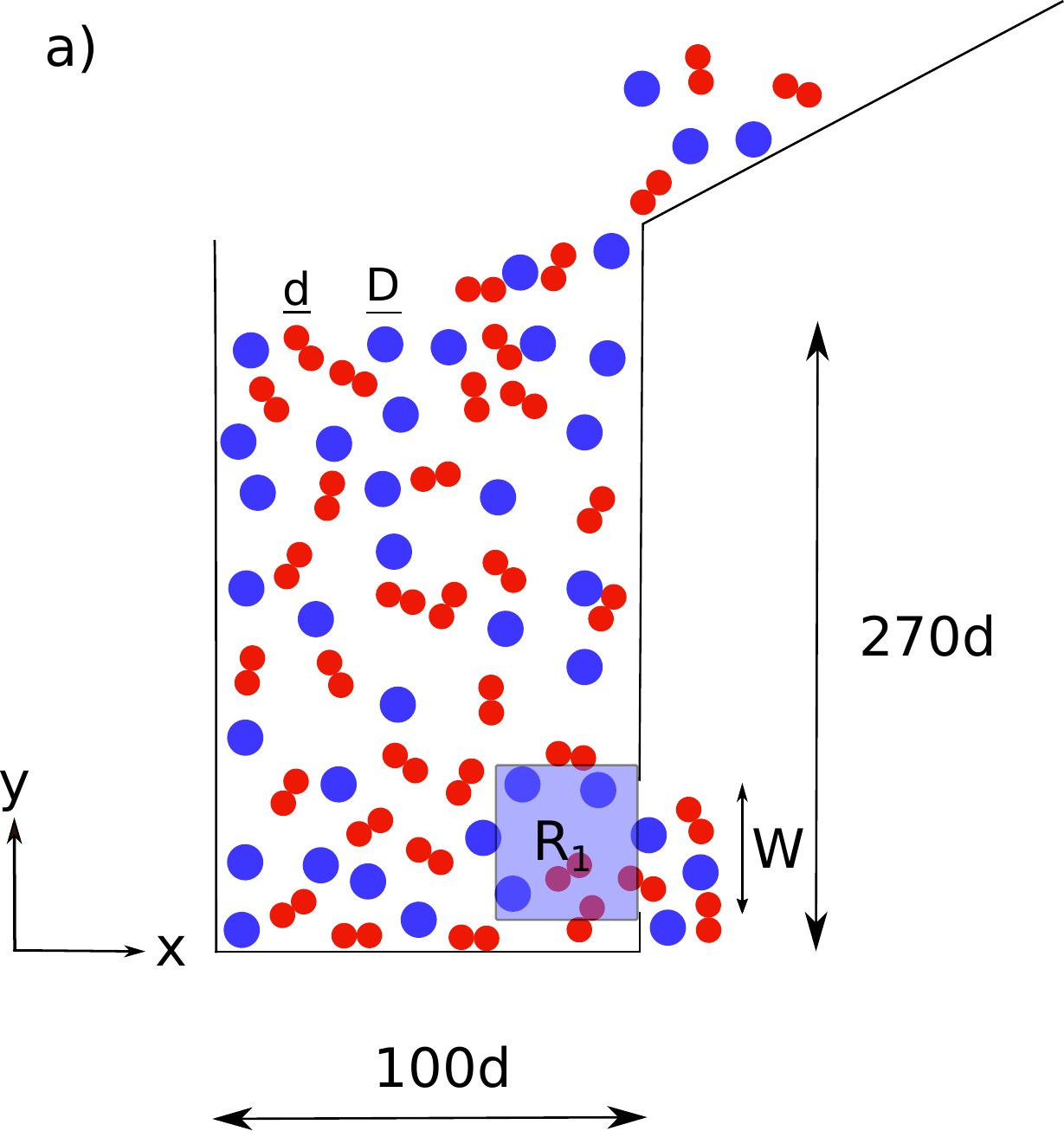}
 \hspace{0.5cm} 
 \includegraphics[width=0.4\linewidth]{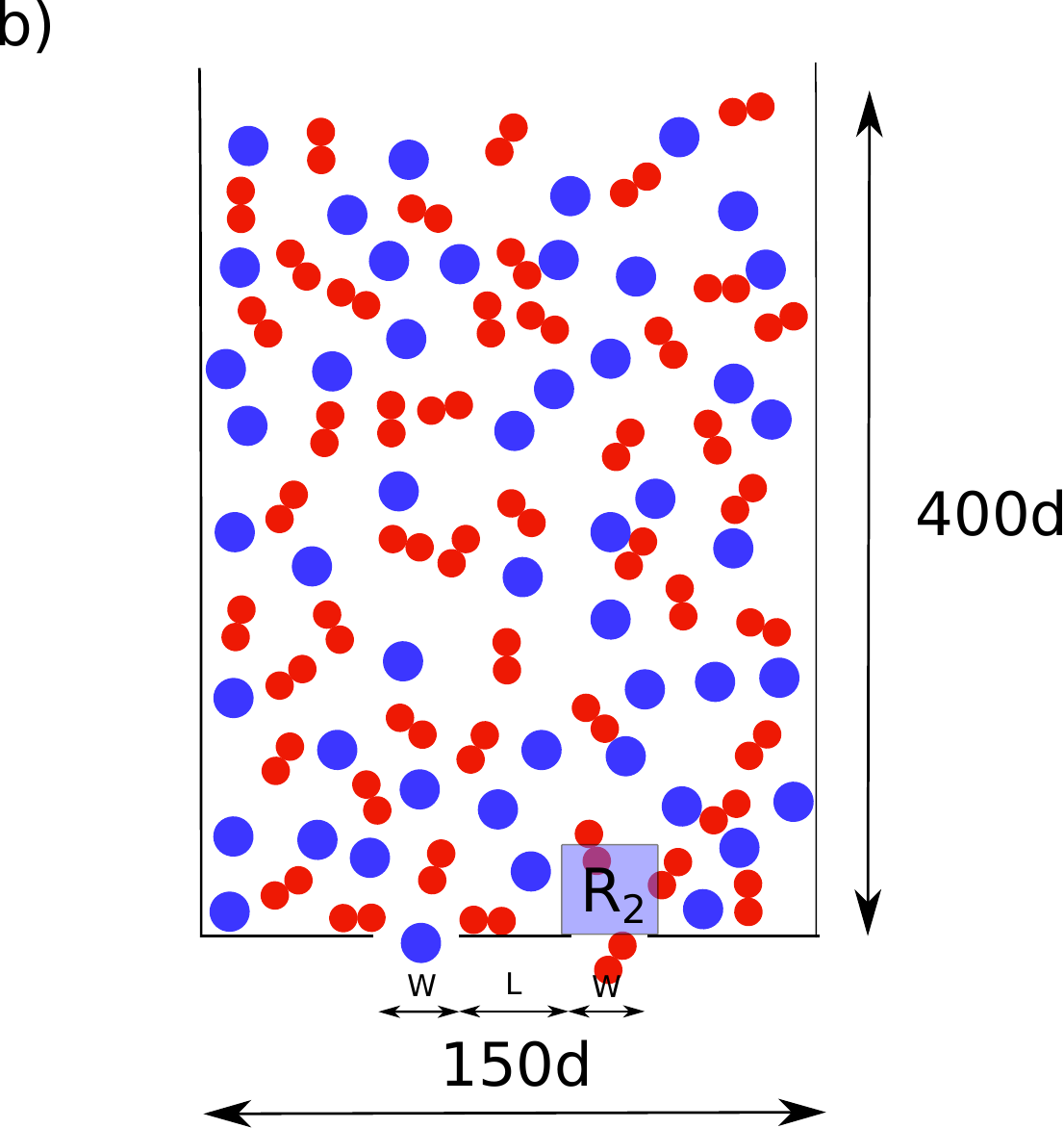}
 \caption{a) Schematic representation of a silo with a) an orifice placed on the \textit{sidewall} and b) \textit{multiple orifices} placed on the silo base. Here, the blue circles indicate discs and the red ones that of dumbbells. We have taken the area of the disc same as that of the dumbbell. The diameter of each circle of a dumbbell is $d$ and the diameter of each disc is $D=\sqrt{2}d$. The orifice width is given by $W$ in both the images and $L$ is the distance between two orifices. Few parameters in our work are analysed in the regions $R_1$ and $R_2$. The region $R_1$ has a length of $5\sqrt{2}d$ in $x$ direction and $W+2\sqrt{2}d$ in y-direction and the region $R_2$ has a length of $W+2\sqrt{2}d$ in $x$ direction and $5\sqrt{2}d$ in y-direction. Please note that the orifice on the $sidewall$ is placed at a height of $3.5d$ from the silo base to avoid the effect of the base. Origin is located at the centre of silo base for both the cases.}
 \label{fig:sim}
\end{figure}

\section{Simulation Methodology\label{sec:Sim}}

\begin{table}[h]
\centering
\small
  \caption{\ The constants used in our numerical simulations}
  \label{tbl:para}
  \begin{tabular*}{0.3\textwidth}{@{\extracolsep{\fill}}ll}
    \hline
    Simulation parameters & Values \\
    \hline
    $K_n$ 		& $2.00\times10^6 \rho dg$	  \\
        $K_t$ 		& $2.45 \times10^6 \rho dg$   \\
        $\gamma_n$ 	& $1000 \sqrt{g/d^3}$ 		  \\
        $\gamma_t$ 	& $1000 \sqrt{g/d^3}$         \\
        $\mu$ 		& $0.5$				  		  \\
        $timestep$ 	& $10^{-4}\sqrt{d/g}$         \\  
    \hline
  \end{tabular*}
\end{table}

We used the discrete element method (DEM) \citep{Dem} to study the influence of the orifice location and the fraction of dumbbells on the dynamics of granular mixtures. Figure  \ref{fig:sim} shows the schematic representation of two-dimensional silos differing in orifice positioning. In the first case, a single orifice is placed on the sidewall whereas, in the other one, two orifices are placed on the silo base. In both the cases, mixtures of discs and dumbbells (two circles are fused to form a rigid body) are analysed. The area of each disc is taken same as that of a dumbbell. Thus, the diameter of the disc is $D=\sqrt{2}d$, where $d$ is the diameter of a single circle of a dumbbell. In figure \ref{fig:sim}a case, $N=15000$ particles are placed at arbitrary locations with random orientations inside a silo confined by the walls at $x=\pm50d$ and $y=0$. We ensured that there are no overlaps among the particles. A gravity of magnitude $g$ is applied in the negative $y$ direction. Consequently, the particles got settled (KE $\approx$ 0.0) and the bed height was found to be close to $270d$. At time $t=0$, an orifice of width $W$ is opened on the sidewall so that particles can discharge out of the silo. Please note that the orifice is placed on the right side wall of the silo and at a height of $3.5d$ to lessen the effect of silo base on the flow dynamics. Periodic boundary conditions (PBC) were applied in $y$ direction and the particles discharging out of the silo were placed on top of the granular bed at random positions with reduced velocities. In our work, few parameters are analysed in the region $R_1$ which is just beside the orifice and having a length of $5\sqrt{2}d$ in $x$-direction and $W+2\sqrt{2}d$ in $y$-direction. In figure \ref{fig:sim}b case, the silo is confined by the walls at $x=\pm75d$ and $y=0$ and it consists of $N=33750$ particles with a bed height close to $400d$. The same procedure is employed for creating the initial configurations of both the silos. The origin is located at the centre of the silo base for both the cases.  At time $t=0$, two orifices on the silo base which are equidistant from the origin and are separated by a distance of $L$ and each of width $W$ were opened and PBC was applied in the $y$ direction. A few parameters are calculated in the region $R_2$ which is having a length of $W+2\sqrt{2}d$ in the $x$ direction and $5\sqrt{2}d$ in $y$ direction and is located just above one of the orifices.

One of the main advantages of the DEM technique is it stores individual data of each particle which helps in understanding particle level dynamics. In this technique, positions and velocities of each particle are updated at regular intervals by integrating equations of motion using the velocity Verlet algorithm. In the equations of motion, gravitational and contact forces are the only forces that are considered. The normal and tangential components $F_{ij}^n$, $F_{ij}^t$ of contact forces on a particle $i$ due to particle $j$ is computed by using contact force model \citep{Brilliantov} as 

$$F_{ij}^n=\sqrt{R_{\textrm{eff}} \delta_{ij}}(K_n \delta_{ij} \boldsymbol{\hat{r}_{ij}}-m_{\textrm{eff}}\gamma_n \boldsymbol{v_{ij}^n} )$$

$$F_{ij}^t=-\textrm{min}\bigg(\sqrt{R_{\textrm{eff}} \delta_{ij}}(K_t \boldsymbol{\Delta s_{ij}}+m_{\textrm{eff}}\gamma_t \boldsymbol{v_{ij}^t} ),\mu F_{ij}^n\textrm{}\bigg)$$

Here, $R_{\textrm{eff}}=\sqrt{\frac{R_iR_j}{R_i+R_j}}$ and $m_{\textrm{eff}}=\sqrt{\frac{m_im_j}{m_i+m_j}}$ are the effective radius and effective mass of the particles $i,j$ in contact where $R_i,R_j$ are radii and $m_i,m_j$ are masses of respective particles. The overlap $\delta_{ij}=R_i+R_j-|R_{ij}|$ must be non-negative for two particles to be in contact. Here, $|R_{ij}|$ is the distance between the centres of two particles. The subscripts or superscripts consisting of $n,t$ represents the normal or tangential components of the respective parameters. The elastic constant and damping coefficient are denoted by $K$ and $\gamma$ and $\boldsymbol{\hat{r}_{ij}}$ is the unit vector in the direction of line joining the centres of two particles. Moreover, $\boldsymbol{v_{ij}}$ is the relative velocity, $\boldsymbol{\Delta s_{ij}}$ is the tangential displacement vector and $\mu$ is the coefficient of friction. The values of various constants used in our numerical simulations are shown in Table \ref{tbl:para}. The positions and velocities of dumbbells are the centre of mass positions and centre of mass velocities. The force on each dumbbell is computed by adding the forces on both the circles of a dumbbell. The torque on each dumbbell is the sum of torques on both the circles of a dumbbell. The force acting on each circle of a dumbbell due to the other circle of the same dumbbell is ignored. All simulations were performed using LAMMPS \citep{lammps} package and the progress of the simulations were visualized using OVITO \citep{Ovito} package.

\section{Results and Discussion\label{sec:results}}

In this section, we will explain the results obtained for a mixture of discs and dumbbells flowing out of a two-dimensional silo. This section consists of two subsections. In Section. \ref{sec:lateral}, we elucidated the effect of fraction of dumbbells on the dynamics of particles flowing through a lateral orifice and in Section. \ref{sec:multi}, we explained how the flow dynamics is influenced by the spacing between the orifices placed on the silo base. In this regard, the flow is characterized by parameters like mass flow rate $Q$, area fraction $\phi$, granular temperature $T_g$ etc. In this paper, wherever we use the term flow rate it means mass flow rate. The flow rate is calculated by using the least-square fitting method on the total mass of particles discharged versus time data. Area fraction is the ratio of the area occupied by the particles in a region of interest and the area of the region. Granular temperature $T_g$, a measurement of velocity fluctuations of particles \citep{grantemp,grantemp1}, in a region of interest is computed as $T_g = \frac{1}{3}\langle m\{(v_x-<v_x>)^2 + (v_y-<v_y>)^2\} + I (\Omega_z - <\Omega_z>)^2 \rangle$. Here, $m$ is mass of a particle, $v_x$ and $v_y$ are instantaneous velocities in $x$ and $y$ directions respectively. Moreover, $I$ is the moment of inertia, $\Omega_z$ is the rotational velocity in $z$ direction and $<.>$ corresponds to a spatio-temporal average over a specified region of interest.

\begin{figure}	
	\centerline{\includegraphics[width=1.0\linewidth]{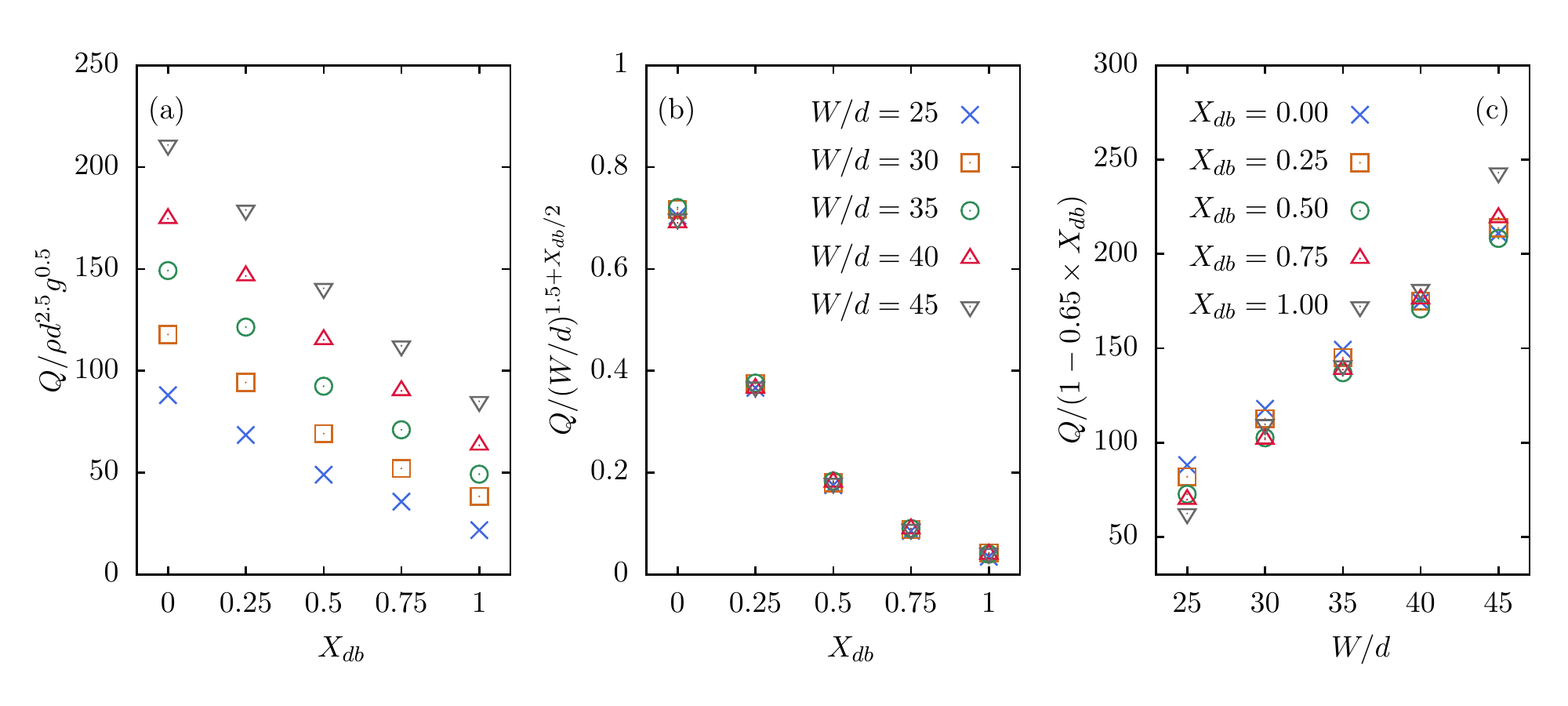}}
	\caption{\label{fig:pfgtl}a) Flow rate $Q$ as a function of the fraction of dumbbells $X_{db}$. b) Scaling of flow rate with the lateral orifice width $W/d$ at different $X_{db}$ and c) scaling of flow rate with the fraction of dumbbells $X_{db}$ at different $W/d$. Here, $d$ corresponds to the diameter of each of the circles of a dumbbell and $\rho$ is the particle density.}
\end{figure}

\begin{figure}
	\centerline{\includegraphics[width=1.0\linewidth]{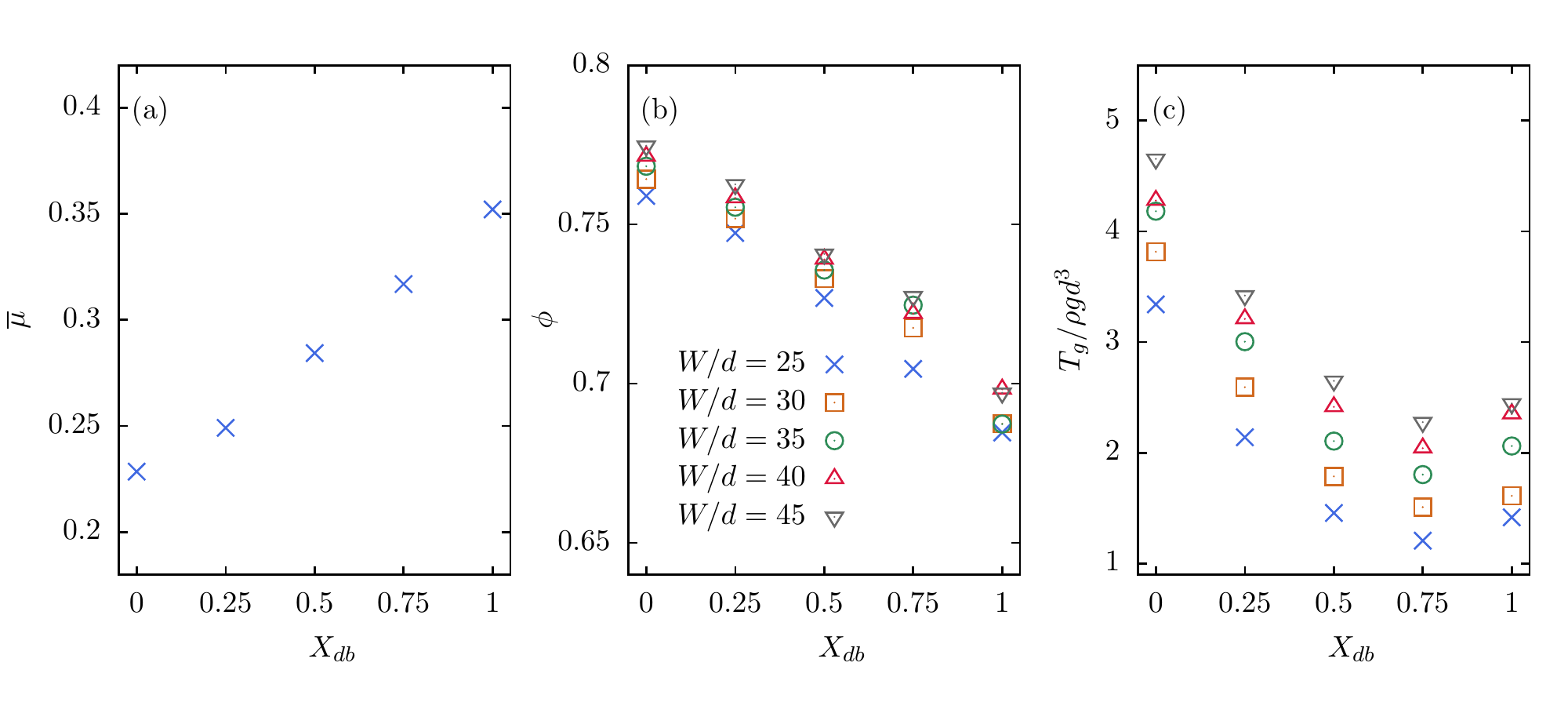}}
		\caption{\label{fig:fric} a) Dynamic friction $\overline{\mu}$, b) area fraction $\phi$ and c) granular temperature $T_g$ as a function of the fraction of dumbbells $X_{db}$ in the region $R_1$ which lies beside the lateral orifice. Note that $\overline{\mu}$ is computed for the lateral orifice case having a width $W/d=25$.}
\end{figure}

\subsection{Lateral orifice\label{sec:lateral}}

In this subsection, we explained how the fraction of dumbbells influences the flow dynamics while particles are flowing out of an orifice placed on the $sidewall$ of a silo. In this regard, the particulate flow is analysed for five different fractions of dumbbells $X_{db}$ ranging from 0.0 to 1.0. Moreover, five different widths of the orifice positioned on the sidewall ranging from $W/d$ = 25 to 45 are considered. Flow rate is observed to decrease with an increase in the fraction of dumbbells at all orifice widths (figure \ref{fig:pfgtl}a). This is due to an increase in the dynamic friction ($\overline{\mu}$ = shear stress/pressure) with an increase in the fraction of dumbbells $X_{db}$ as shown in figure \ref{fig:fric}a. The increase in the geometrical interlocking among the particles with an increase of $X_{db}$ might be the reason behind an increase in $\overline{\mu}$. The shear stress and pressure are obtained from the time-average flow fields which are explained in detail in the next section and the magnitude of $\overline{\mu}$ is spatio-temporal average over the region beside the orifice $R_1$. \citet{rodsphere} found a decrease in the flow rate with an increase in the number of rods in a rod-sphere mixture discharging through an orifice in the silo base. Beverloo's law  \cite{Beverloo_1961} states that the mass flow rate scales with $(W-kd)^{5/2}$ for a system of spherical particles flowing out of a three-dimensional silo, where $W,k,d$ are orifice width, shape coefficient and diameter of the disc. For a two-dimensional case, it can be derived that the mass flow rate scales with $(W-kd)^{3/2}$ and $k$ was found to be 1 for spherical or disc particles. In our case which involves a mixture of dumbbells and discs we tried to find whether the flow rate scales with orifice width or the fraction of dumbbells. We noticed that Beverloo's law was fitting reasonably well only when $X_{db}=0.0$ which involves only discs. So, we tried with a modified Beverloo's law to collapse the data. The flow rate is found to scale with $(W/d)^{1.5+X_{db}/2}$ at all fractions of dumbbells $X_{db}$ as shown in figure \ref{fig:pfgtl}b. Moreover, we noticed that flow rate scales with $1-0.65 \times X_{db}$ at different orifice widths as shown in figure \ref{fig:pfgtl}c.

\begin{figure}
\centerline{\includegraphics[width=1.0\linewidth]{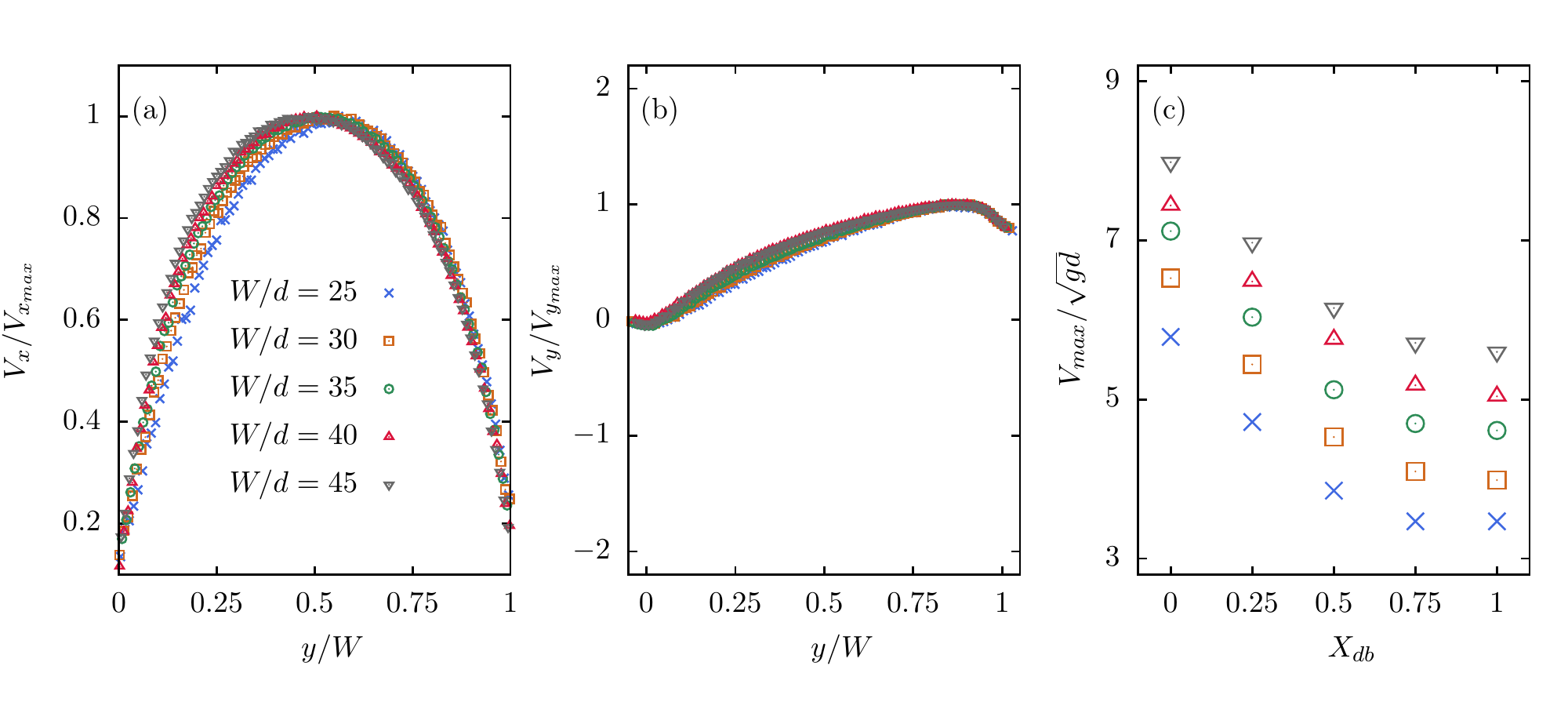}}
	\caption{\label{fig:flowl}a) Normalized horizontal $V_x/V_{x_{max}}$ and b) normalized vertical $V_y/V_{y_{max}}$ velocities as a function of normalized vertical positions $y/W$ where, $W$ is orifice width on the sidewall of the silo. c) Maximum velocity $V_{max}$ as a function of the fraction of dumbbells $X_{db}$ at different orifice widths $W/d$.}
\end{figure}

\begin{figure}[h]
\centering
\includegraphics[width=\linewidth]{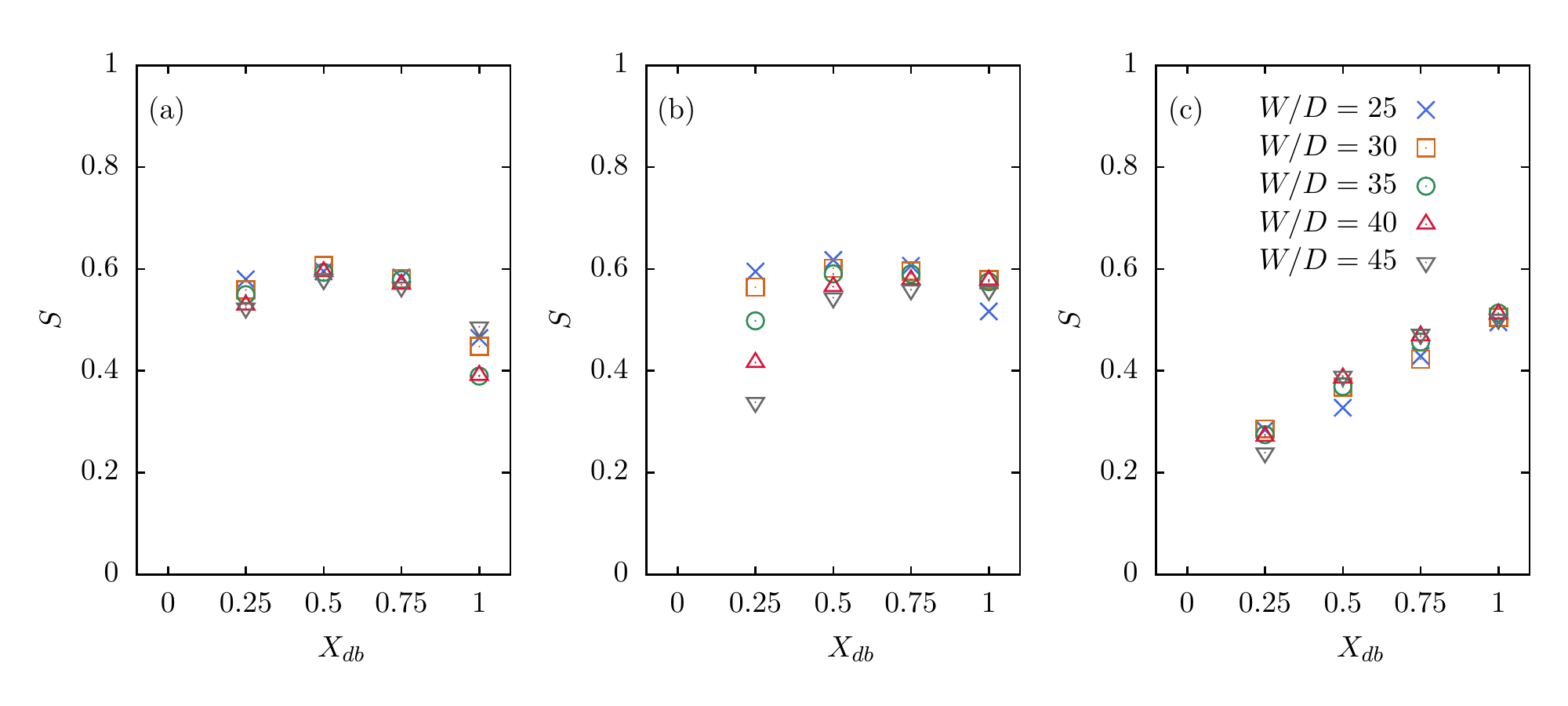}
\caption{\label{fig:ori} Orientational order parameter $S$ as a function of the fraction of dumbbells $X_{db}$ in the region a) $35<x<45$ and $10<y<W-5$, b) $15<x<25$ and $W-5<y<W+5$, c) $5<x<15$ and $60<y<70$ at different lateral orifice widths $W/d$.}
\end{figure}

Area fraction and granular temperature are analysed in the region $R_1$ which lies just beside the orifice as shown in figure \ref{fig:sim}a. With an increase in the fraction of dumbbells, the voids formed among the particles increases resulting in a decrease in the area fraction (figure \ref{fig:fric}b). Recently, \citet{rodsphere} noticed a similar result where the packing fraction was found to decrease with an increase in the fraction of rod-like particles of a spherical-rod mixture flowing through the silo base. Granular temperature is found to decrease with an increase in $X_{db}$ as shown in figure \ref{fig:fric}c due to a decrease in the velocity fluctuations of the particles in the region $R_1$. This can be explained by a decrease in the velocities of colliding particles with an increase in $X_{db}$ due to geometrical interlocking among the particles. Self-similar velocity profiles are noticed when normalized horizontal and normalized vertical velocities are plotted against normalized vertical position (figure \ref{fig:flowl}a, \ref{fig:flowl}b). Zhou \textit{et al.} \cite{selfsimilar1} noticed self-similar velocity profiles for a system of bidisperse spherical particles discharging out of an orifice placed on the silo base. Moreover, Janda \textit{et al.} \cite{selfsimilar2} reported self-similar velocity profiles in a system of monodispersed spherical beads discharging out of a two-dimensional silo from an aperture on a silo base. The stagnant zone or a set of almost stationary particles that are present below the lateral orifice \citep{Zhou2017} hinders the free-flow of particles. Moreover, with an increase in the fraction of dumbbells the stagnant zone offers more resistance to the flow  resulting in a decrease in the flowing zone (Please refer to the supplementary information$\dag$). This yields in a decrease in the maximum velocity $V_{max}$ (figure \ref{fig:flowl}c) in the region beside the orifice ($R_1$) with an increase in $X_{db}$. This result complements a decrease in the flow rate of particles with an increase of $X_{db}$ (figure \ref{fig:pfgtl}a). We computed orientational order parameter to check whether the particles are aligned in a particular direction at three different regions. The first region is beside the orifice ($35<x<45$ and $10<y<W-5$), the second one is slightly away from the orifice ($15<x<25$ and $W-5<y<W+5$) and the third one is taken in the bulk ($5<x<15$ and $60<y<70$). Please note that the origin ($x=0,y=0$) is located at the centre of the silo base. The orientational order parameter $S$ is computed as $S=2<cos^2\theta> -1$ where $\theta$ is the difference of the orientation of dumbbell and the director vector. The orientation of dumbbell is computed as the angle between the larger axis of the dumbbell and the horizontal axis $y=0$. The director vector indicates the flow direction. The orientational order parameter is noticed to decrease with an increase in the fraction of dumbbells in the region close to the orifice (figure \ref{fig:ori}a). A disordered packing of elongated particles result in more voids and consequently lower area fraction. In our case, this decrease in the area fraction might be another reason for a decrease in the flow rate with an increase in the fraction of dumbbells. In the bulk, as the particles are closely packed, the orientational order parameter is found to increase with an increase in the fraction of dumbbells (figure \ref{fig:ori}c). However, in the intermediate region, ordering of particles is almost unaffected by the fraction of dumbbells. To explore the flow dynamics at various lateral orifice widths $W/d$ we have analysed spatial flow fields of various parameters by using coarse-graining method \citep{weinhart,coarsegrain} in the next subsection.

\subsubsection{Flow fields at various lateral orifice widths $W/d$ \label{sec:coarsew}}

We have generated continuous macroscopic flow fields by using discrete microscopic data like positions, velocities etc,. of the individual particles. Here we employeed coarse-graining technique as suggested by \citep{coarsegrain} for a two-dimensional silo. The area fraction $\phi(t)$, rotational velocity $\Omega(t)$, fluctuations in rotational velocity $\Omega_{fl}(t)$, velocity $\bm{v}(t)$, granular temperature $T_g(t)$, stress tensor $\sigma_{ij}(t)$ and pressure $P(t)$ at any point of time $t$ and at any position $p$ having a position vector $\bm{r_p}$ is computed as follows:

\begin{equation}
\phi(t) = \left[ \sum_{i=1}^{n} \frac{\rho \pi d_i^2}{4} \mathcal{W}(\bm{r_p}-\bm{r_i(t)}) \right]/\rho
\end{equation}

\begin{equation}
\Omega(t) = \left[ \sum_{i=1}^{n} \frac{\rho \pi d_i^2}{4} \Omega_{z_i} \mathcal{W}(\bm{r_p}-\bm{r_i(t)}) \right]/\rho \phi 
\end{equation}

\begin{equation} 
\Omega_{fl}(t) = \left[ \sum_{i=1}^{n} \frac{\rho \pi d_i^2}{4} (\Omega_{z_i}-\Omega)^2 \mathcal{W}(\bm{r_p}-\bm{r_i(t)})  \right]/\rho \phi 
\end{equation}

\begin{equation}
\bm{v}(t) = \left[\sum_{i=1}^{n} \frac{\rho \pi d^2_i}{4} \bm{v_i} \mathcal{W}(\bm{r_p}-\bm{r_i(t)}) \right]/\rho \phi
\end{equation}

\begin{equation}
T_g(t) = \frac{\sum_{i=1}^{n} \frac{\rho \pi d^2_i}{4}|\bm{v_i}-\bm{v}|^2 \mathcal{W}(\bm{r_p}-\bm{r_i(t)})}{2\rho \phi}
\end{equation}

\begin{equation}
\bm{\sigma_{ij}}(t)= \sum_{i=1}^{n}\sum_{j=i+1}^{n}(\bm{F^{ij} \times r_{ij}}) \int_{s=0}^{1} \mathcal{W}(\bm{r_p}-\bm{r_i(t)}+s\bm{r_{ij}})ds
\end{equation}

\begin{equation}
P(t) =\frac{-tr(\sigma_{ij}(t))}{2}
\end{equation}

\begin{equation}
\mathcal{W}(\bm{r})=\frac{1}{\pi w^2}e^{-\bm{r}^2/w^2} 
\end{equation}

Here, $\rho$, $d_i$ and $\bm{r_i}$ are density, diameter and position vector of the $i^{th}$ particle and  $\mathcal{W}(\bm{r})$ is the coarse-graining function which weighs the parameters over space from discrete data with $w=1.414$. At any position $p$ with position vector $\bm{r_p}$, the parameters are evaluated only when $|\bm{r_p}-\bm{r_i}|<3w$ where $\bm{r_i}$ is the postion vector of $i^{th}$ particle. Moreover, $\phi$, $\Omega$, $\Omega_{fl}$, $\bm{v}$, $T_g$, $\sigma_{ij}$ and $P$ are the time-averaged quantities of the respective parameters. All the flow fields demonstrated in this subsection are averaged over 2500 frames and they corresponds to the region: $-38.5\le x \le 48.5$ and $1.5\le y \le 148.5$. For each parameter we have produced five flow fields each corresponding to a different lateral orifice width $W/d$ ranging from 25 to 45. The fraction of dumbbells is $X_{db}$ = 0.5 for all the flow fields demonstrated in this subsection.

\begin{figure}	
	\centering
	\subfloat[][]{\includegraphics[clip,trim=5.15cm 1.2cm 4.65cm 0.6cm,width=0.14\linewidth]{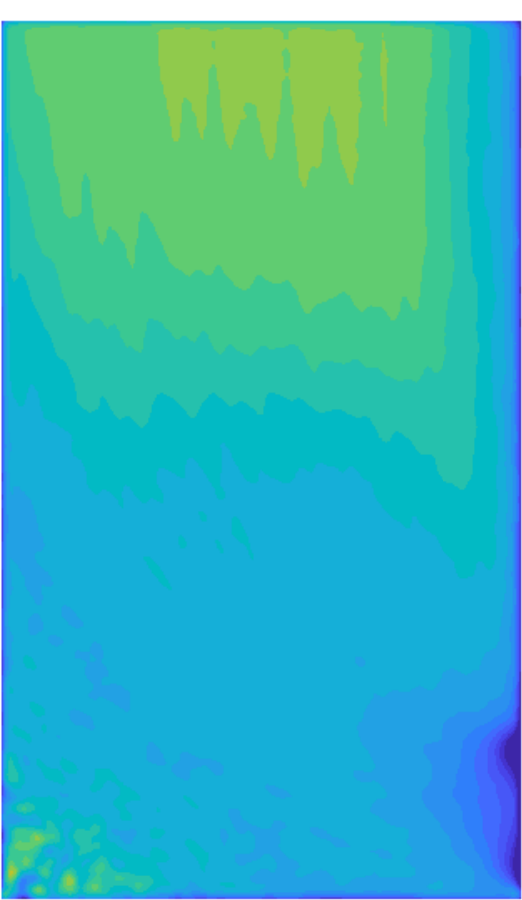}}
	\hspace{0.25cm}
	\subfloat[][]{\includegraphics[clip,trim=5.15cm 1.2cm 4.65cm 0.6cm,width=0.14\linewidth]{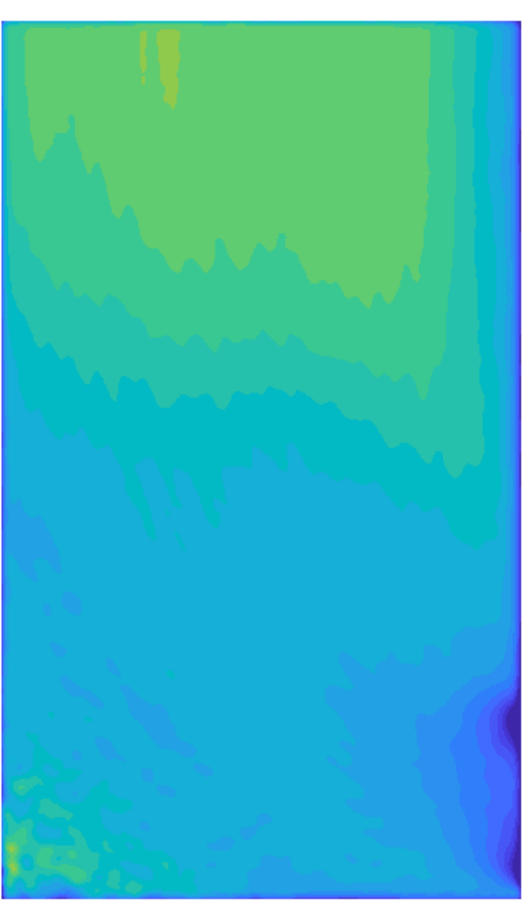}}
	\hspace{0.25cm}
	\subfloat[][]{\includegraphics[clip,trim=5.15cm 1.2cm 4.65cm 0.6cm,width=0.14\linewidth]{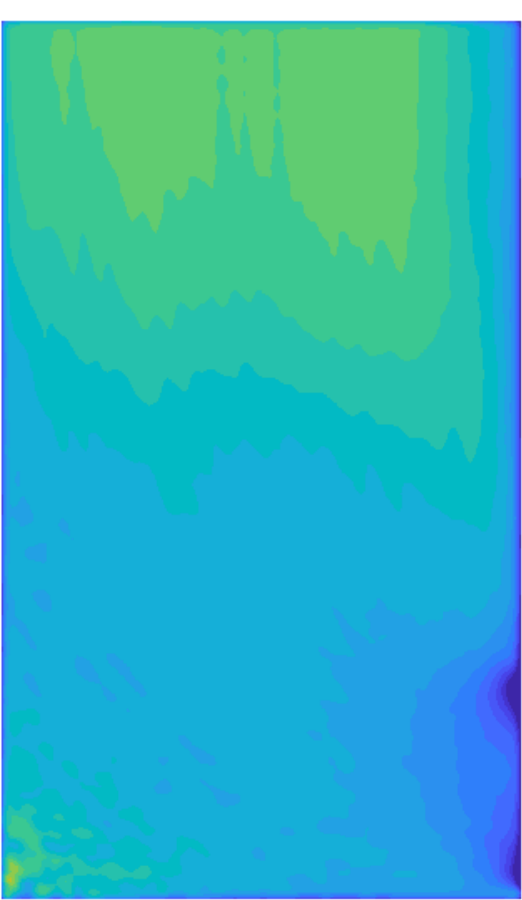}}
	\hspace{0.25cm}
	\subfloat[][]{\includegraphics[clip,trim=5.15cm 1.2cm 4.65cm 0.6cm,width=0.14\linewidth]{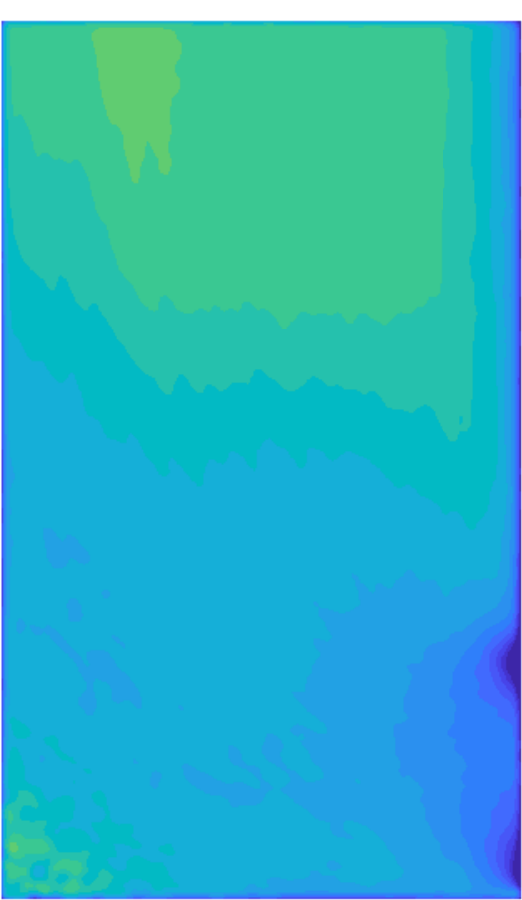}}
	\hspace{0.25cm}
	\subfloat[][]{\includegraphics[clip,trim=5.15cm 1.2cm 4.65cm 0.6cm,width=0.14\linewidth]{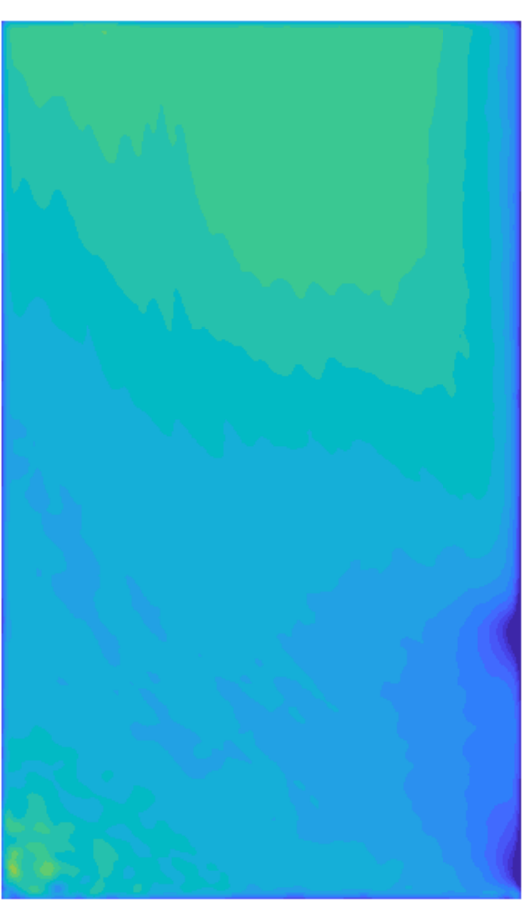}}
	\hspace{0.25cm} 
	\includegraphics[width=0.07\linewidth]{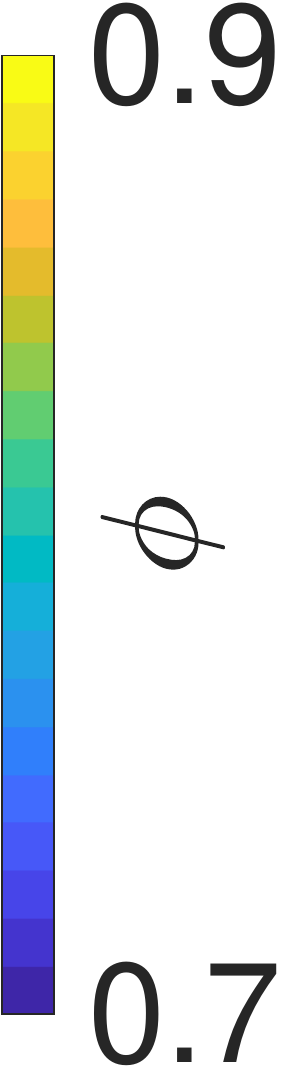}
	\caption{\label{fields:pf}Spatial distribution of area fraction $\phi $ at an orifice width $W/d$ = a) 25, b) 30, c) 35, d) 40 and e) 45 on the sidewall with fraction of dumbbells $X_{db}=0.5$.}
\end{figure}

\begin{figure}	
	\centering
	\subfloat[][]{\includegraphics[clip,trim=5.15cm 1.2cm 4.65cm 0.6cm,width=0.14\linewidth]{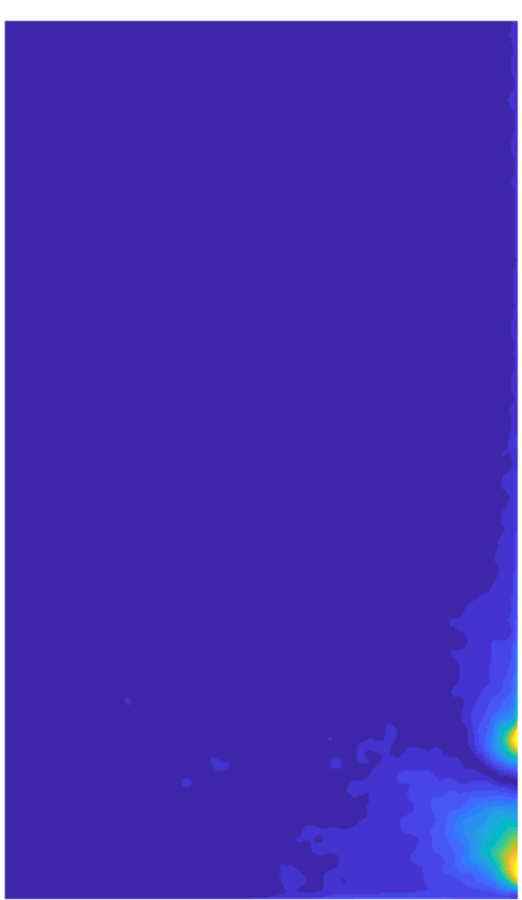}}
	\hspace{0.25cm}
	\subfloat[][]{\includegraphics[clip,trim=5.15cm 1.2cm 4.65cm 0.6cm,width=0.14\linewidth]{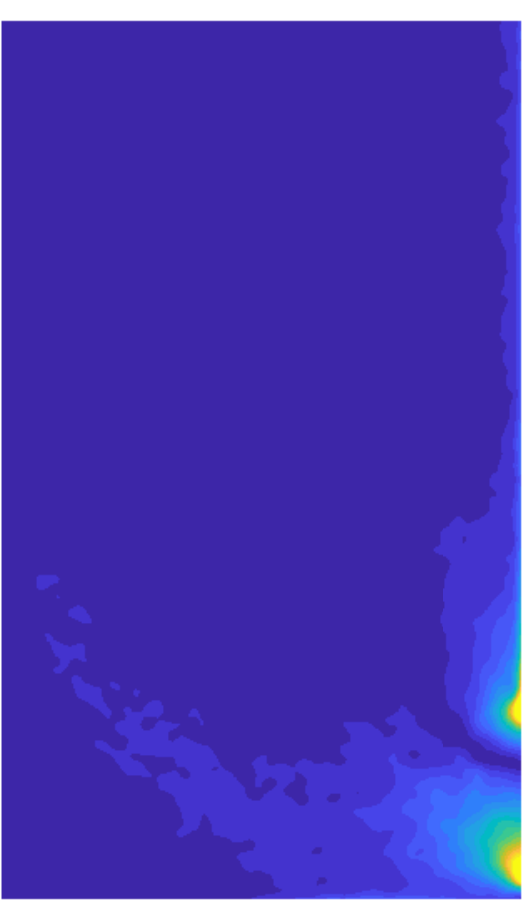}}
	\hspace{0.25cm}
	\subfloat[][]{\includegraphics[clip,trim=5.15cm 1.2cm 4.65cm 0.6cm,width=0.14\linewidth]{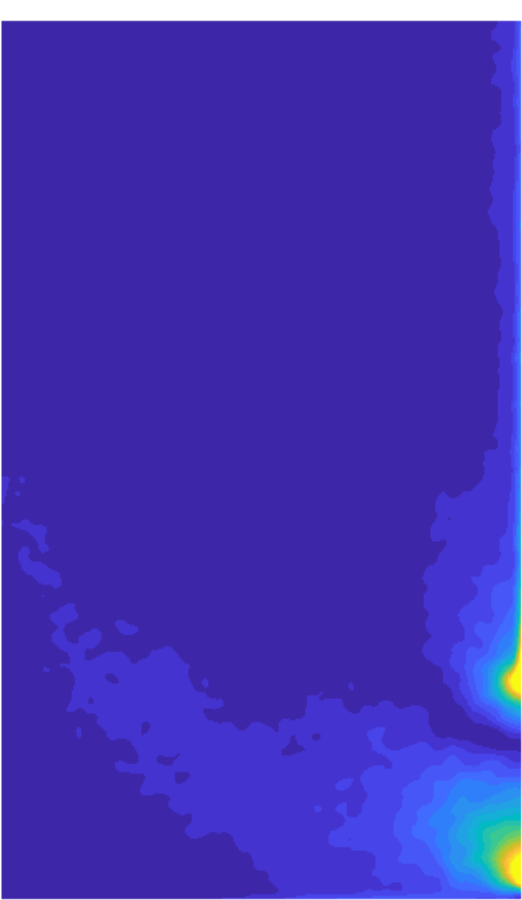}}
	\hspace{0.25cm}
	\subfloat[][]{\includegraphics[clip,trim=5.15cm 1.2cm 4.65cm 0.6cm,width=0.14\linewidth]{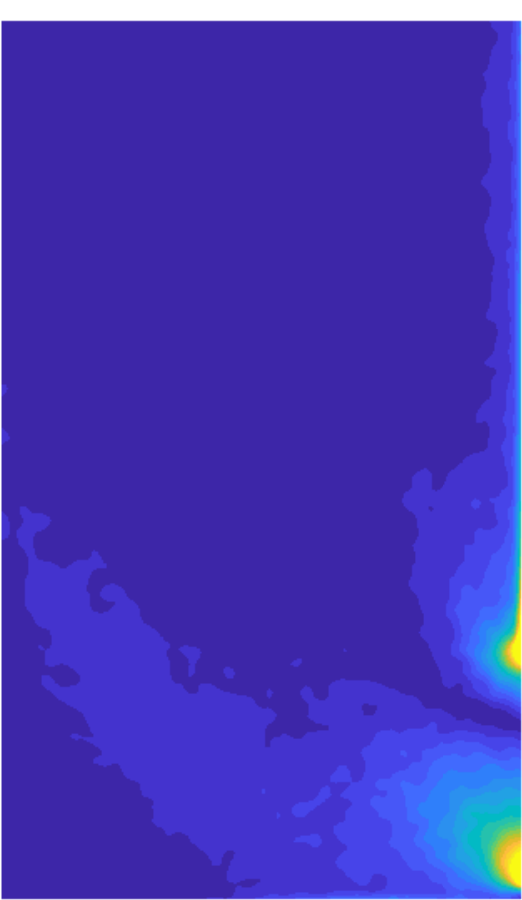}}
	\hspace{0.25cm}
	\subfloat[][]{\includegraphics[clip,trim=5.15cm 1.2cm 4.65cm 0.6cm,width=0.14\linewidth]{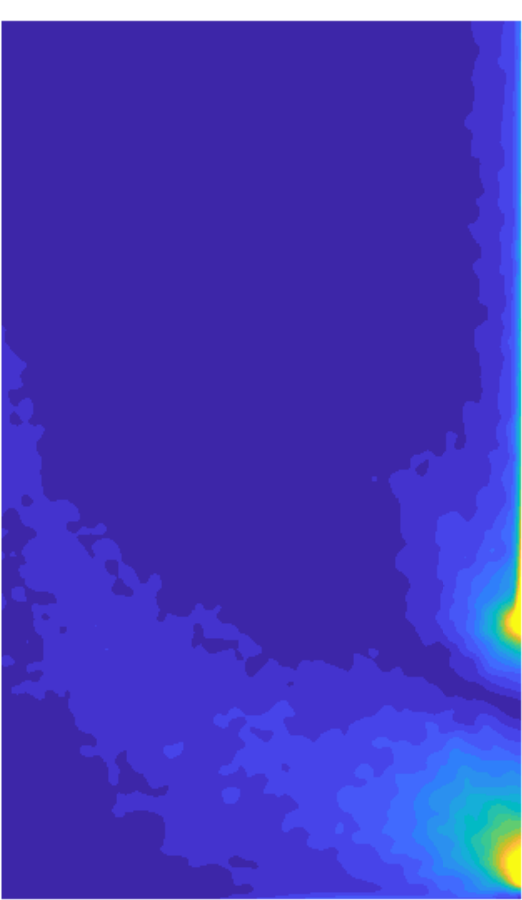}}
	\hspace{0.25cm} 
	\includegraphics[width=0.1\linewidth]{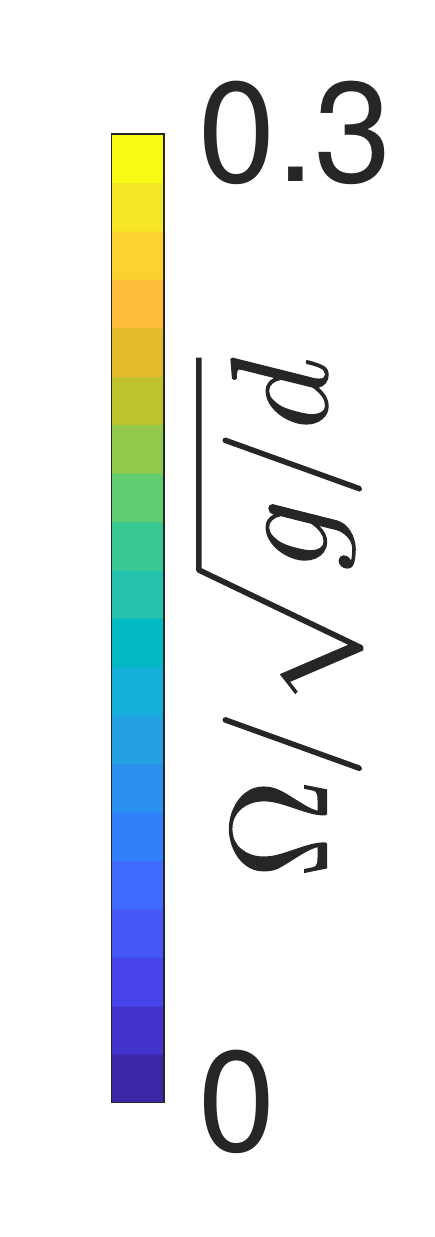}
	\caption{\label{fields:omeg}Spatial distribution of rotational velocity $\Omega $ at an orifice width $W/d$ = a) 25, b) 30, c) 35, d) 40 and e) 45 on the sidewall with fraction of dumbbells $X_{db}=0.5$.}
\end{figure}

\begin{figure}	
	\centering
	\subfloat[][]{\includegraphics[clip,trim=5.15cm 1.2cm 4.65cm 0.6cm,width=0.14\linewidth]{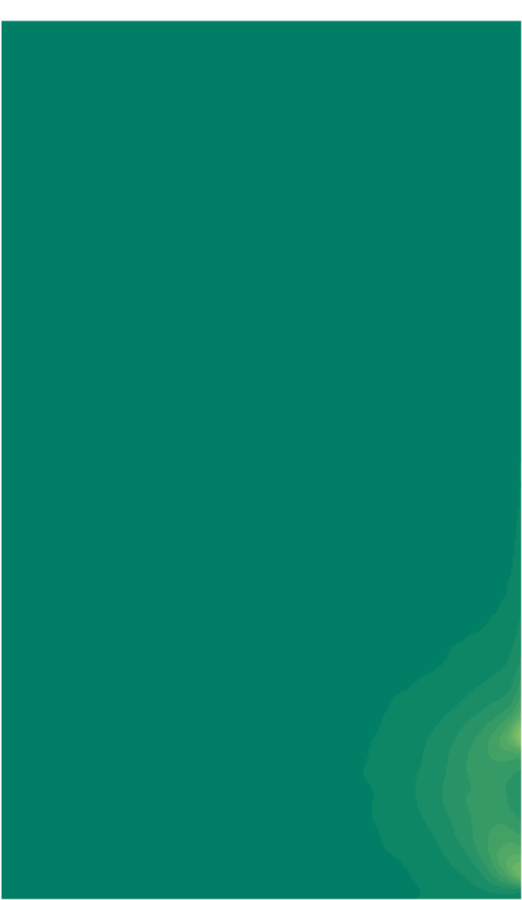}}
	\hspace{0.25cm}
	\subfloat[][]{\includegraphics[clip,trim=5.15cm 1.2cm 4.65cm 0.6cm,width=0.14\linewidth]{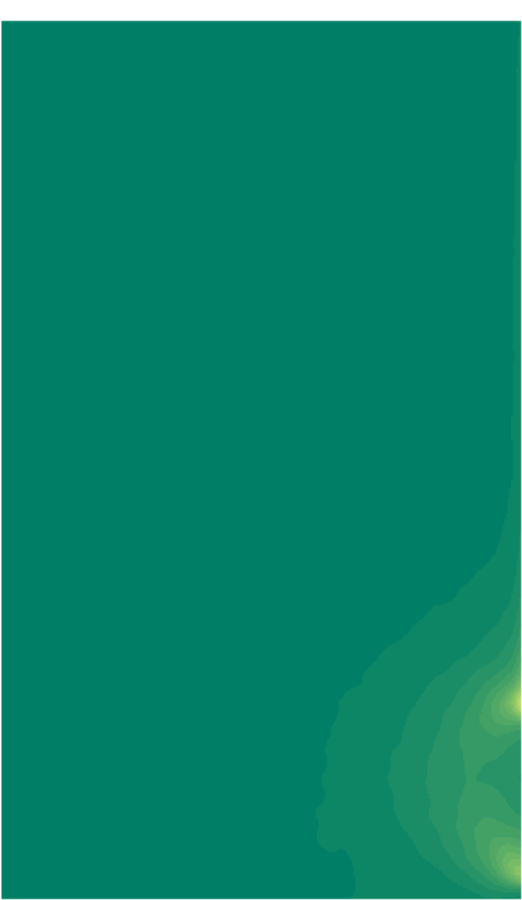}}
	\hspace{0.25cm}
	\subfloat[][]{\includegraphics[clip,trim=5.15cm 1.2cm 4.65cm 0.6cm,width=0.14\linewidth]{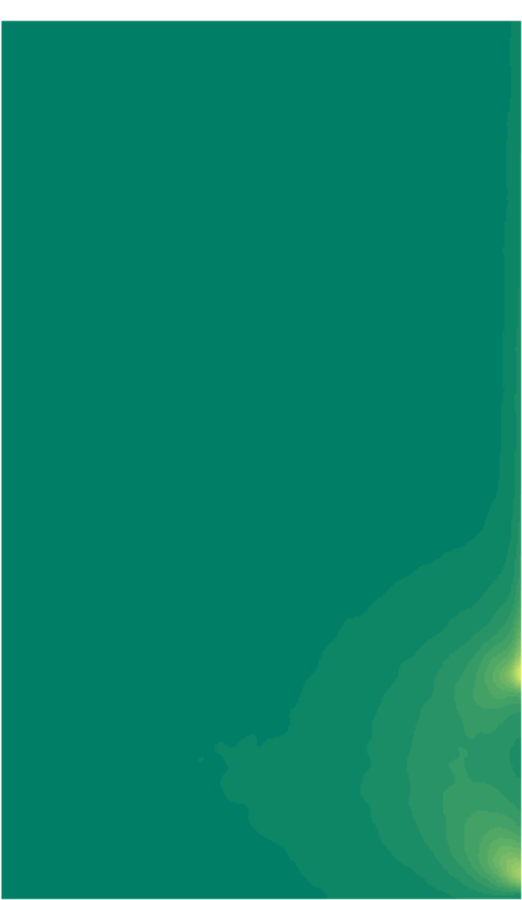}}
	\hspace{0.25cm}
	\subfloat[][]{\includegraphics[clip,trim=5.15cm 1.2cm 4.65cm 0.6cm,width=0.14\linewidth]{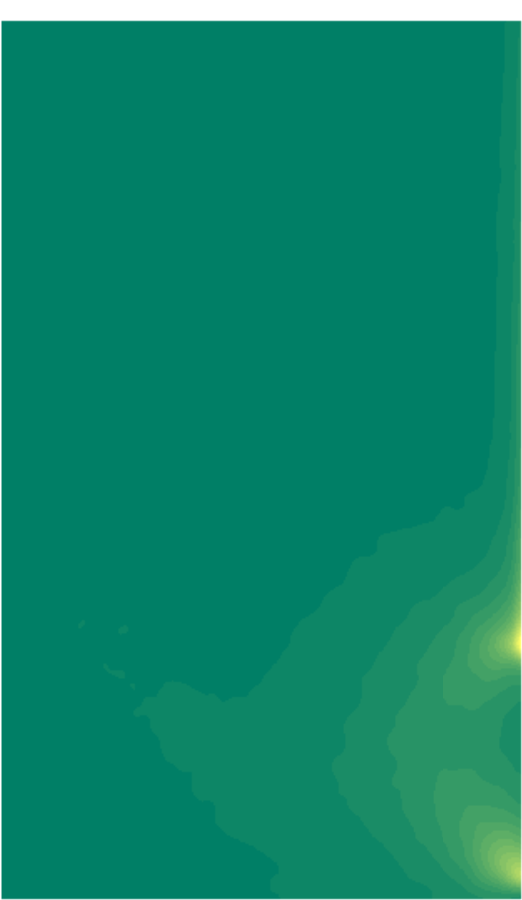}}
	\hspace{0.25cm}
	\subfloat[][]{\includegraphics[clip,trim=5.15cm 1.2cm 4.65cm 0.6cm,width=0.14\linewidth]{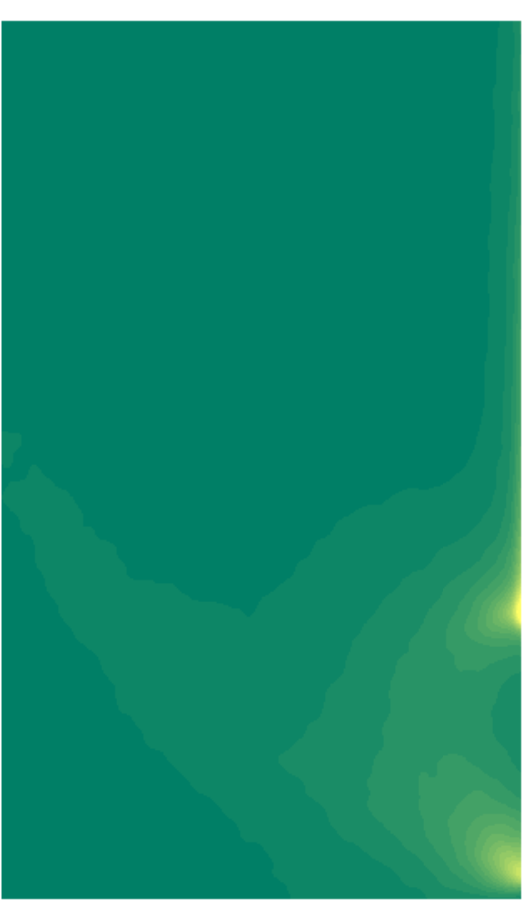}}
	\hspace{0.25cm} 
	\includegraphics[width=0.1\linewidth]{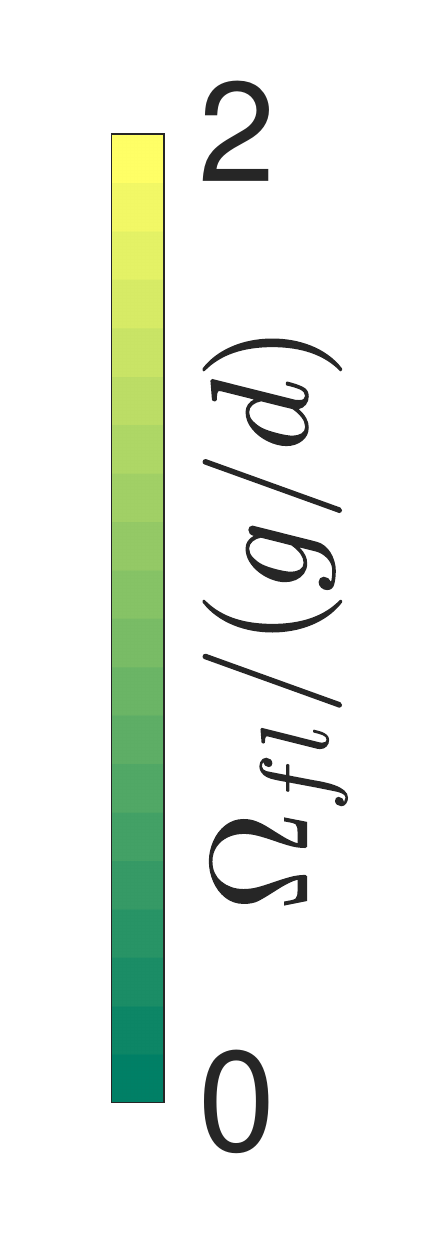}
	\caption{\label{fields:omegfluc}Spatial distribution of fluctuations in rotational velocity $\Omega_{fl}$ at an orifice width $W/d$ = a) 25, b) 30, c) 35, d) 40 and e) 45 on the sidewall with fraction of dumbbells $X_{db}=0.5$.}
\end{figure}

\begin{figure}	
	\centering
	\subfloat[][]{\includegraphics[clip,trim=5.15cm 1.2cm 4.65cm 0.6cm,width=0.14\linewidth]{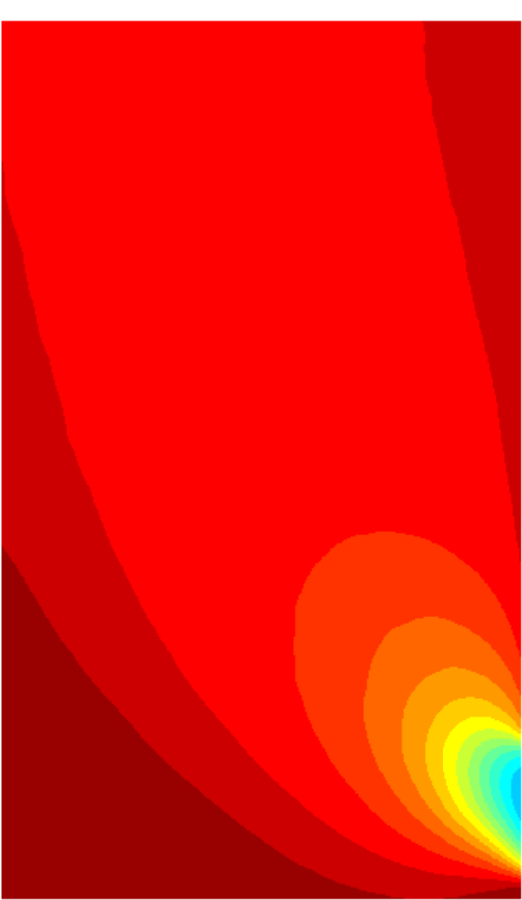}}
	\hspace{0.25cm}
	\subfloat[][]{\includegraphics[clip,trim=5.15cm 1.2cm 4.65cm 0.6cm,width=0.14\linewidth]{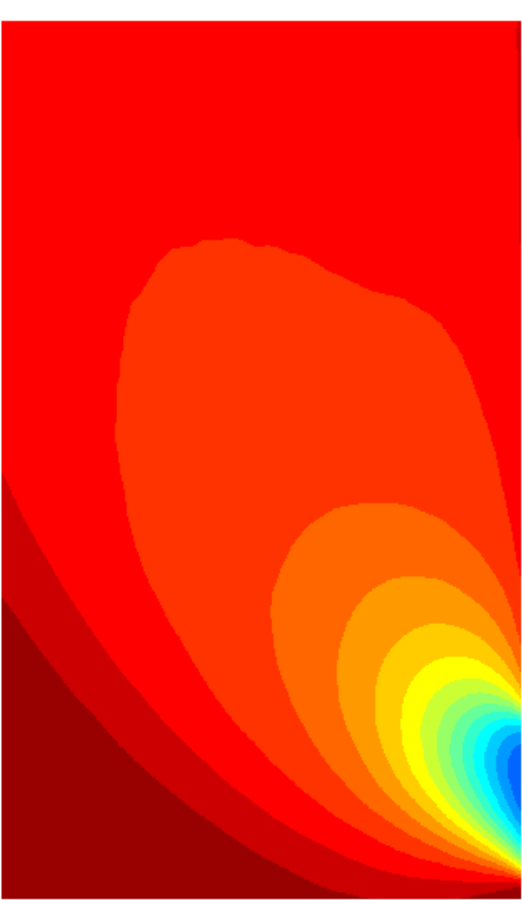}}
	\hspace{0.25cm}
	\subfloat[][]{\includegraphics[clip,trim=5.15cm 1.2cm 4.65cm 0.6cm,width=0.14\linewidth]{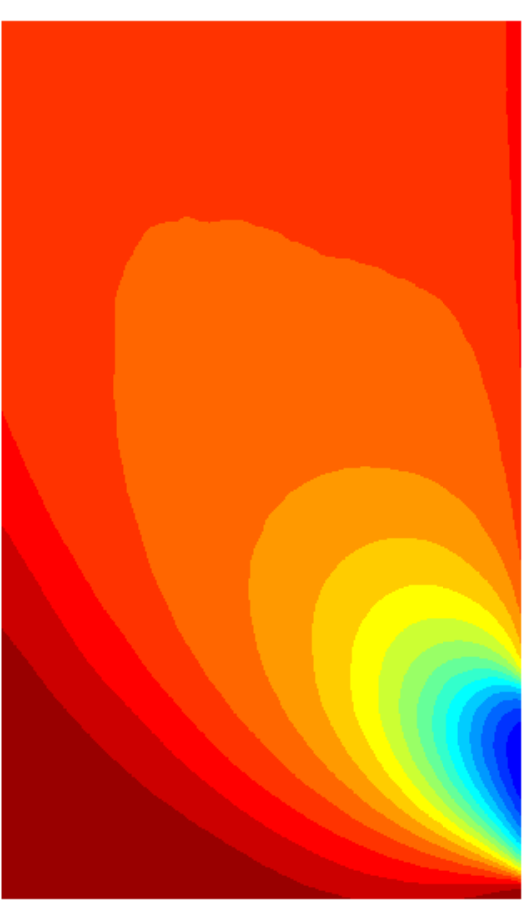}}
	\hspace{0.25cm}
	\subfloat[][]{\includegraphics[clip,trim=5.15cm 1.2cm 4.65cm 0.6cm,width=0.14\linewidth]{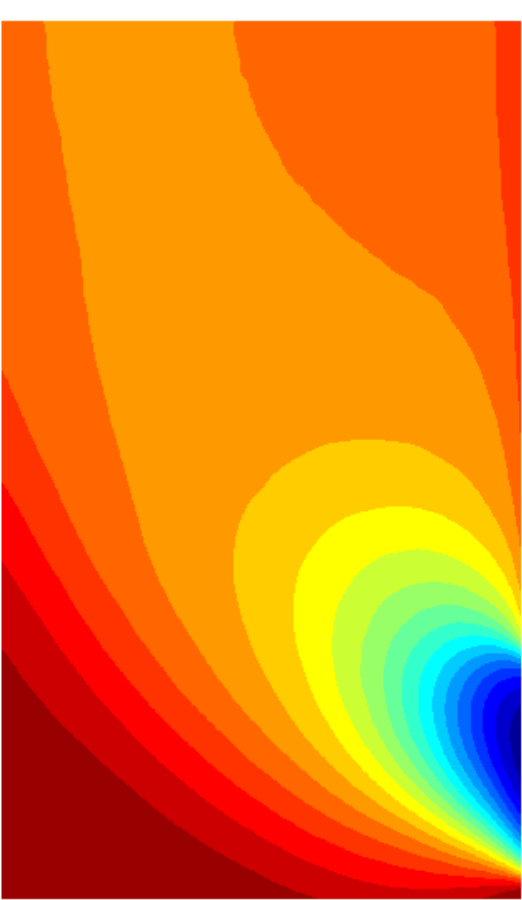}}
	\hspace{0.25cm}
	\subfloat[][]{\includegraphics[clip,trim=5.15cm 1.2cm 4.65cm 0.6cm,width=0.14\linewidth]{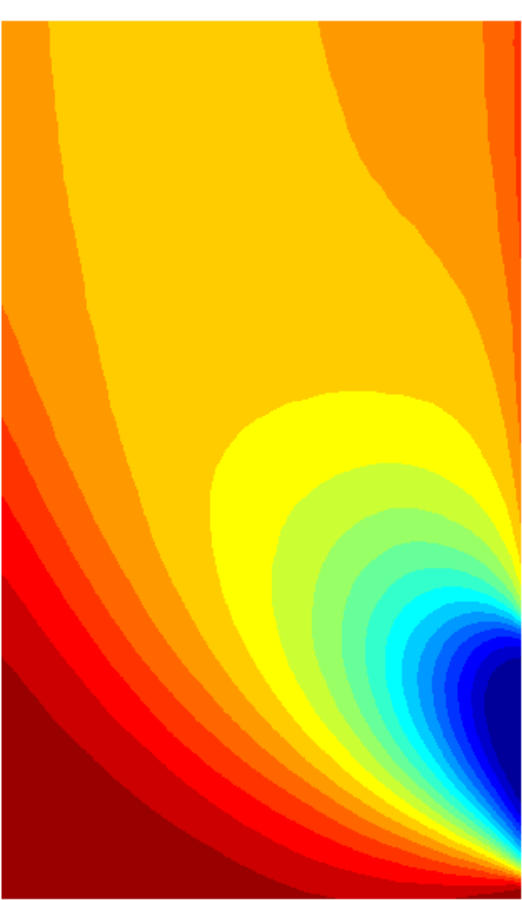}}
	\hspace{0.25cm} 
	\includegraphics[width=0.1\linewidth]{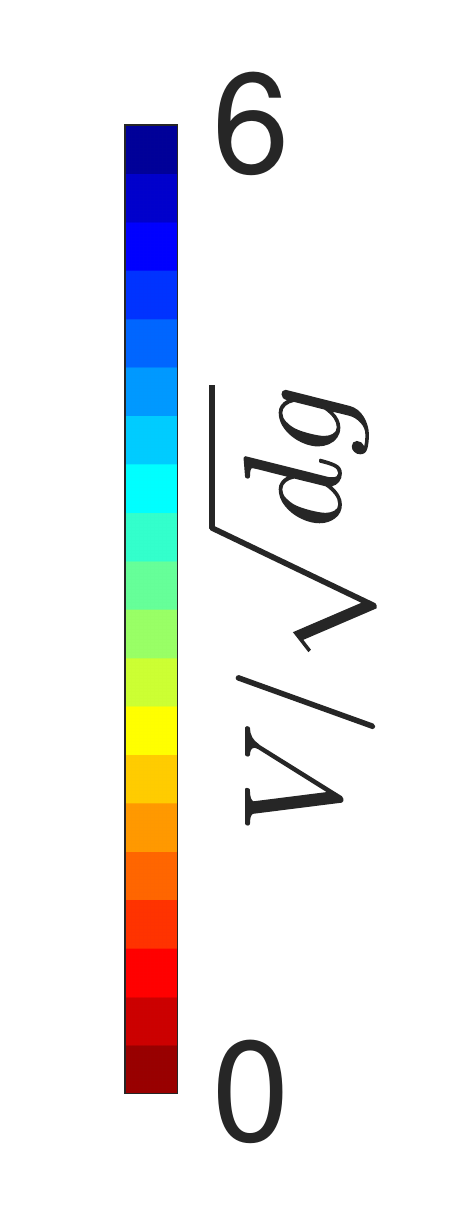}
	\caption{\label{fields:vel}Spatial distribution of velocity $V$ at an orifice width $W/d$ = a) 25, b) 30, c) 35, d) 40 and e) 45 on the sidewall with fraction of dumbbells $X_{db}=0.5$.}
\end{figure}

\begin{figure}	
	\centering
	\subfloat[][]{\includegraphics[clip,trim=5.15cm 1.2cm 4.65cm 0.6cm,width=0.14\linewidth]{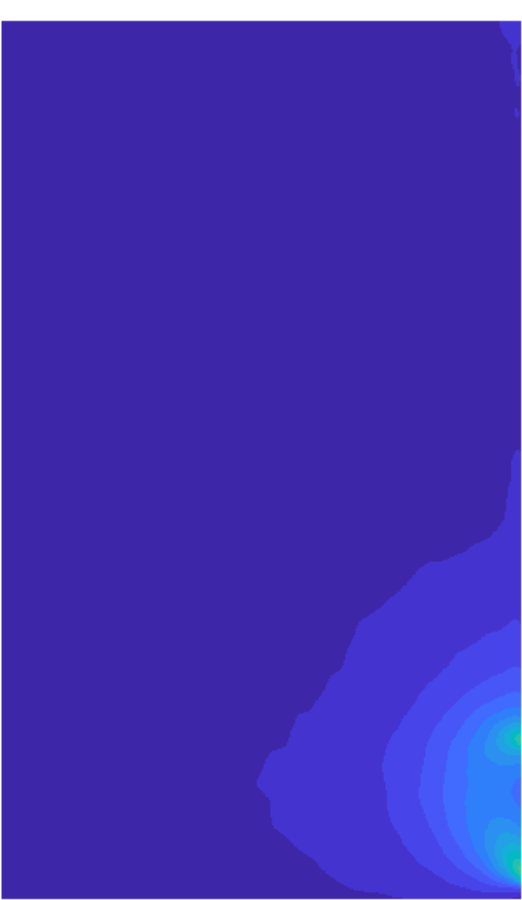}}
	\hspace{0.25cm}
	\subfloat[][]{\includegraphics[clip,trim=5.15cm 1.2cm 4.65cm 0.6cm,width=0.14\linewidth]{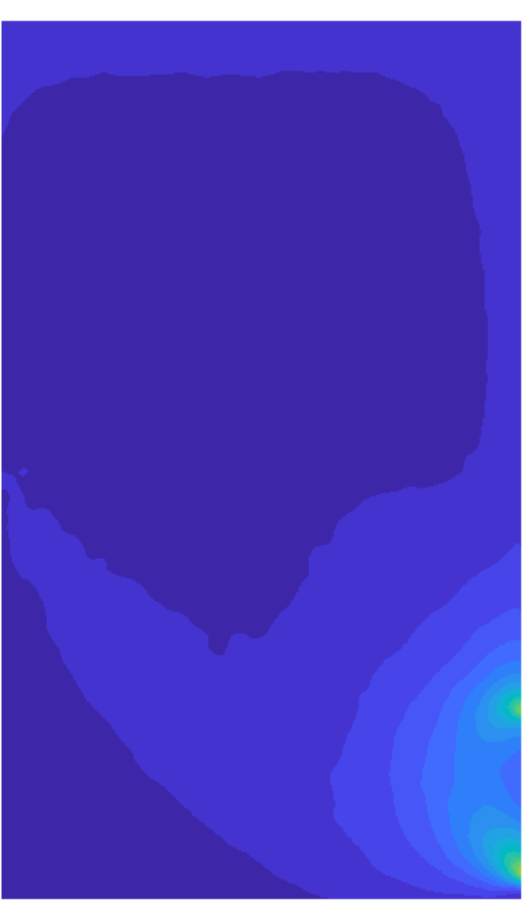}}
	\hspace{0.25cm}
	\subfloat[][]{\includegraphics[clip,trim=5.15cm 1.2cm 4.65cm 0.6cm,width=0.14\linewidth]{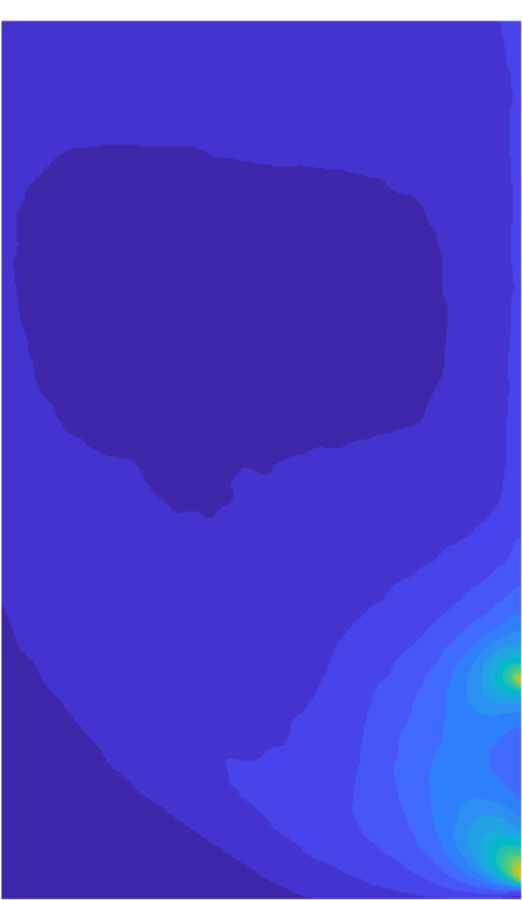}}
	\hspace{0.25cm}
	\subfloat[][]{\includegraphics[clip,trim=5.15cm 1.2cm 4.65cm 0.6cm,width=0.14\linewidth]{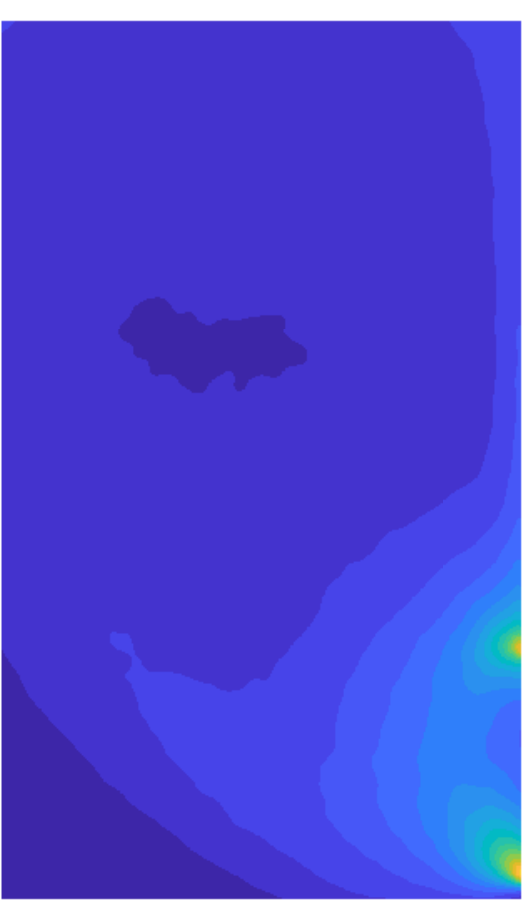}}
	\hspace{0.25cm}
	\subfloat[][]{\includegraphics[clip,trim=5.15cm 1.2cm 4.65cm 0.6cm,width=0.14\linewidth]{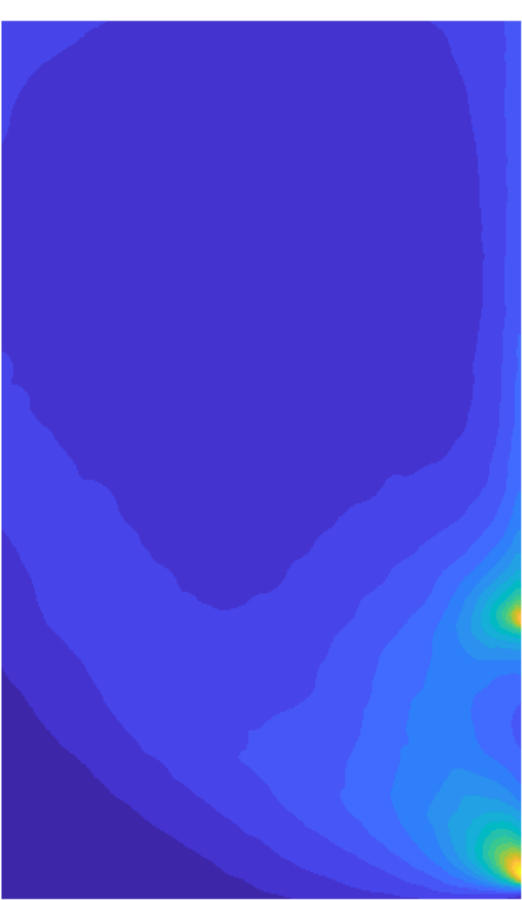}}
	\hspace{0.25cm} 
	\includegraphics[width=0.1\linewidth]{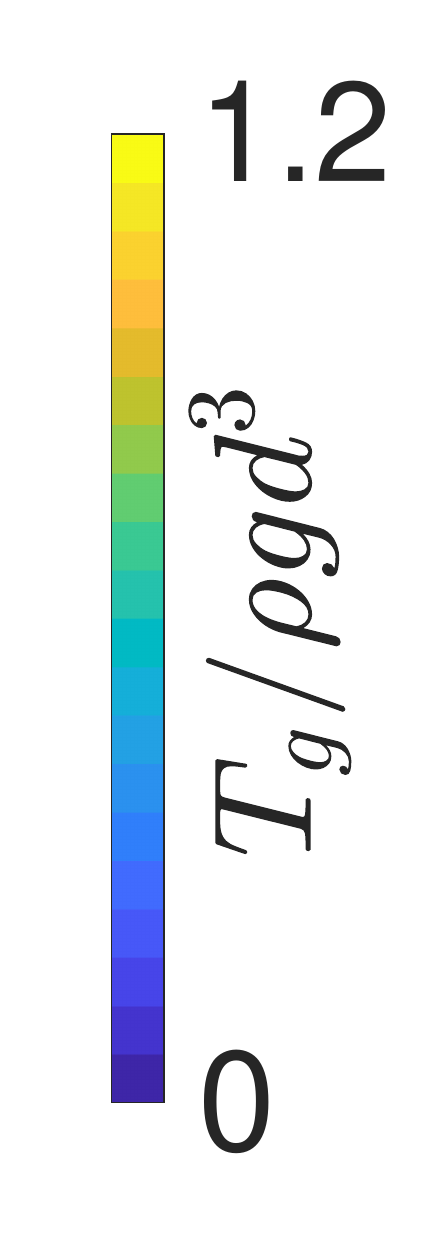}
	\caption{\label{fields:temp}Spatial distribution of granular temperature $T_g$ at an orifice width $W/d$ = a) 25, b) 30, c) 35, d) 40 and e) 45 on the sidewall with fraction of dumbbells $X_{db}=0.5$.}
\end{figure}

Area fraction $\phi$ is found to vary slightly with an increase in the width $W$ of the lateral orifice  (figure \ref{fields:pf}). The area fraction is least in the region beside the orifice due to shear-induced dilation \citep{tamas2016}. Moreover, with an increase of $W$, the region of dilation is noticed to increase. In the bulk as well as in the left side corner of the silo, $\phi$ is noticed to be slightly less than that of the random close packing (0.84) at all $W$. Figure \ref{fields:omeg} displays the spatial distribution of rotational velocity $\Omega$ at various lateral orifice widths. The flowing solid particles tend to rotate while crossing the edges \citep{rotation}, thus $\Omega$ is found to be maximum at both edges of the orifice for all the cases. In the bulk, as the particles are closely packed they hardly rotate resulting in a negligible $\Omega$. However, in the region beside the orifice, the particles are loosely packed (figure \ref{fields:pf}) thus the particle collisions might yield in their rotations. Moreover, the particle collisions result in the fluctuations of rotational velocities $\Omega_{fl}$ in the region beside the orifice as seen in figure \ref{fields:omegfluc}. This region is noticed to expand with the lateral orifice width due to an increase in the flowing zone of the particles. Figure \ref{fields:vel} displays velocity $V$ fields at various lateral orifice widths $W$. A stagnant zone is observed at the left side of the silo as well as on the base where velocity is found to be almost zero. This zone is found to vary slightly with an increase in the lateral orifice width. The stagnant zone hinders the movement of particles flowing adjacent to it and its influence is negligible at locations far away from it. Moreover, the flowing zone reaches the left wall at a certain height due to the presence of the stagnant zone. The presence of stagnant zone on the entire silo base as well as until certain height of the left side wall was observed previously \cite{Zhou2017} though in their study they presented velocity fields at various widths of the silo. As particle velocities are almost constant in the bulk, granular temperature $T_g$ or fluctuations in velocities are found to be almost negligible (figure \ref{fields:temp}) in the bulk. However, in the region beside the orifice, $T_g$ is present due to two types of particle collisions. The first one is between the particles of the flowing zone and those present above the orifice. The other type is between the particles in the flowing zone and those present in the stagnant zone. Moreover, $T_g$ is found to increase with an increase in $W/d$ in the region beside the orifice due to an increase in the velocity fluctuations. This can be explained by an increase in the particle collisions as well as particle velocities (figure \ref{fields:vel}) due to an increase in the orifice width. 

\begin{figure}	
	\centering
	\subfloat[][]{\includegraphics[clip,trim=5.15cm 1.2cm 4.65cm 0.6cm,width=0.14\linewidth]{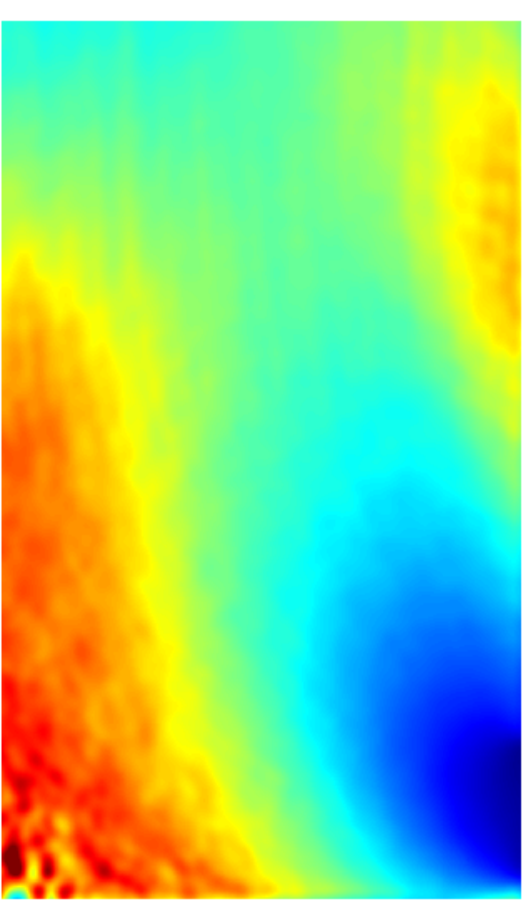}}
	\hspace{0.25cm}
	\subfloat[][]{\includegraphics[clip,trim=5.15cm 1.2cm 4.65cm 0.6cm,width=0.14\linewidth]{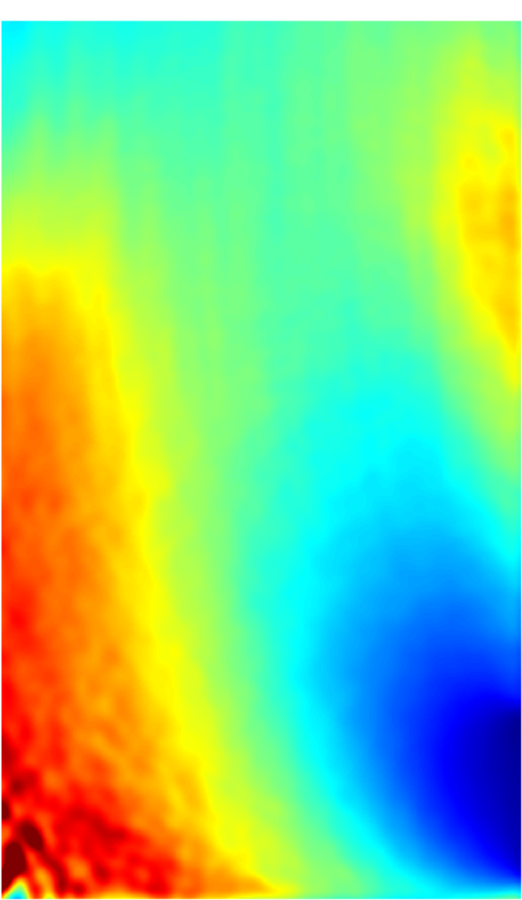}}
	\hspace{0.25cm}
	\subfloat[][]{\includegraphics[clip,trim=5.15cm 1.2cm 4.65cm 0.6cm,width=0.14\linewidth]{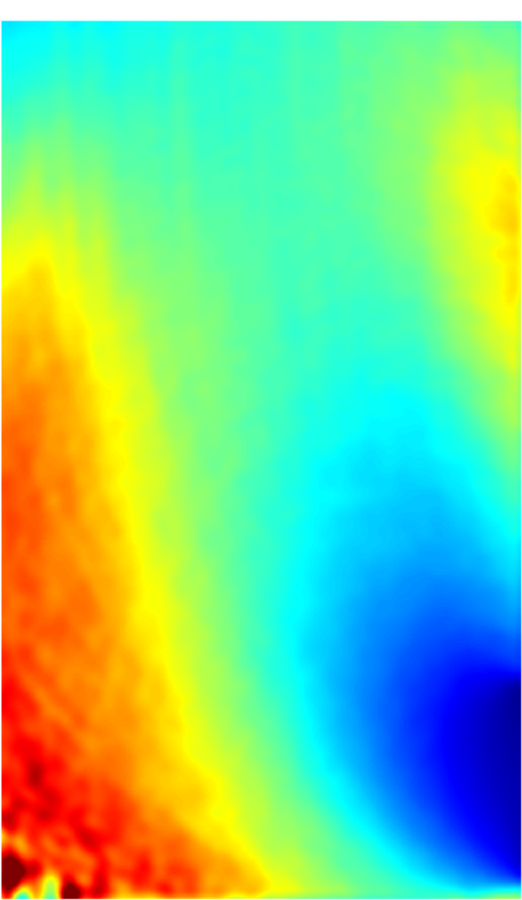}}
	\hspace{0.25cm}
	\subfloat[][]{\includegraphics[clip,trim=5.15cm 1.2cm 4.65cm 0.6cm,width=0.14\linewidth]{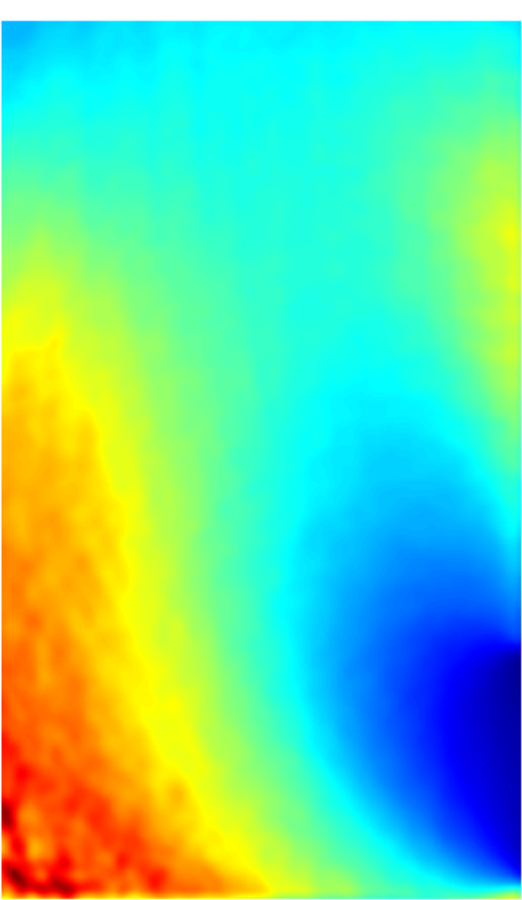}}
	\hspace{0.25cm}
	\subfloat[][]{\includegraphics[clip,trim=5.15cm 1.2cm 4.65cm 0.6cm,width=0.14\linewidth]{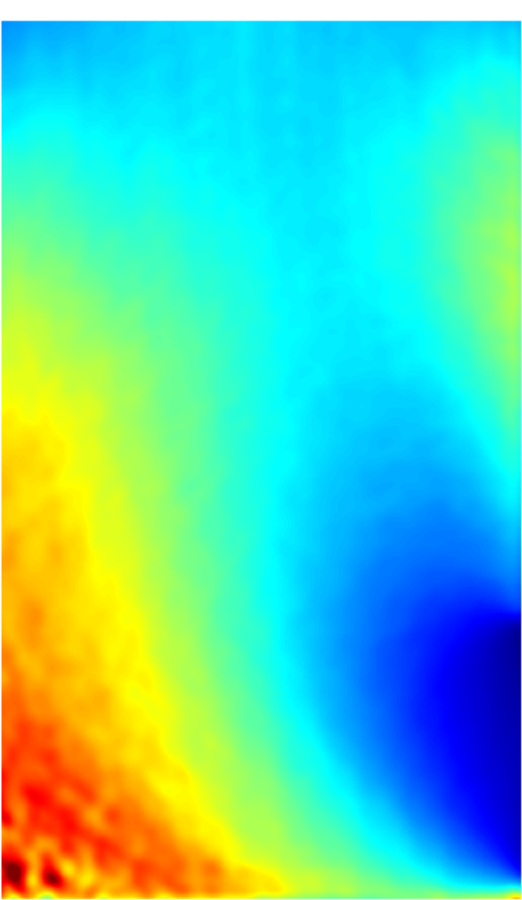}}
	\hspace{0.25cm} 
	\includegraphics[width=0.1\linewidth]{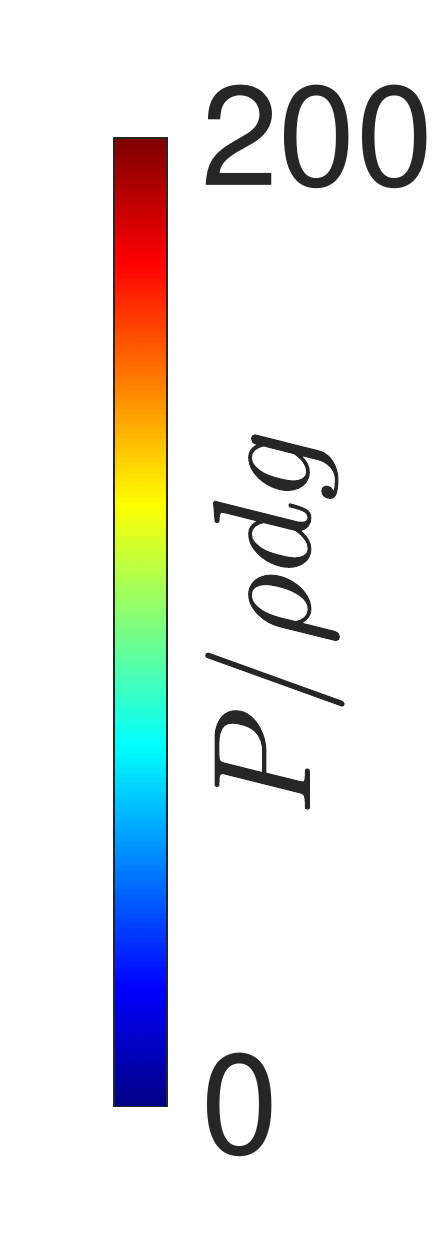}
	\caption{\label{fields:press}Spatial distribution of pressure $P$ at an orifice width $W/d$ = a) 25, b) 30, c) 35, d) 40 and e) 45 on the sidewall with fraction of dumbbells $X_{db}=0.5$.} 
\end{figure}

\begin{figure}	
	\centering
	\subfloat[][]{\includegraphics[clip,trim=5.15cm 1.2cm 4.65cm 0.6cm,width=0.14\linewidth]{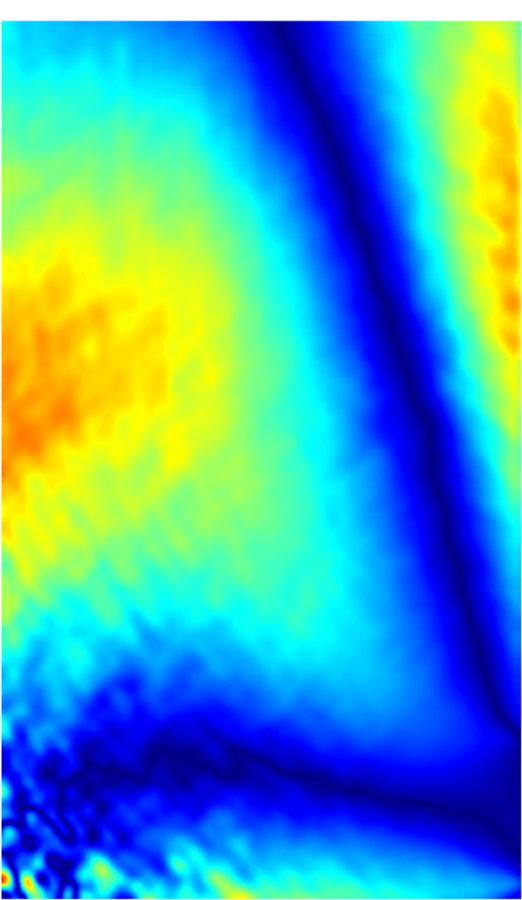}}
	\hspace{0.25cm}
	\subfloat[][]{\includegraphics[clip,trim=5.15cm 1.2cm 4.65cm 0.6cm,width=0.14\linewidth]{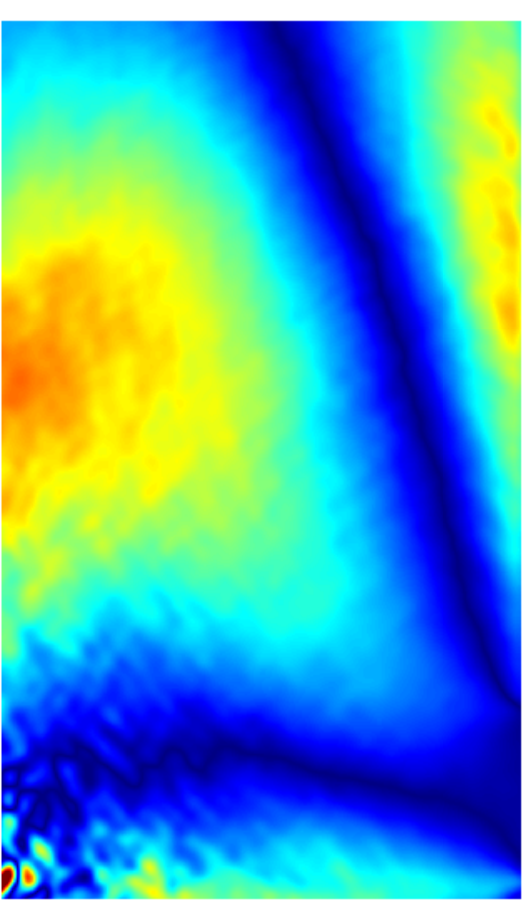}}
	\hspace{0.25cm}
	\subfloat[][]{\includegraphics[clip,trim=5.15cm 1.2cm 4.65cm 0.6cm,width=0.14\linewidth]{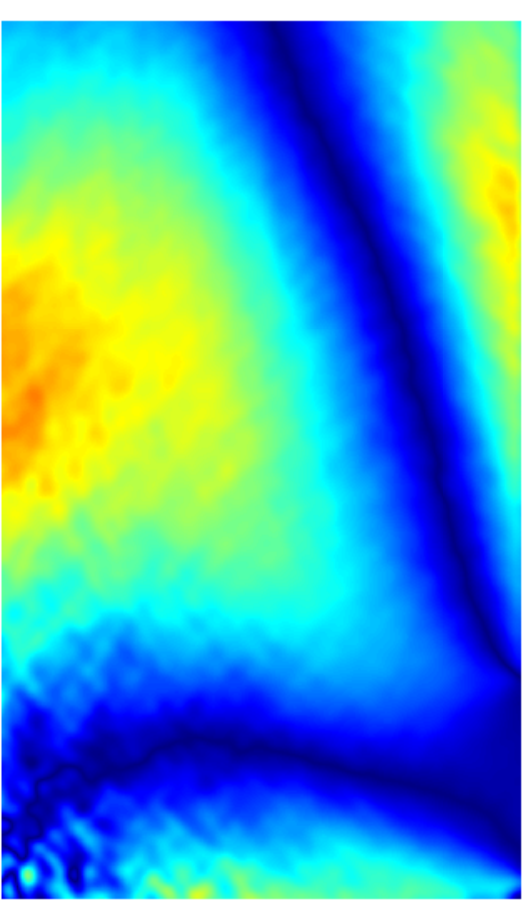}}
	\hspace{0.25cm}
	\subfloat[][]{\includegraphics[clip,trim=5.15cm 1.2cm 4.65cm 0.6cm,width=0.14\linewidth]{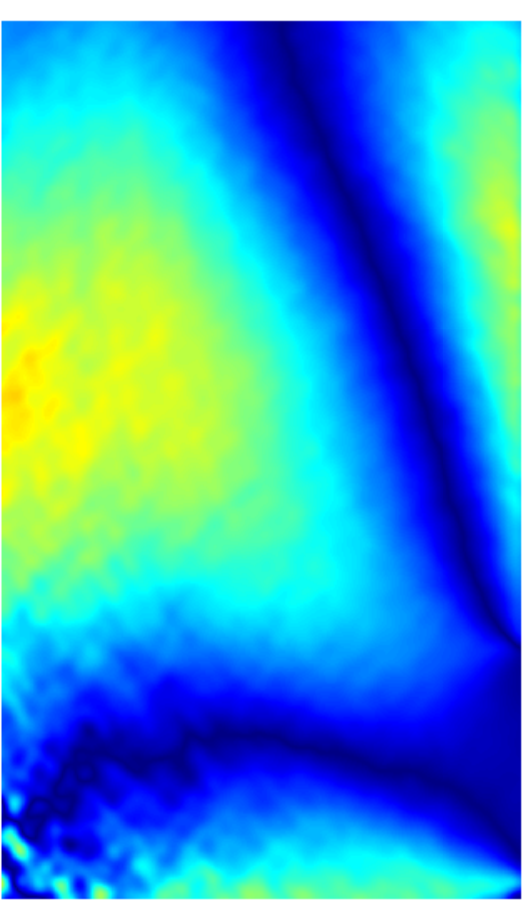}}
	\hspace{0.25cm}
	\subfloat[][]{\includegraphics[clip,trim=5.15cm 1.2cm 4.65cm 0.6cm,width=0.14\linewidth]{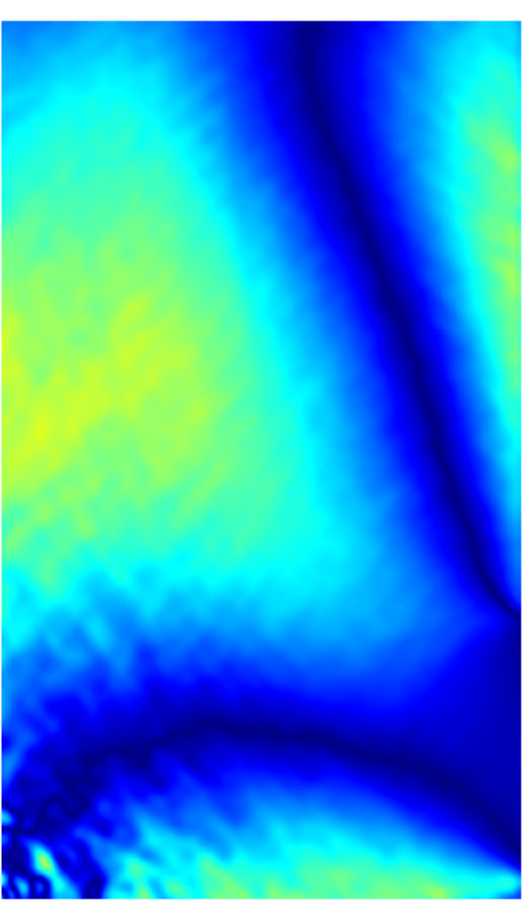}}
	\hspace{0.25cm} 
	\includegraphics[width=0.1\linewidth]{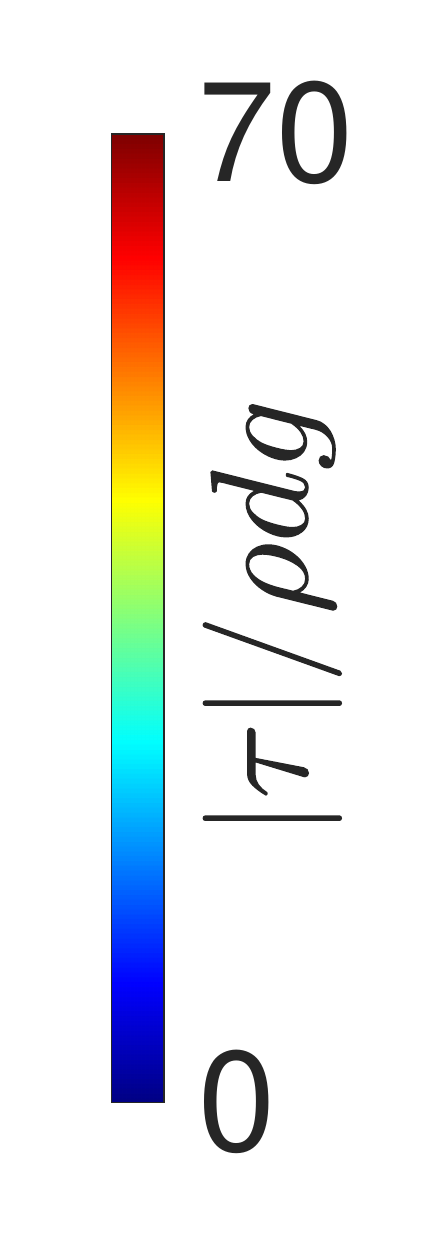}
	\caption{\label{fields:shear}Spatial distribution of shear stress $|\tau|$ at an orifice width $W/d$ = a) 25, b) 30, c) 35, d) 40 and e) 45 on the sidewall with fraction of dumbbells $X_{db}=0.5$.}
\end{figure}

\begin{figure}	
	\centerline{\includegraphics[width=1.0\linewidth]{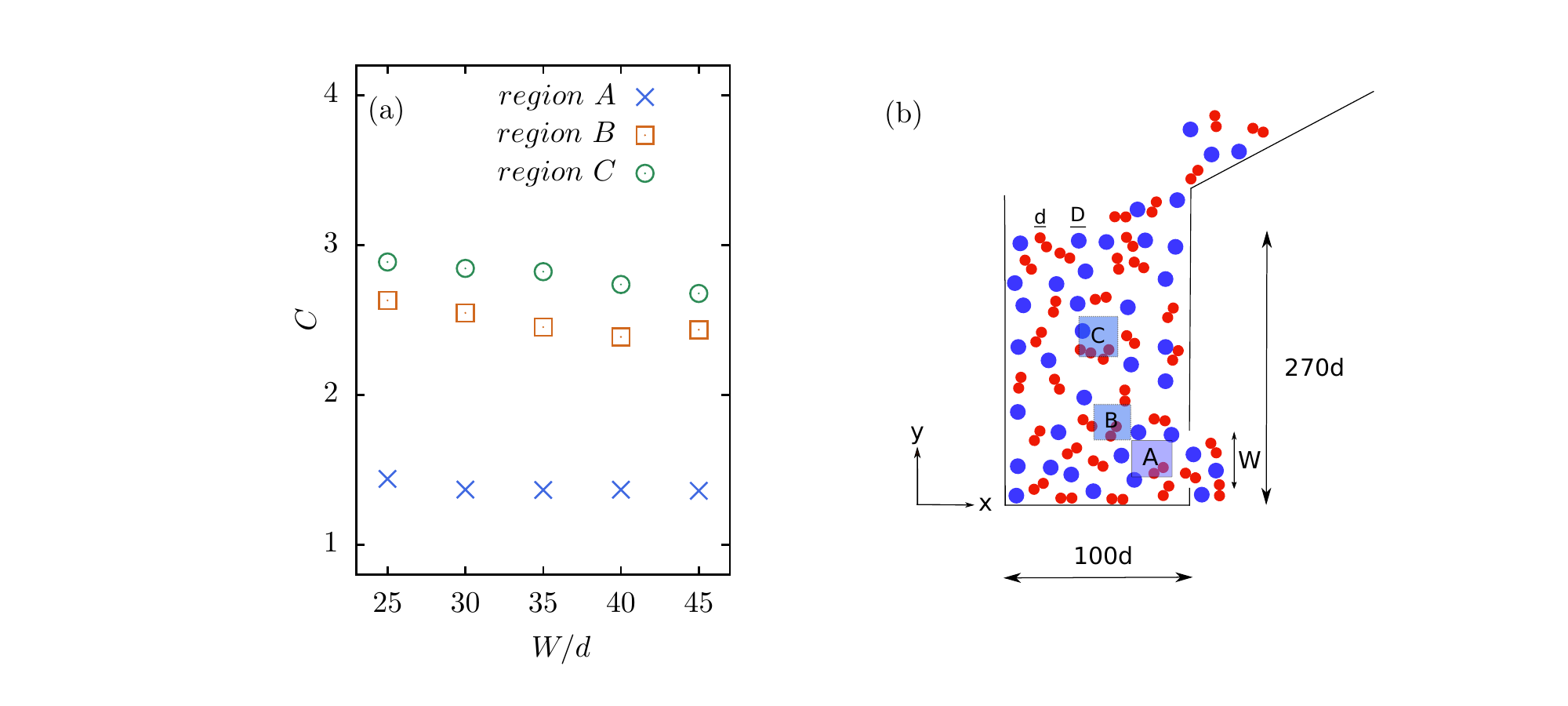}}
	\caption{\label{fig:coords}a) Average coordination number $C$ as a function of lateral orifice width $W/d$ for the fraction of dumbbells $X_{db}=0.5$ at regions $A$, $B$ and $C$. b) The regions of interest $A$, $B$ and $C$ are shown. Here, region $A$: $35<x<45$ and $10<y<W-5$, region $B$: $15<x<25$ and region $C$: $W-5<y<W+5$, c) $5<x<15$ and $60<y<70$}
\end{figure}

Pressure $P$ flow fields are illustrated in figure \ref{fields:press}. In a granular media, stress is transmitted through a network of contacts namely force chains \citep{forcechains}. Thus, the number of contacts or coordination number plays a crucial role in determining the magnitude of stress experienced by a particle. The average coordination number $C$ is found to be less in the region beside the orifice as compared to that of the bulk (figure \ref{fig:coords}). This is the reason for a lesser pressure in the region beside the orifice as compared to that of the bulk. With an increase in the orifice width, pressure decreases slightly in the region beside the orifice due to a slight decrease in the average coordination number. Towards the left side of the silo, the pressure is maximum due to the presence of stagnant zone and the force exerted by the particles above it which corresponds to the flowing zone. Figure \ref{fields:shear} displays the shear stress $|\tau|$ and it is least in the region beside the orifice as it is flowing zone. $|\tau|$ is observed to be maximum for all the cases at the left side of the silo as it is the region that lies between the wall and the flowing zone. This kind of behaviour is similar to that observed in the liquids. Moreover, as there is hardly any flow in the region above the base of the silo, shear stress is least in this region. 

\subsection{Two orifices on a silo base\label{sec:multi}}

\begin{figure}	
	\centerline{\includegraphics[width=1.0\linewidth]{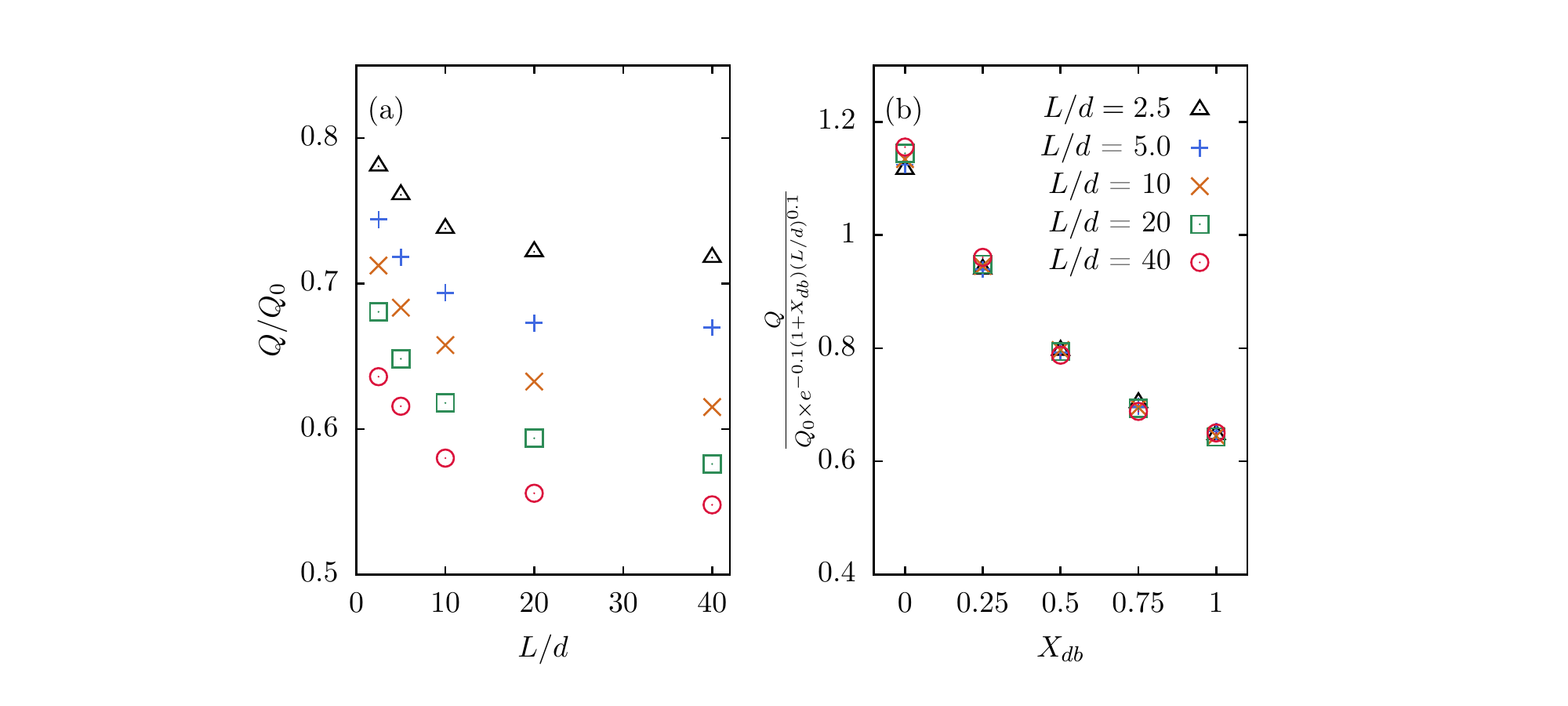}}
	\caption{\label{fig:scalingm}a) Normalized flow rate $Q/Q_{0}$ as a function of inter-orifice distances $L/d$ at different fraction of dumbbells $X_{db}$ and b) scaling of flow rate with inter-orifice distance $L/d$. Here, the width of each of the two orifices is $W/d=20$ and $Q_0$ is the flow rate at $L/d=0$.}
\end{figure}

In this subsection, we are going to elucidate how the flow dynamics of a mixture of discs and dumbbells is affected by the separation distance between the two orifices that are placed on the silo base. In this regard, we have analysed the flow at six different separation lengths $L/d$ ranging from $0$ to $40$ between the two orifices each of width $W/d=20$. Moreover, we have considered five different fractions of dumbbells $X_{db}$ from 0.0 to 1.0. When granular particles are discharging through an orifice on a flat bottomed silo base, a stagnant zone is present on either side of the orifice \citep{tamas2016}. The stagnant zone usually hinders the movement of particles flowing adjacent to it. In the case of a silo with multiple orifices, an additional stagnant zone is present in between the orifices \cite{multiori} along with the one that exists beside the sidewalls. As the distance between the orifices increases, the stagnant zone formed between them expands and the hindrance to the flow increases thus decreasing the flow rate $Q$. This is shown in figure \ref{fig:scalingm}a, where $Q/Q_{0}$ decreases with $L/d$ for all fractions of dumbbells $X_{db}$ until $L/d=20$ and then it gets saturated. Here, $Q$ is the flowrate of particles exiting through both the orifices and $Q_{0}$ represents flow rate when the inter-orifice distance is zero. \citet{multiori} reported a similar result to that of ours where they noticed a gradual decrease followed by saturation in $Q$ with an increase in the inter-orifice distance for a system of spherical particles. We observed a decrease in the flow rate with an increase in the fraction of dumbbells $X_{db}$, similar to the result observed in figure \ref{fig:pfgtl}a. The flow rates corresponding to different inter-orifice distances for various fractions of dumbbells are collapsed into a single curve. The flow rate $Q$ scales with the inter-orifice distance $L/d$ as $Q \propto Q_0\times e^{-0.1(1+X_{db})(L/d)^{0.1}}$ (figure \ref{fig:scalingm}).

\begin{figure}	
	\centerline{\includegraphics[width=1.0\linewidth]{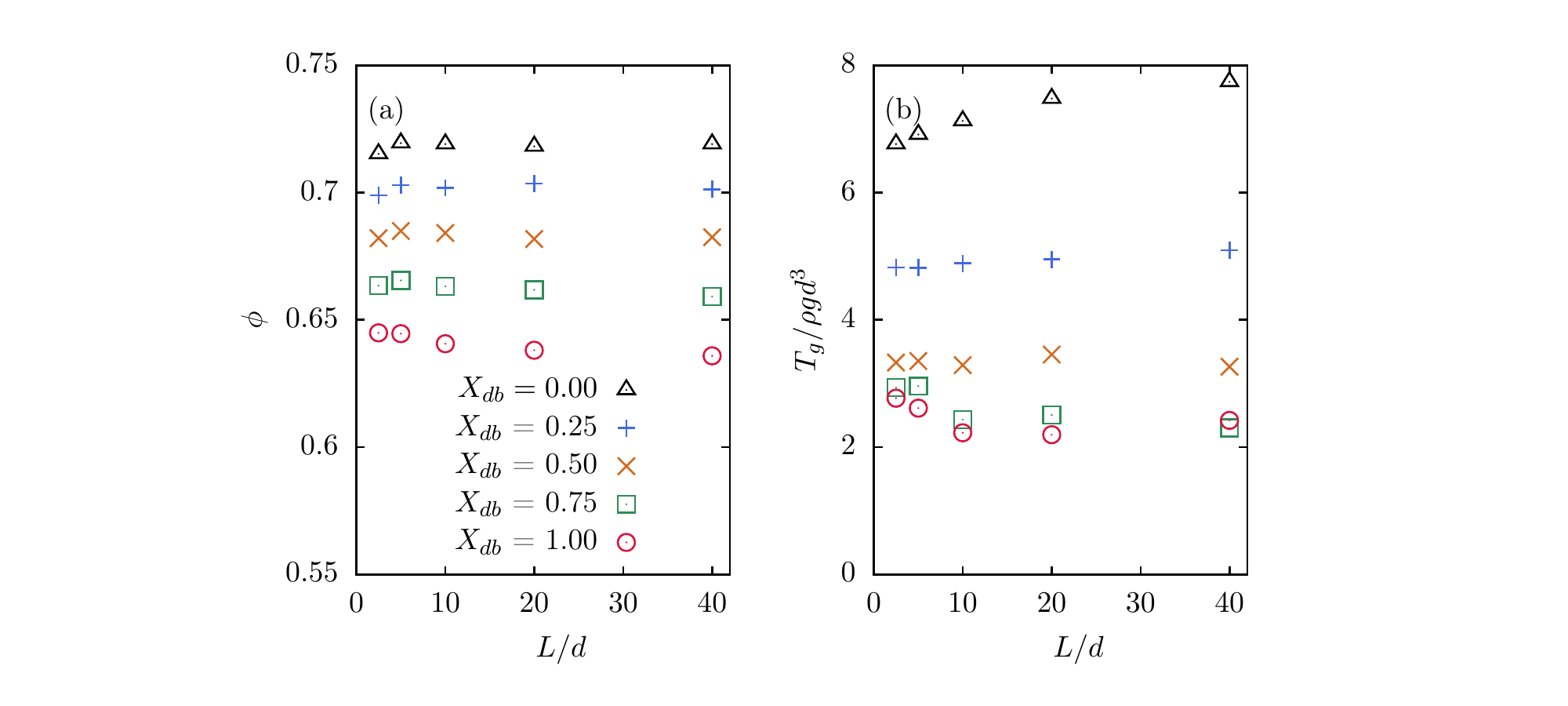}}
	\caption{\label{fig:pfgtm}a) Area fraction $\phi$ and b) granular temperature $T_g$ at different spacings $L/d$ between the two orifices, each of width $W/d=20$.}
\end{figure}

\begin{figure}
\centerline{\includegraphics[width=1.0\linewidth]{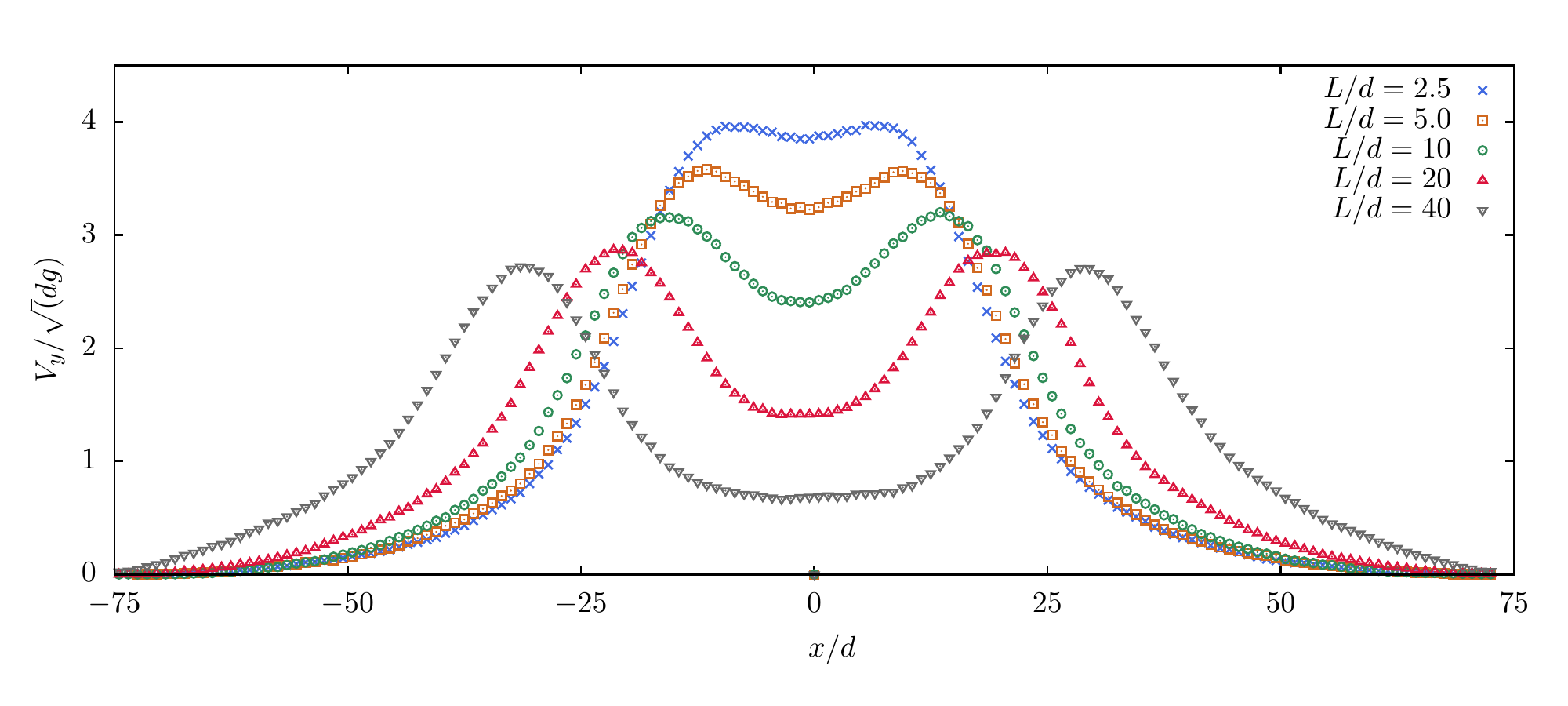}}
	\caption{\label{fig:flowm}Vertical velocity $V_y$ as a function of horizontal position $x$ at various spacings $L$ between the two orifices, each of width $W/d=20$.}
\end{figure}

Area fraction and granular temperature are computed in the region $R_2$ lying just above one of the orifices as shown in the figure \ref{fig:sim}b. Area fraction $\phi$ is found to vary slightly with the inter-orifice distance $L/d$ however it decreases with an increase in $X_{db}$ (figure \ref{fig:pfgtm}). Moreover, $\phi$, in this case, is less as compared to that of the lateral orifice case because the particles lying above the lateral orifice flow into the region beside the orifice due to gravity and thus results in higher $\phi$. Granular temperature $T_g$ is found to increase with an increase in $L/d$ for a system of discs $X_{db}=0.0$ and it decreases with an increase in $L/d$ for that of dumbbells $X_{db}=1.0$. For the mixtures of dumbbells and discs $T_g$ is noticed to remain almost constant. Moreover, $T_g$ is noticed to decrease with an increase in $X_{db}$ at all $L/d$. With the addition of dumbbells, velocity fluctuations in the region above the orifice decreases due to decrease in the particle collisions and particle velocities as dumbbells have more affinity to interlock due to its geometry. Figure \ref{fig:flowm} displays the vertical velocity profiles as a function of horizontal position. At a lower inter-orifice distance $L/d\le10$, the flow through an orifice is found to influence the flow of particles through another orifice. This interaction zone between the orifices is responsible for a significant difference in the magnitude of the particle velocities at $L/d=2.5$ and $L/d=20$ in the region above the orifice. The interaction zone is noticed to be almost absent at $L/d\ge20$.  As the distance between the orifices increases, the velocities of the particles lying between the two orifices decreases due to an increase in the stagnant zone. The silo flows are usually characterised by the spatio-temporal heterogeneities \citep{mehta}. However, the two orifices are observed to have almost similar velocity profiles because $V_y$ is averaged over a certain time. The maximum vertical velocity of the particles is noticed to decrease with an increase in $L/d$ supporting the result of a decrease in the flow rate with $L/d$ in figure \ref{fig:pfgtm}a.

\begin{figure}	
	\centering
	\subfloat[][]{\includegraphics[clip,trim=3.15cm 1.2cm 2.65cm 0.6cm,width=0.14\linewidth]{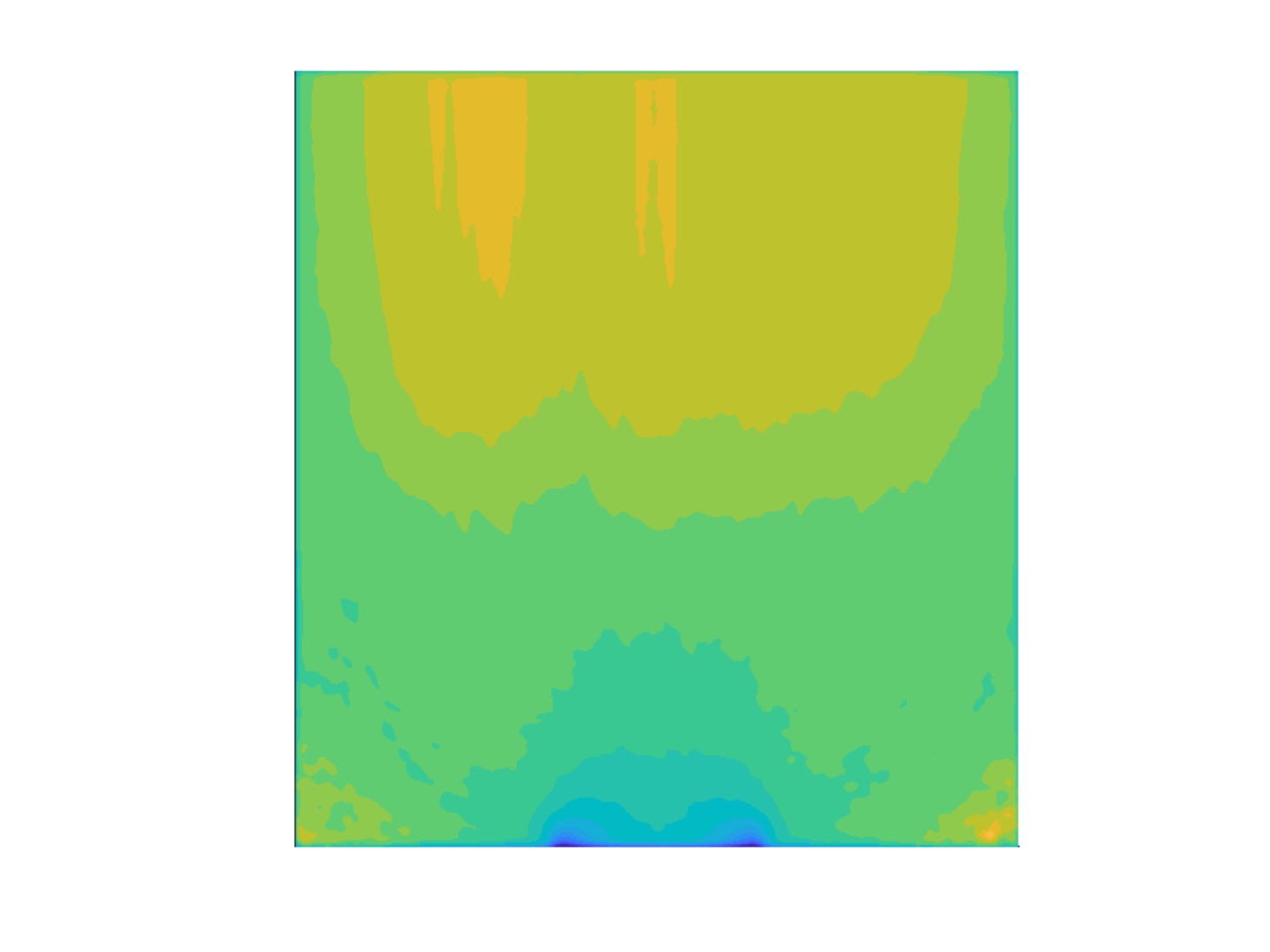}}
	\hspace{0.01cm}
	\subfloat[][]{\includegraphics[clip,trim=3.15cm 1.2cm 2.65cm 0.6cm,width=0.14\linewidth]{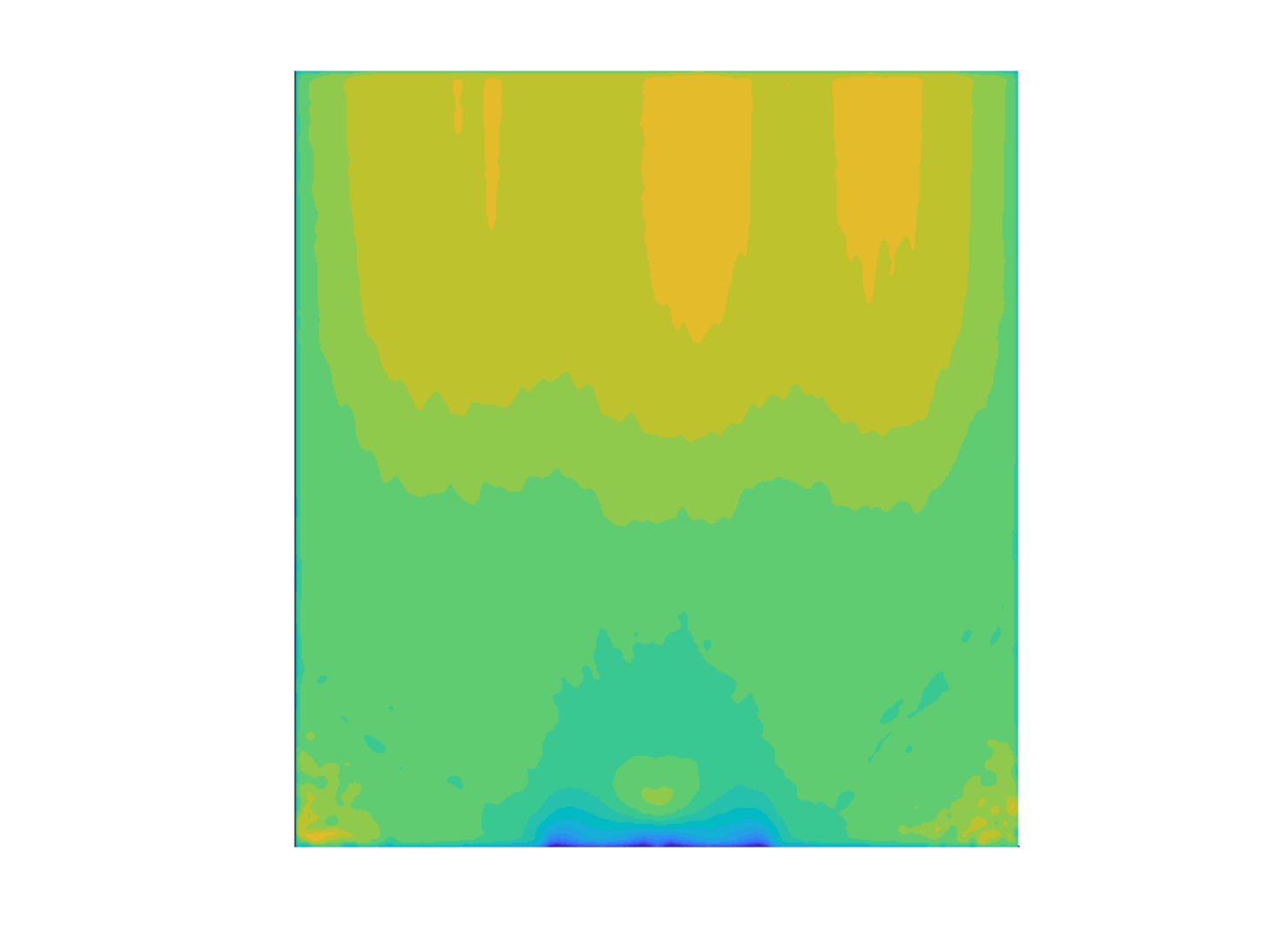}}
	\hspace{0.01cm}
	\subfloat[][]{\includegraphics[clip,trim=3.15cm 1.2cm 2.65cm 0.6cm,width=0.14\linewidth]{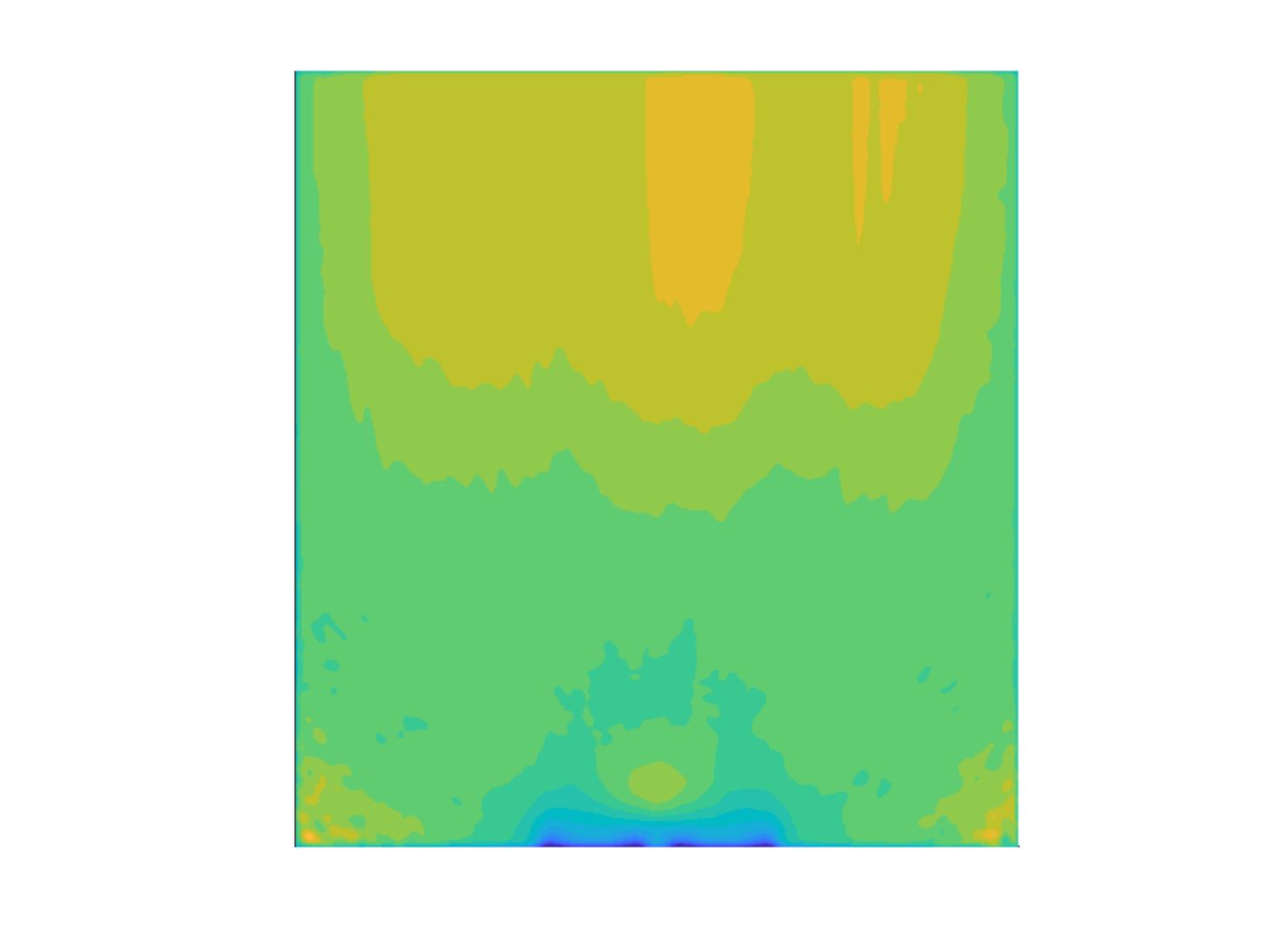}}
	\hspace{0.01cm}
	\subfloat[][]{\includegraphics[clip,trim=3.15cm 1.2cm 2.65cm 0.6cm,width=0.14\linewidth]{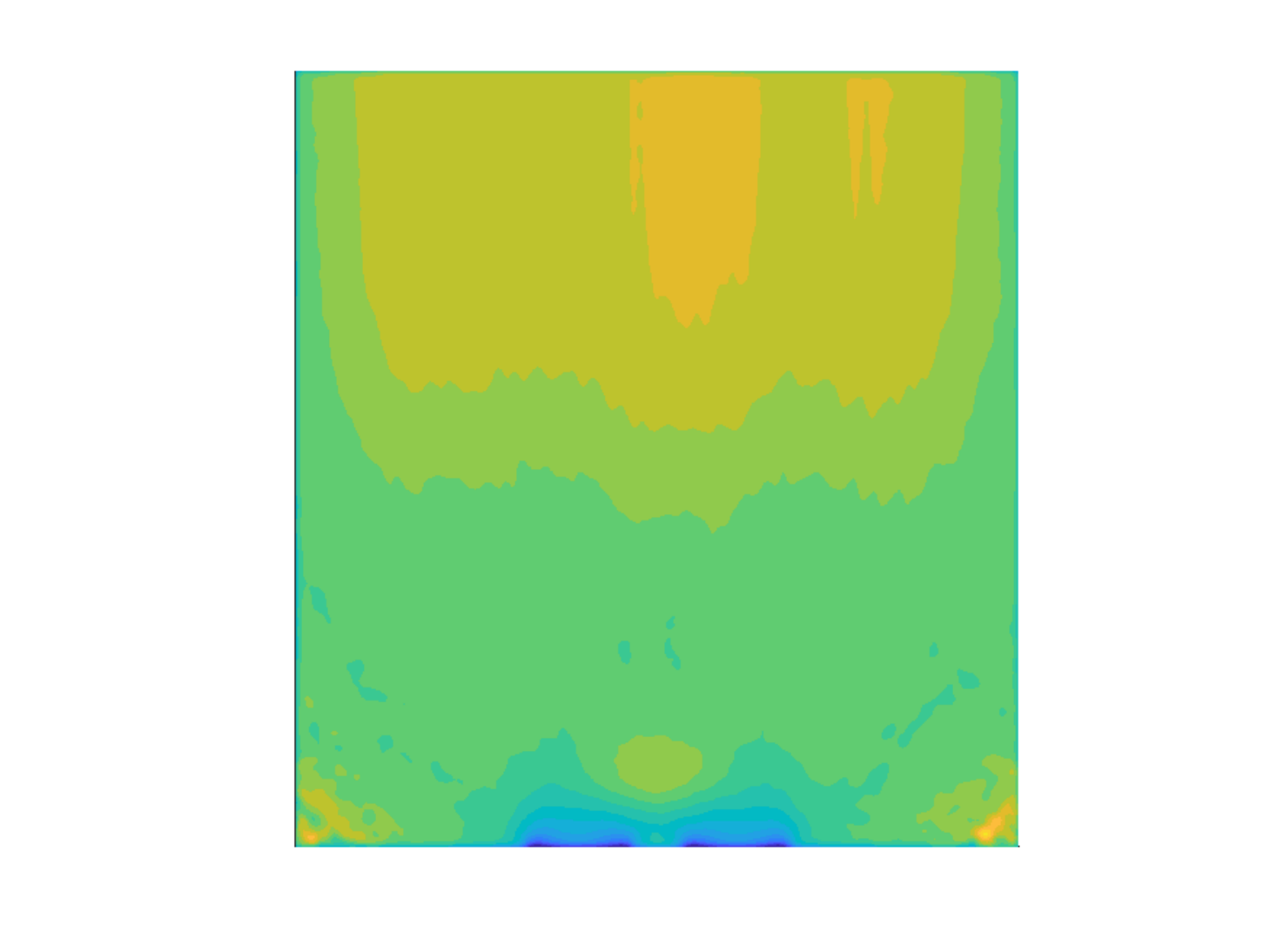}}
	\hspace{0.01cm}
	\subfloat[][]{\includegraphics[clip,trim=3.15cm 1.2cm 2.65cm 0.6cm,width=0.14\linewidth]{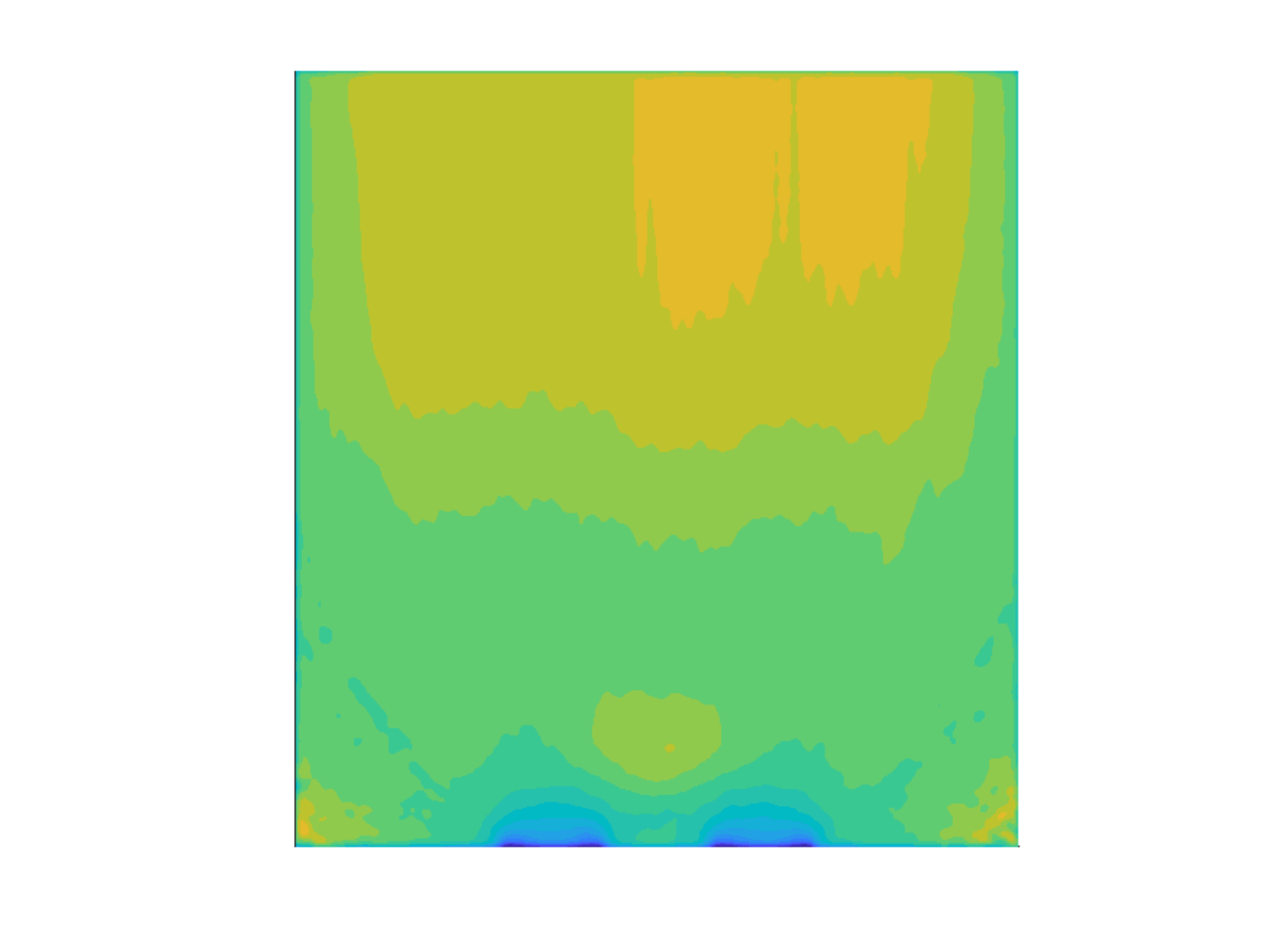}}
	\hspace{0.01cm} 
	\subfloat[][]{\includegraphics[clip,trim=3.15cm 1.2cm 2.65cm 0.6cm,width=0.14\linewidth]{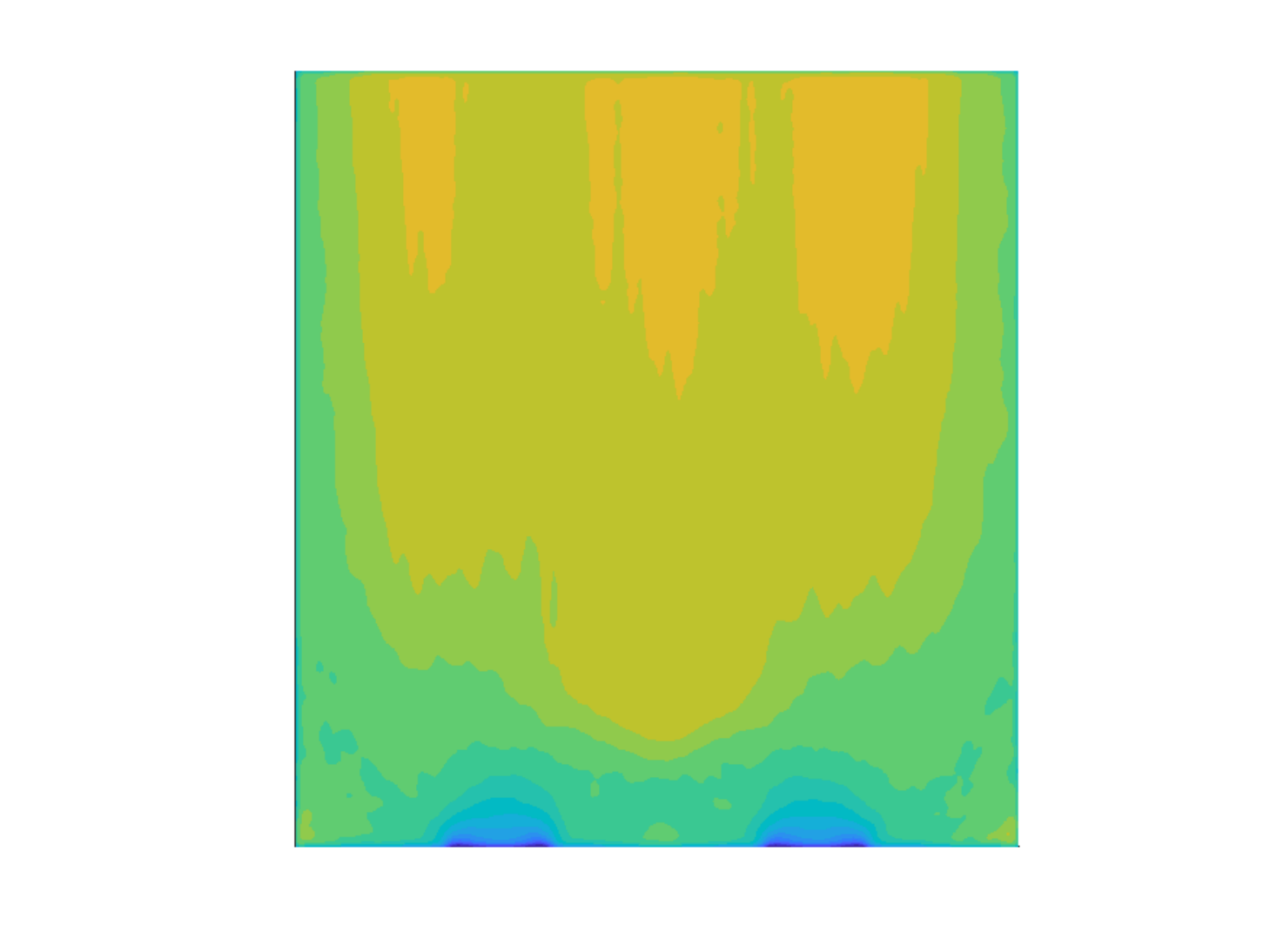}}
	\hspace{0.01cm}
	\includegraphics[width=0.06\linewidth]{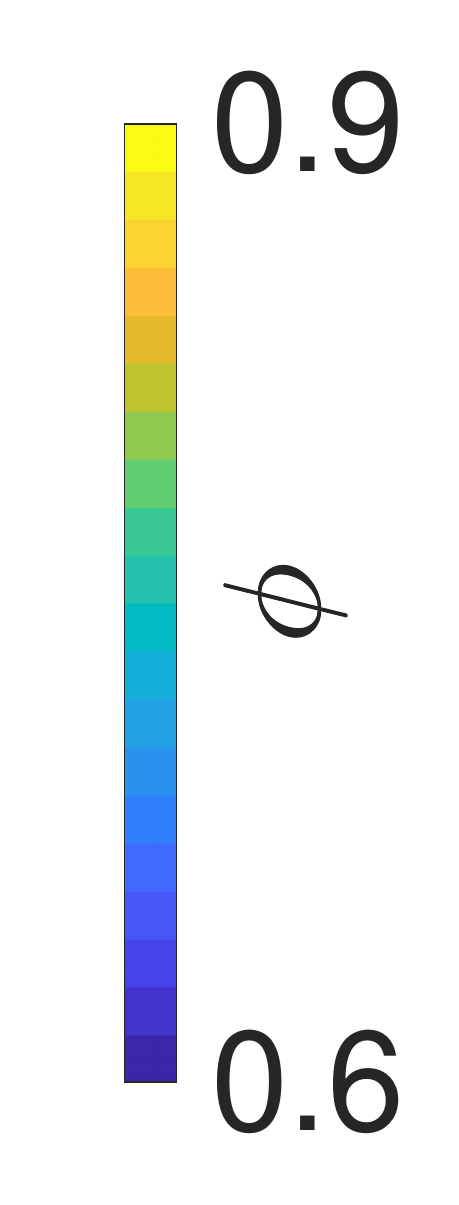}
	\caption{\label{fields:pfm}Spatial distribution of area fraction $\phi $ at different spacings $L/d$ = a) 0.0, b) 2.5, c) 5, d) 10, e) 20 and f) 40 between the orifices, each of width $W/d=20$, placed on the silo base. The fraction of dumbbells for all the cases is $X_{db}=0.5$. }
\end{figure}

\begin{figure}	
	\centering
	\subfloat[][]{\includegraphics[clip,trim=3.15cm 1.2cm 2.65cm 0.6cm,width=0.14\linewidth]{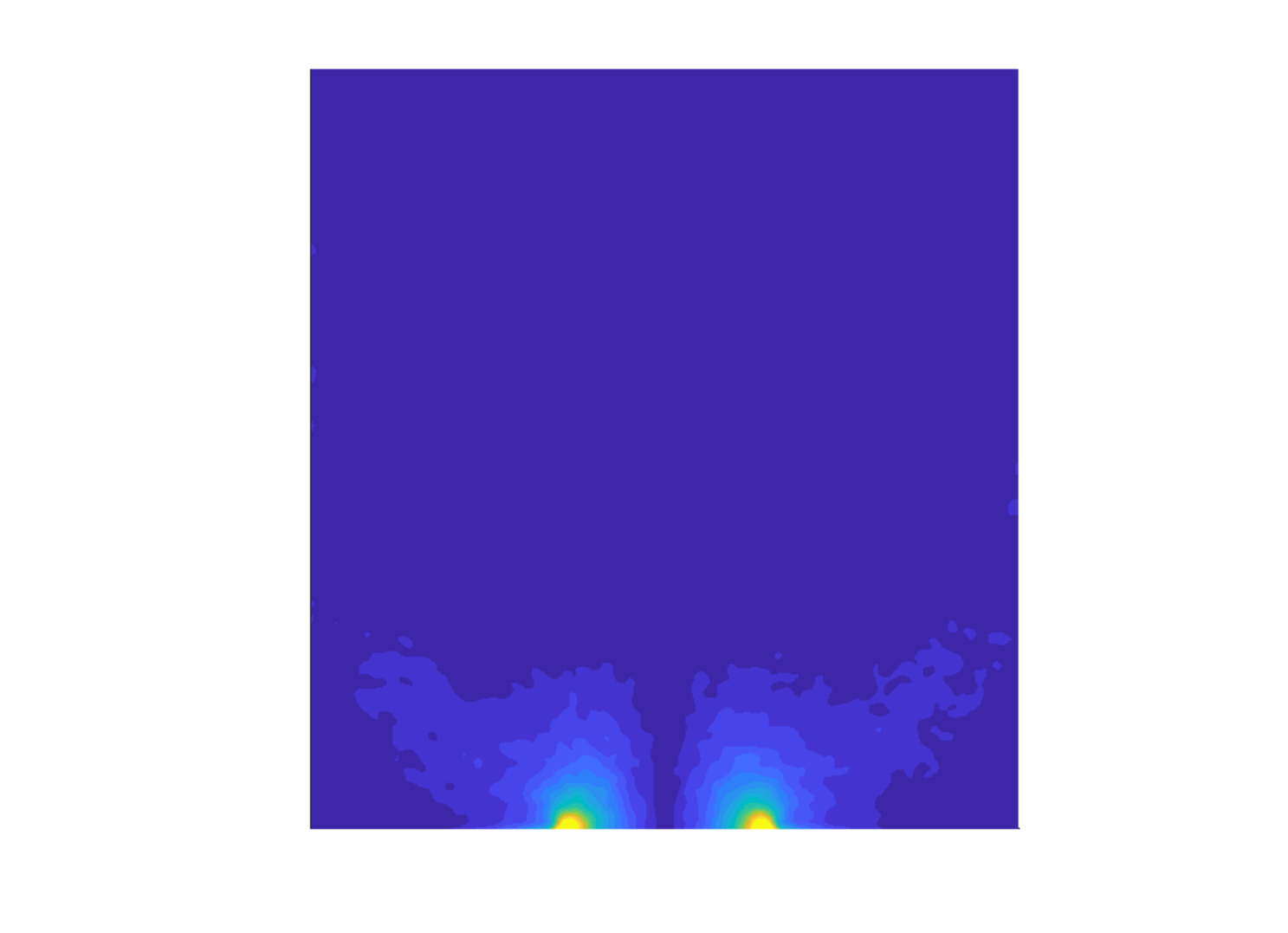}}
	\hspace{0.01cm}
	\subfloat[][]{\includegraphics[clip,trim=3.15cm 1.2cm 2.65cm 0.6cm,width=0.14\linewidth]{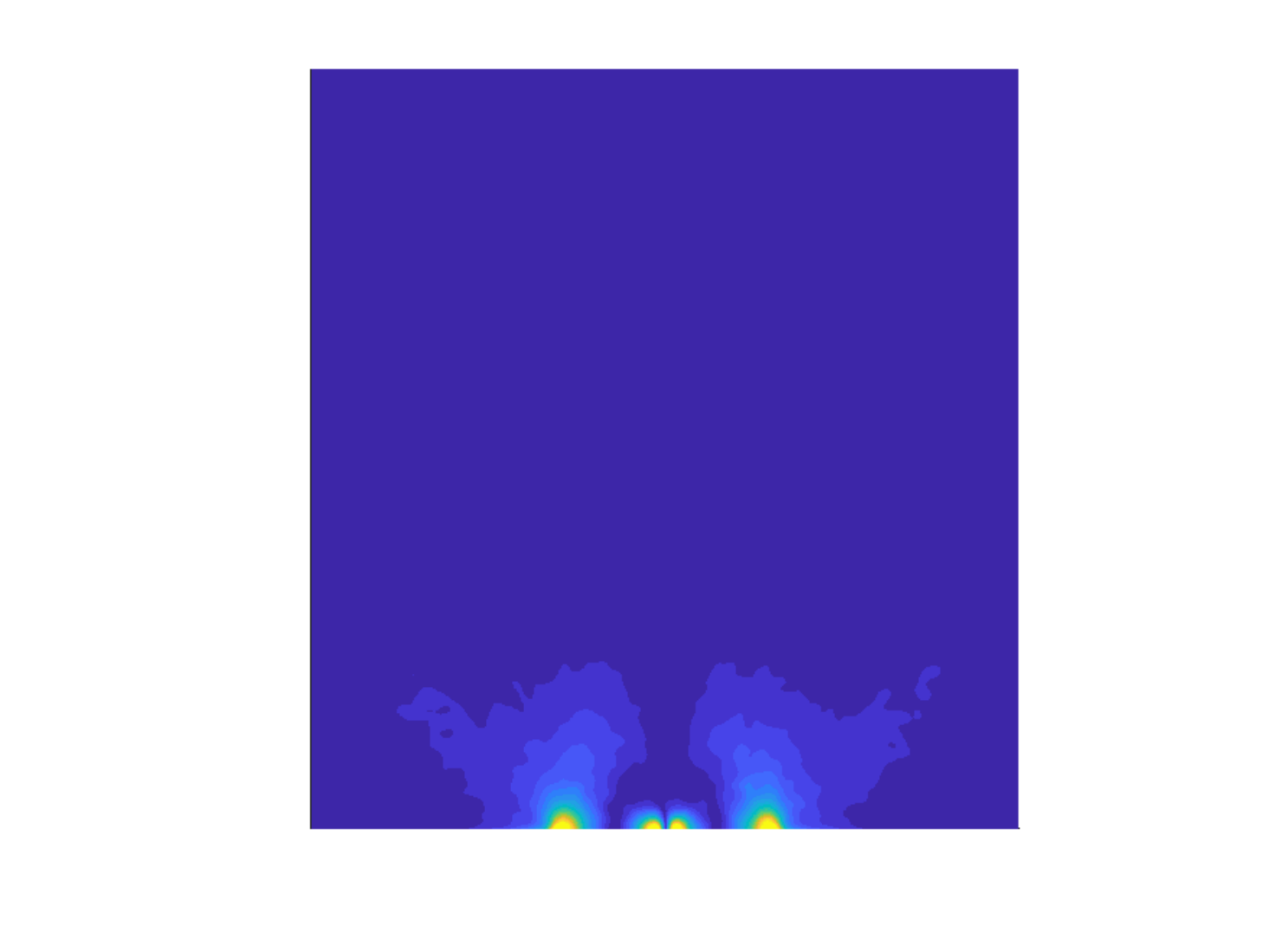}}
	\hspace{0.01cm}
	\subfloat[][]{\includegraphics[clip,trim=3.15cm 1.2cm 2.65cm 0.6cm,width=0.14\linewidth]{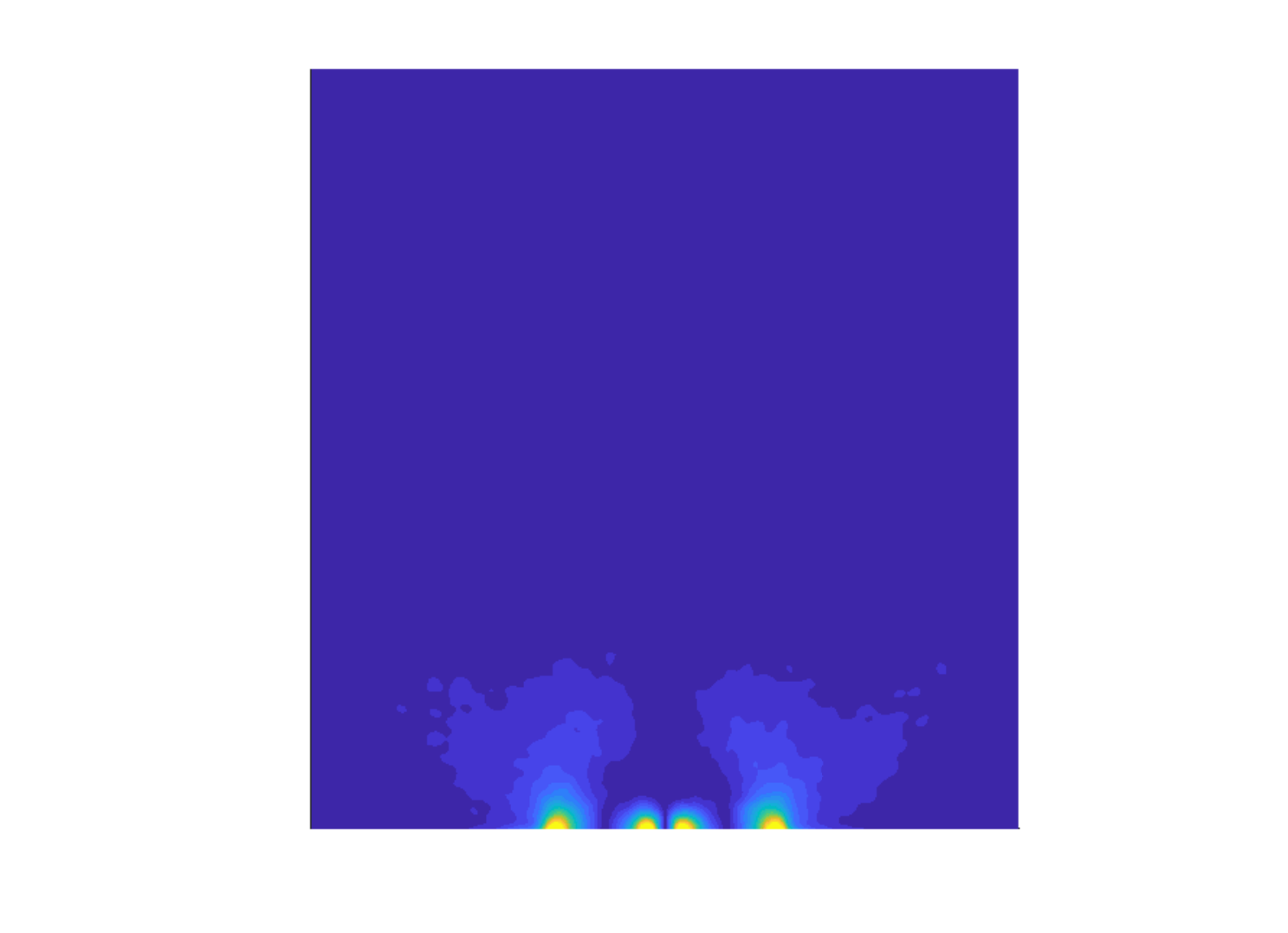}}
	\hspace{0.01cm}
	\subfloat[][]{\includegraphics[clip,trim=3.15cm 1.2cm 2.65cm 0.6cm,width=0.14\linewidth]{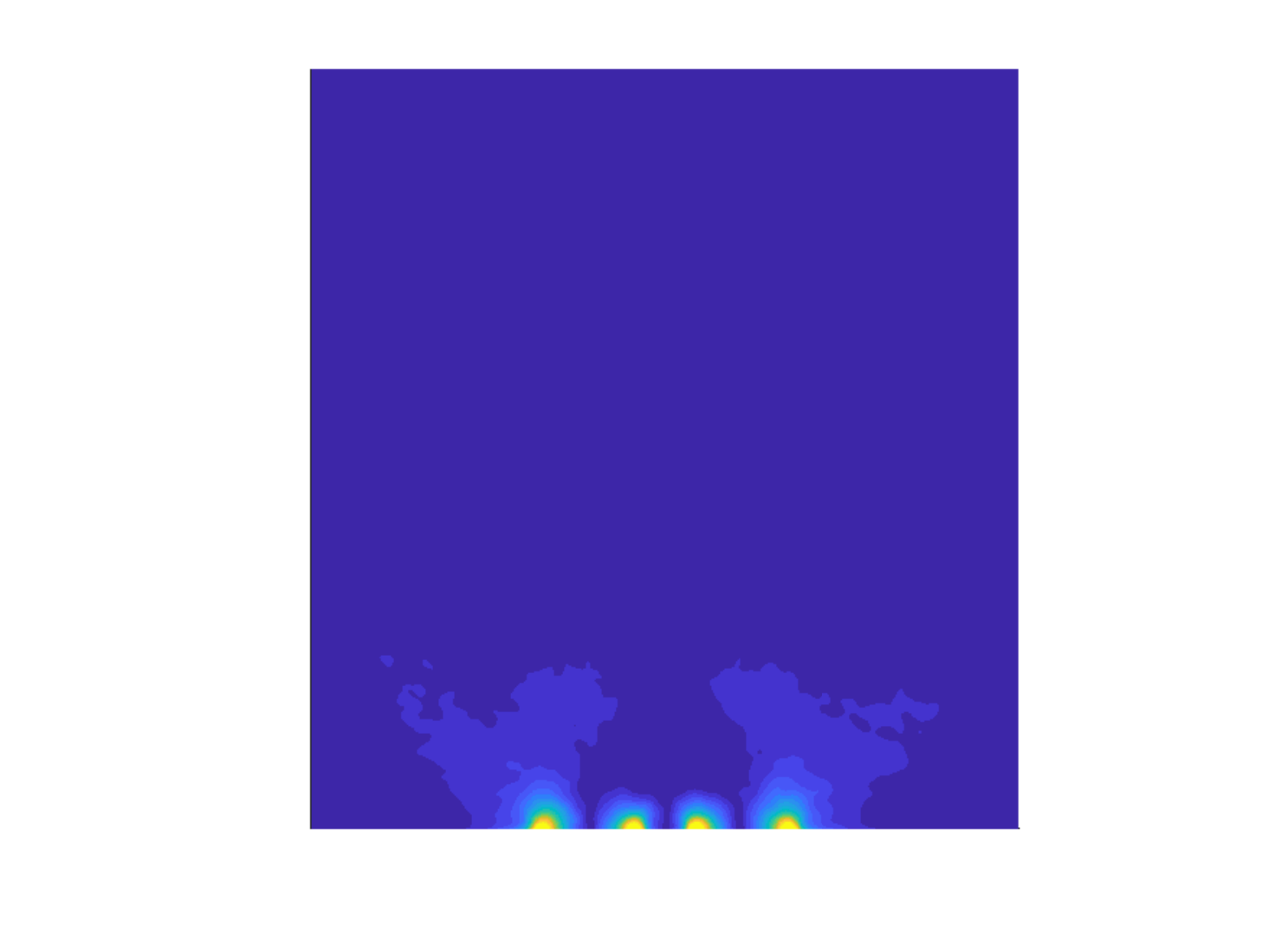}}
	\hspace{0.01cm}
	\subfloat[][]{\includegraphics[clip,trim=3.15cm 1.2cm 2.65cm 0.6cm,width=0.14\linewidth]{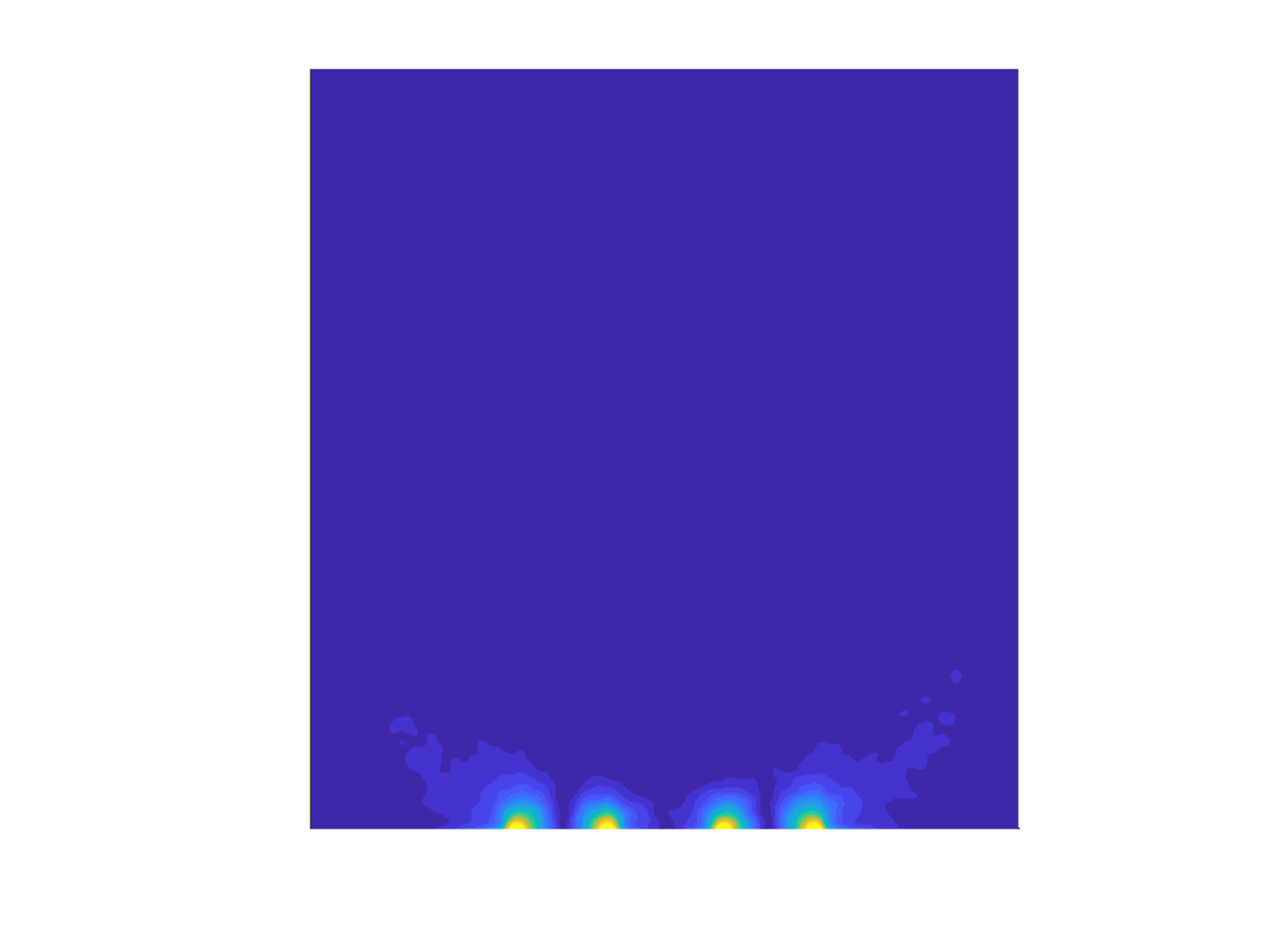}}
	\hspace{0.01cm} 
	\subfloat[][]{\includegraphics[clip,trim=3.15cm 1.2cm 2.65cm 0.6cm,width=0.14\linewidth]{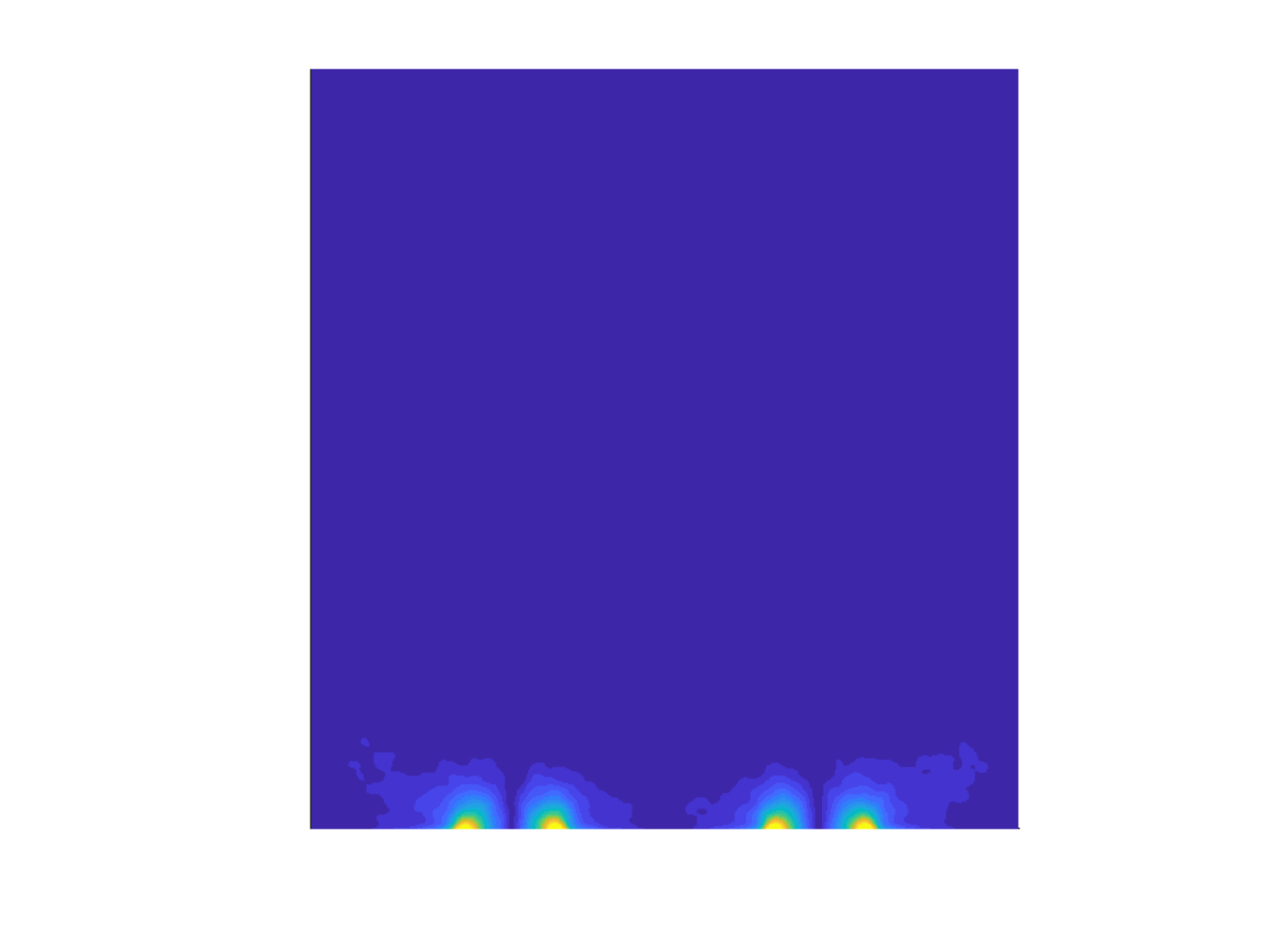}}
	\hspace{0.01cm}
	\includegraphics[width=0.054\linewidth]{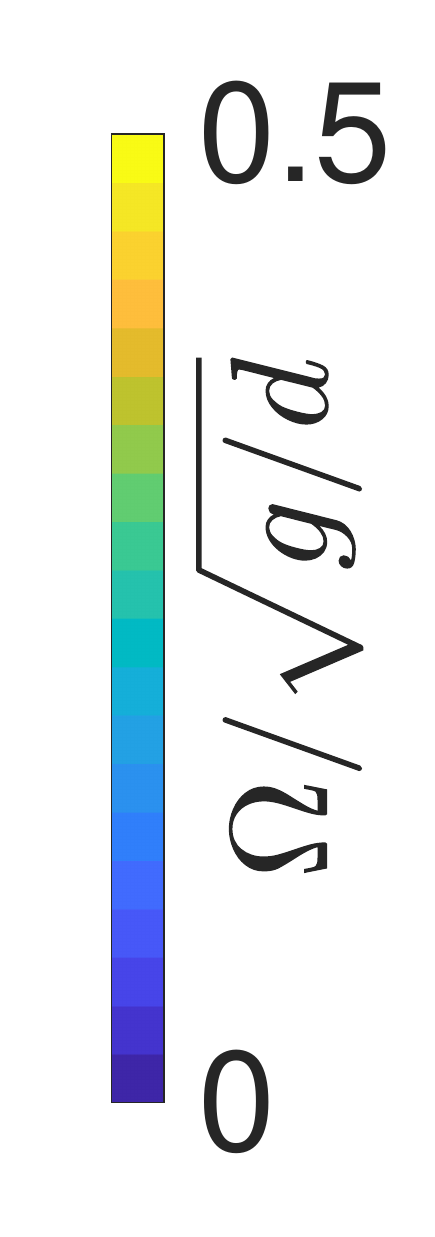}
	\caption{\label{fields:omegm}Spatial distribution of rotational velocity $\Omega $ at different spacings $L/d$ = a) 0.0, b) 2.5, c) 5, d) 10, e) 20 and f) 40 between the orifices, each of width $W/d=20$, placed on the silo base. The fraction of dumbbells for all the cases is $X_{db}=0.5$. }
\end{figure}

\subsubsection{Flow fields at various $L/d$ \label{sec:coarsesepa}}

\begin{figure}	
	\centering
	\subfloat[][]{\includegraphics[clip,trim=3.15cm 1.2cm 2.65cm 0.6cm,width=0.14\linewidth]{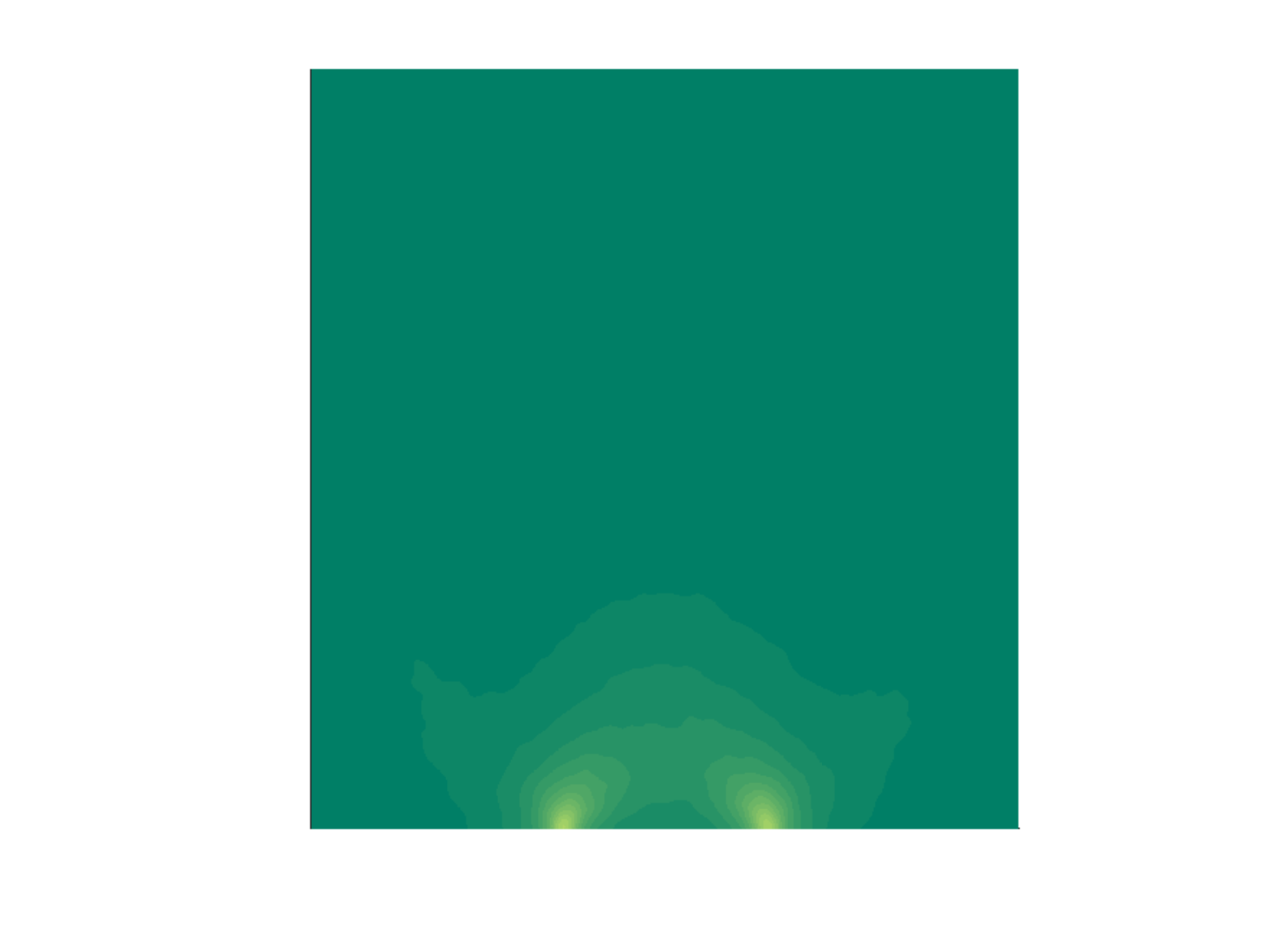}}
	\hspace{0.01cm}
	\subfloat[][]{\includegraphics[clip,trim=3.15cm 1.2cm 2.65cm 0.6cm,width=0.14\linewidth]{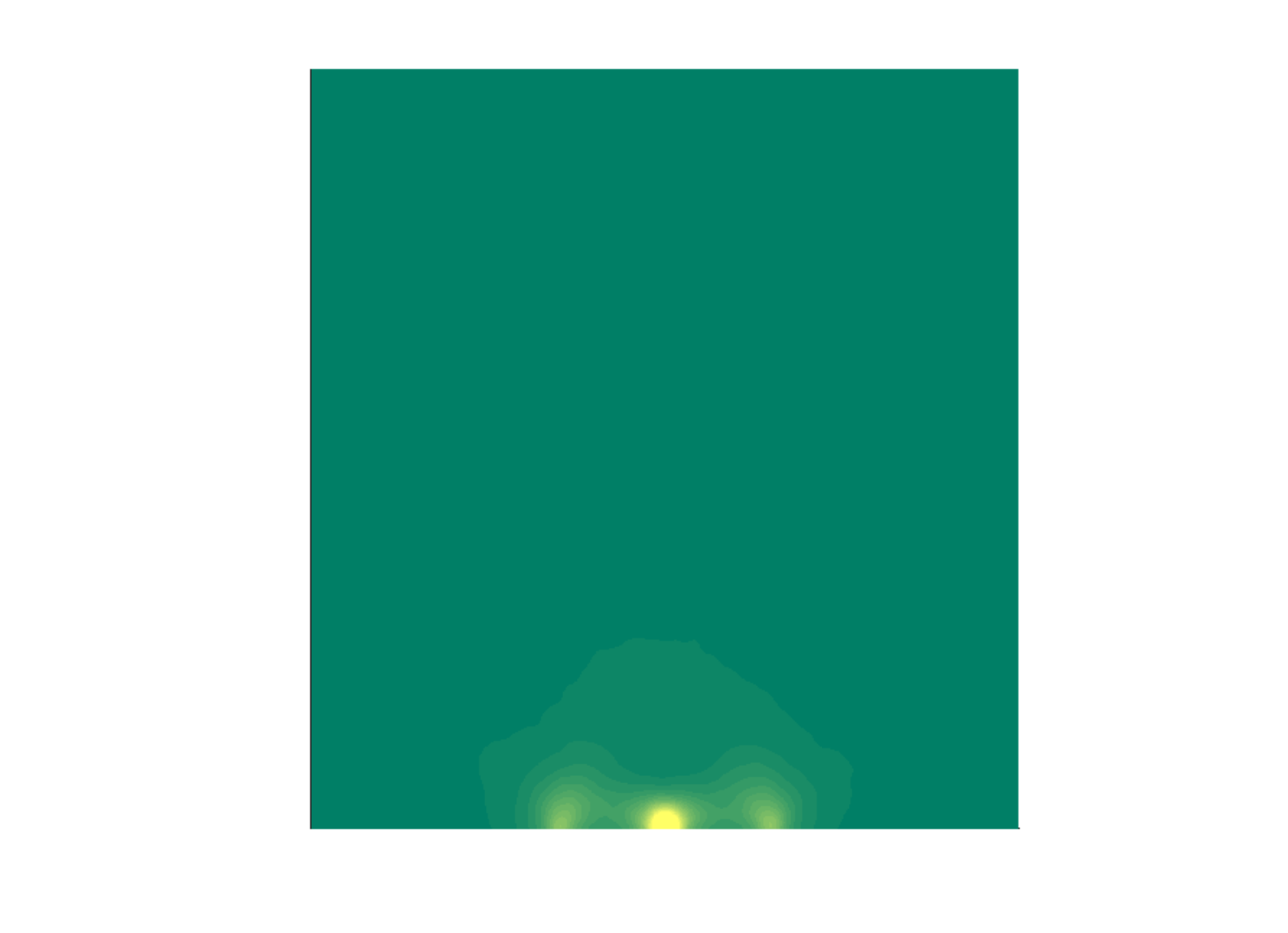}}
	\hspace{0.01cm}
	\subfloat[][]{\includegraphics[clip,trim=3.15cm 1.2cm 2.65cm 0.6cm,width=0.14\linewidth]{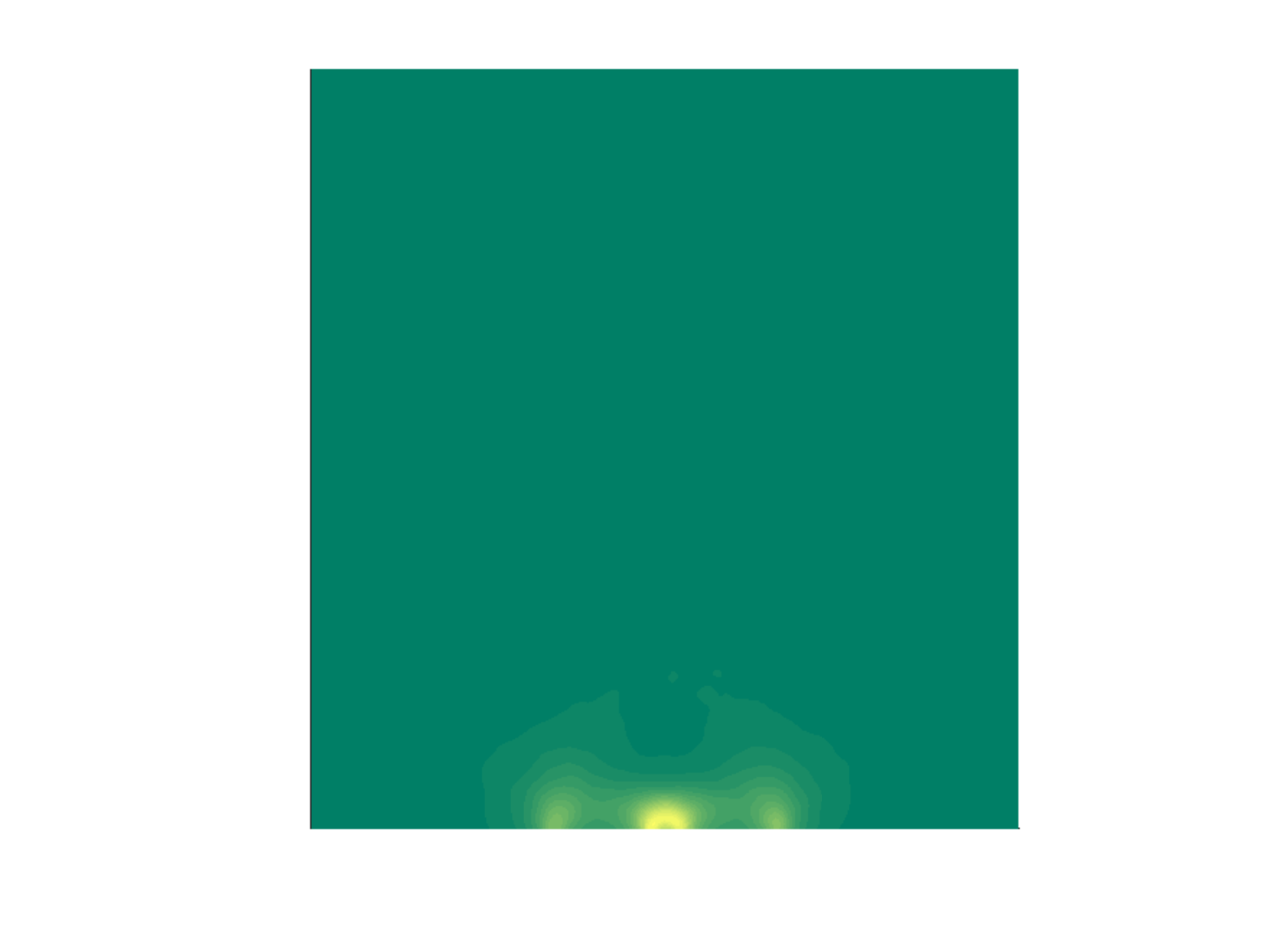}}
	\hspace{0.01cm}
	\subfloat[][]{\includegraphics[clip,trim=3.15cm 1.2cm 2.65cm 0.6cm,width=0.14\linewidth]{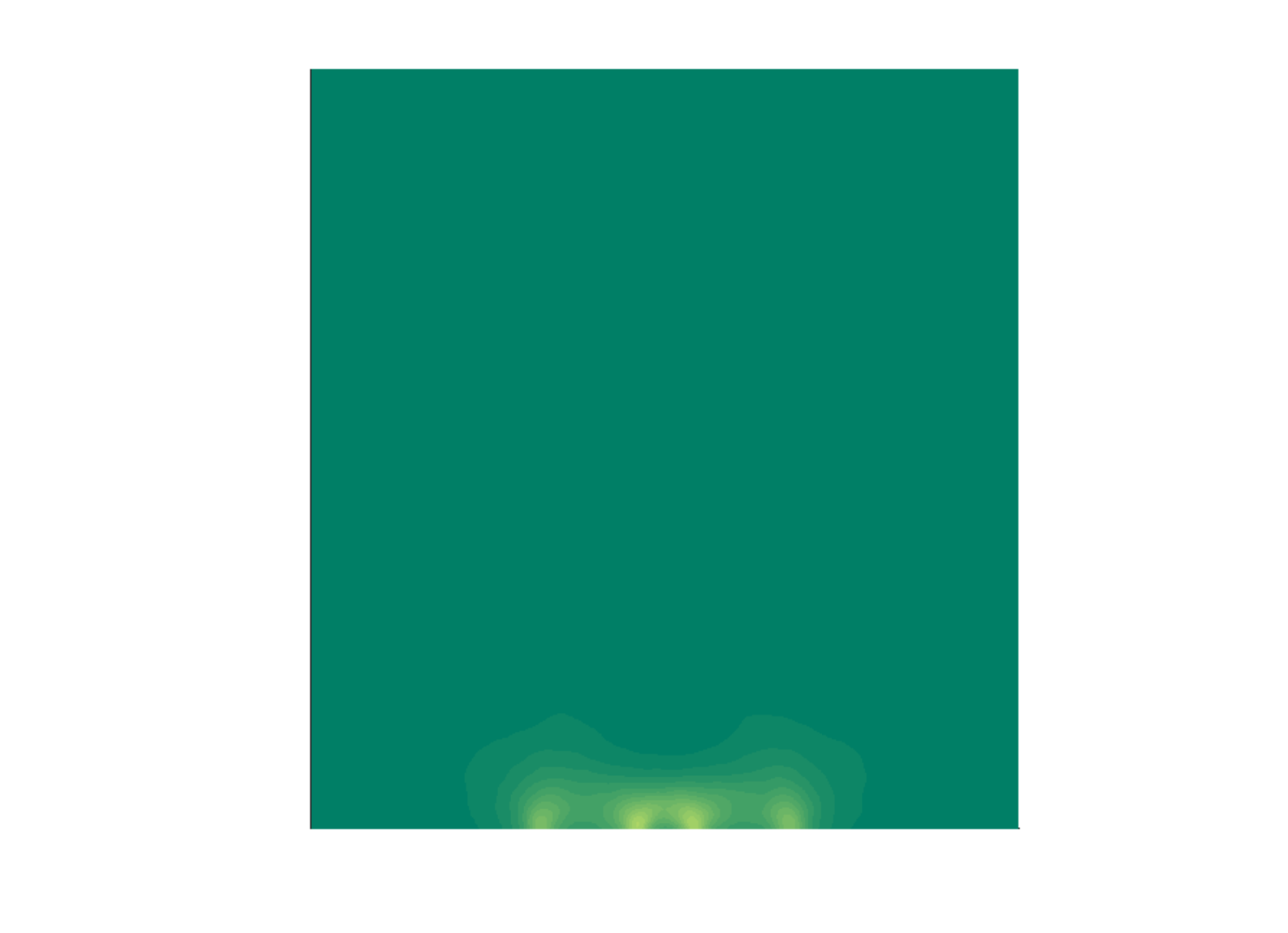}}
	\hspace{0.01cm}
	\subfloat[][]{\includegraphics[clip,trim=3.15cm 1.2cm 2.65cm 0.6cm,width=0.14\linewidth]{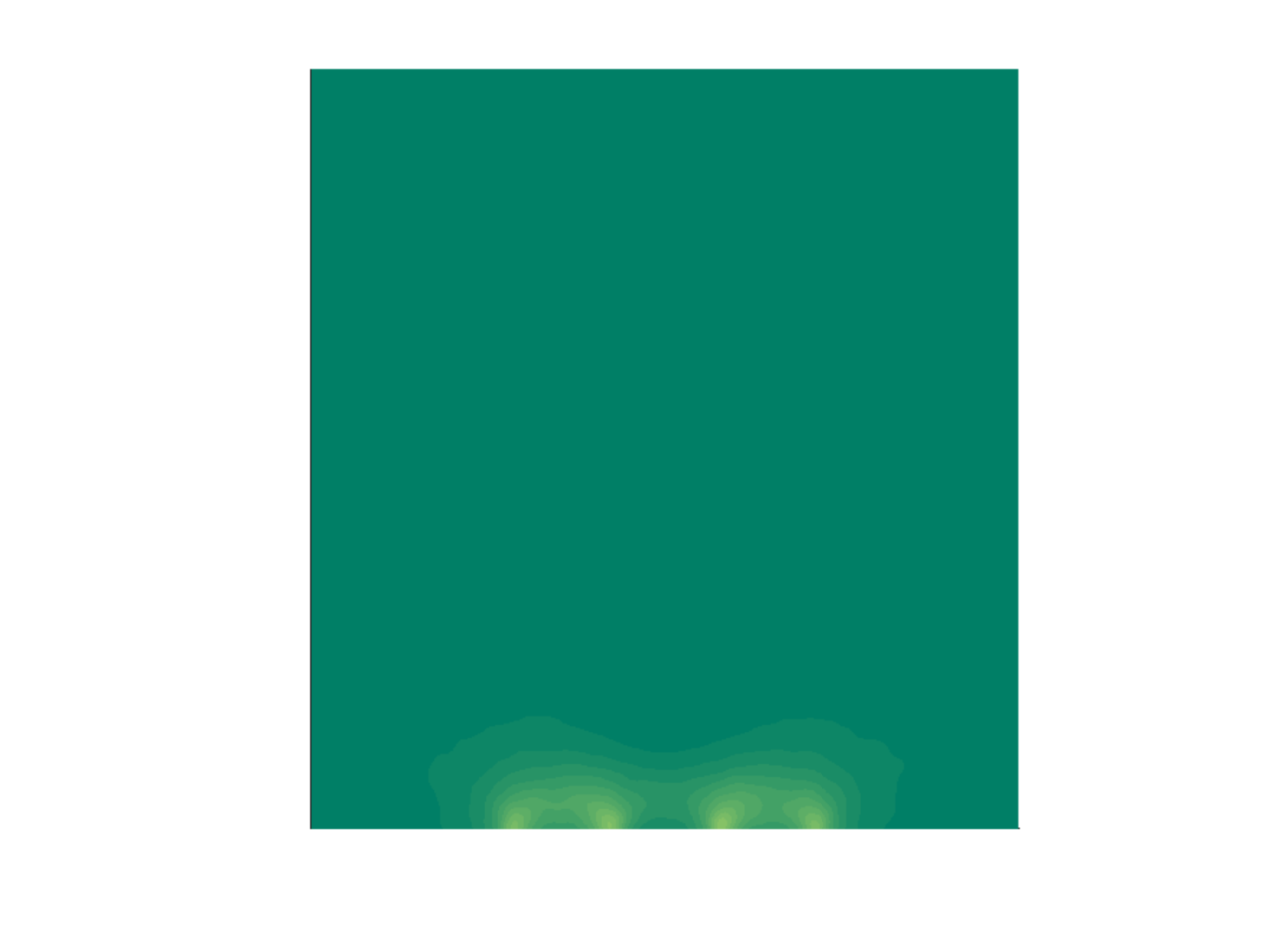}}
	\hspace{0.01cm} 
	\subfloat[][]{\includegraphics[clip,trim=3.15cm 1.2cm 2.65cm 0.6cm,width=0.14\linewidth]{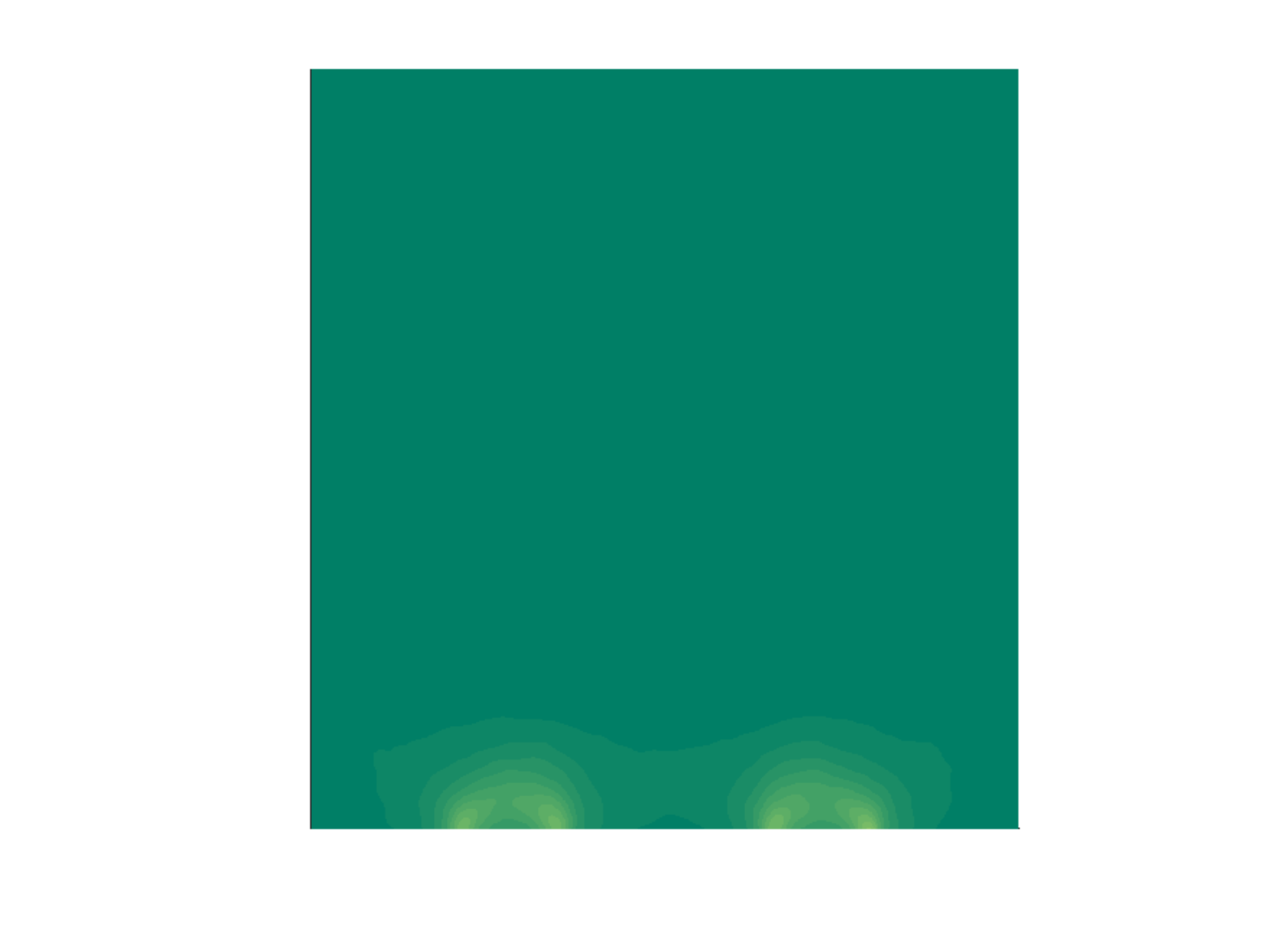}}
	\hspace{0.01cm}
	\includegraphics[width=0.054\linewidth]{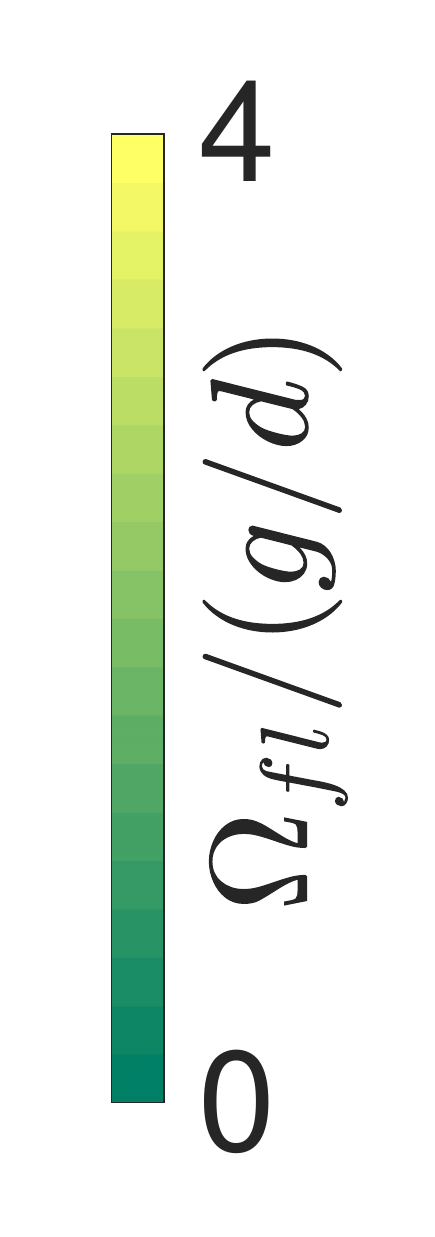}
	\caption{\label{fields:omegflucm}Spatial distribution of fluctuations in rotational velocity $\Omega_{fl} $ at different spacings $L/d$ = a) 0.0, b) 2.5, c) 5, d) 10, e) 20 and f) 40 between the orifices, each of width $W/d=20$, placed on the silo base. The fraction of dumbbells for all the cases is $X_{db}=0.5$. }
\end{figure}

We employed the coarse-graining technique as explained in section \ref{sec:coarsew} to comprehend the influence of the inter-orifice distance		 on the flow dynamics. Here, we demonstrated flow fields of area fraction $\phi$, rotational velocity $\Omega$, fluctuations in rotational velocity $\Omega_{fl}$, velocity $V$, granular temperature $T_g$, pressure $P$ and shear stress $|\tau|$. Each parameter is plotted at six different separation distances $L/d$ ranging from 0.0 to 40.0 in the region: $-68.5\le x \le 68.5$ and $1.5 \le y \le 148.5$. All the flow fields correspond to the fraction of dumbbells $X_{db}=0.5$ and width of each of the orifice is $W/d=20$. The variation in $L/d$ has little effect on $\phi$ in the bulk (figure \ref{fields:pfm}) as the fraction of dumbbells is same for all the cases. In the region above the orifice due to shear-induced dilation, area fraction is noticed to be less. With an increase in $L/d$, the region of dilation decreases gradually and for $L/d\ge10$, it is confined to small regions just above the orifices. Rotational velocity $\Omega$ is almost negligible in the bulk as the particle rotations are not possible because the particles are closely packed. As $L/d$ increases, $\Omega$ is present on either side of the two orifices. Until $L/d=10$, rotational velocity varied slightly from that of single big orifice case ($L/d=0$), however at $L/d>10$, the $\Omega$ at each orifice is found to be independent of the other orifice. Fluctuations in rotational velocity are almost negligible in the bulk (figure \ref{fields:omegflucm}) as the rotational velocity is almost constant in the bulk as observed in figure \ref{fields:omegm}. In the region above the orifice, $\Omega_{fl}$ is found to be less at inter-orifice distance $L/d=0$ and $L/d\ge20$. However, $\Omega_{fl}$ is found to be more in the region between the two orifices for $2.5 \le L/d \le 10$ as the particles present in between the orifices tends to discharge through either of the orifices.

\begin{figure}	
	\centering
	\subfloat[][]{\includegraphics[clip,trim=3.15cm 1.2cm 2.65cm 0.6cm,width=0.14\linewidth]{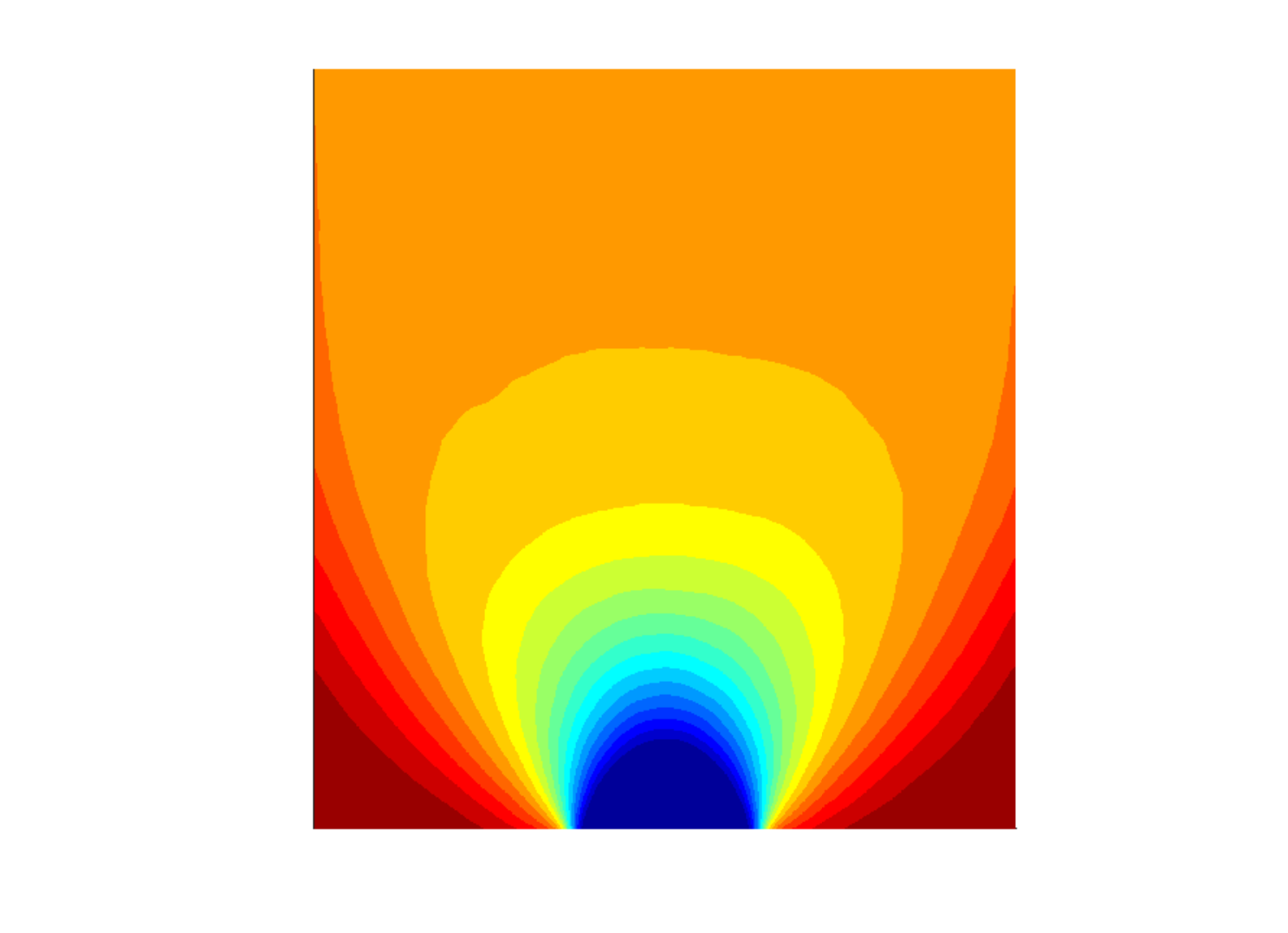}}
	\hspace{0.01cm}
	\subfloat[][]{\includegraphics[clip,trim=3.15cm 1.2cm 2.65cm 0.6cm,width=0.14\linewidth]{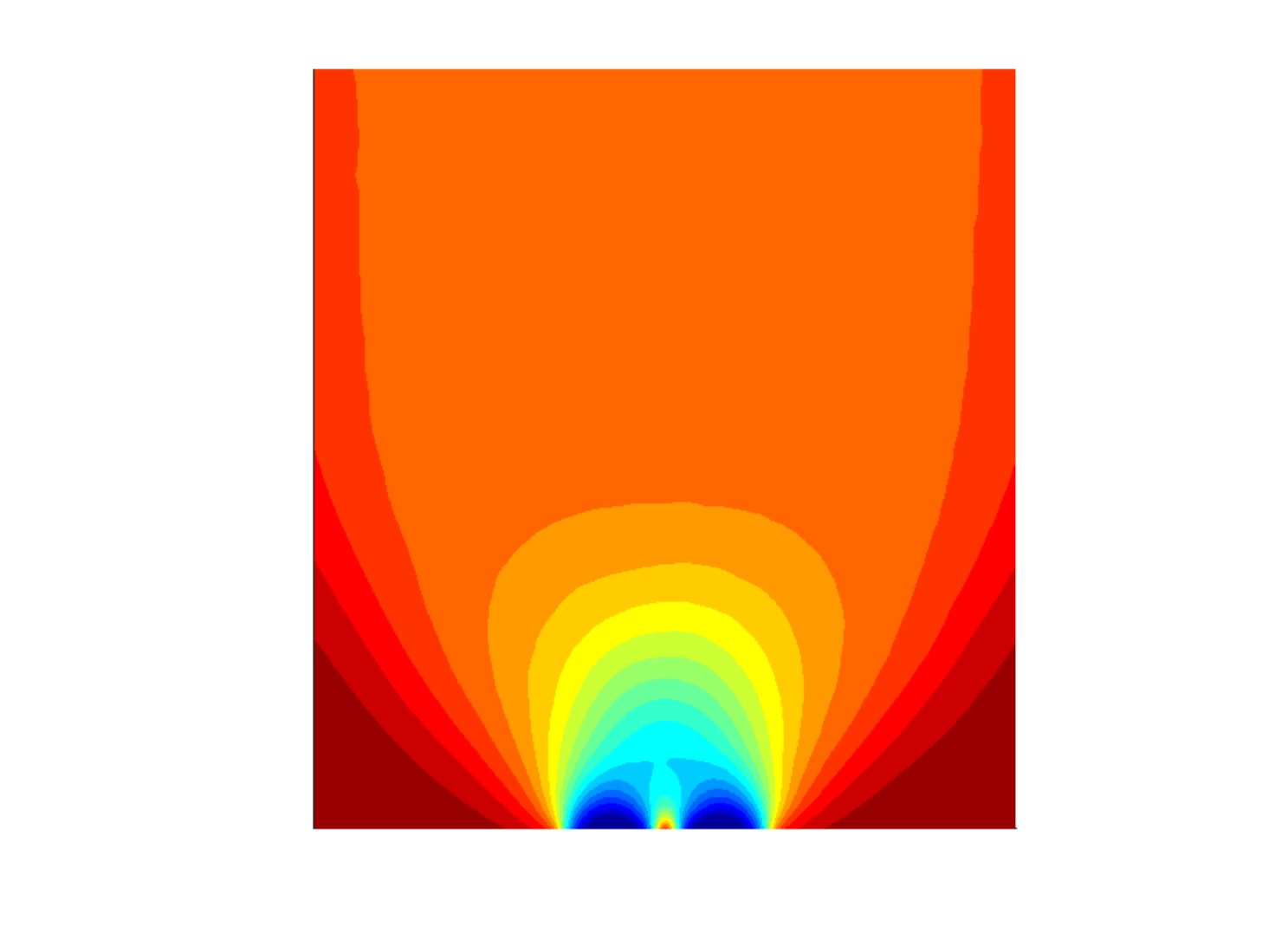}}
	\hspace{0.01cm}
	\subfloat[][]{\includegraphics[clip,trim=3.15cm 1.2cm 2.65cm 0.6cm,width=0.14\linewidth]{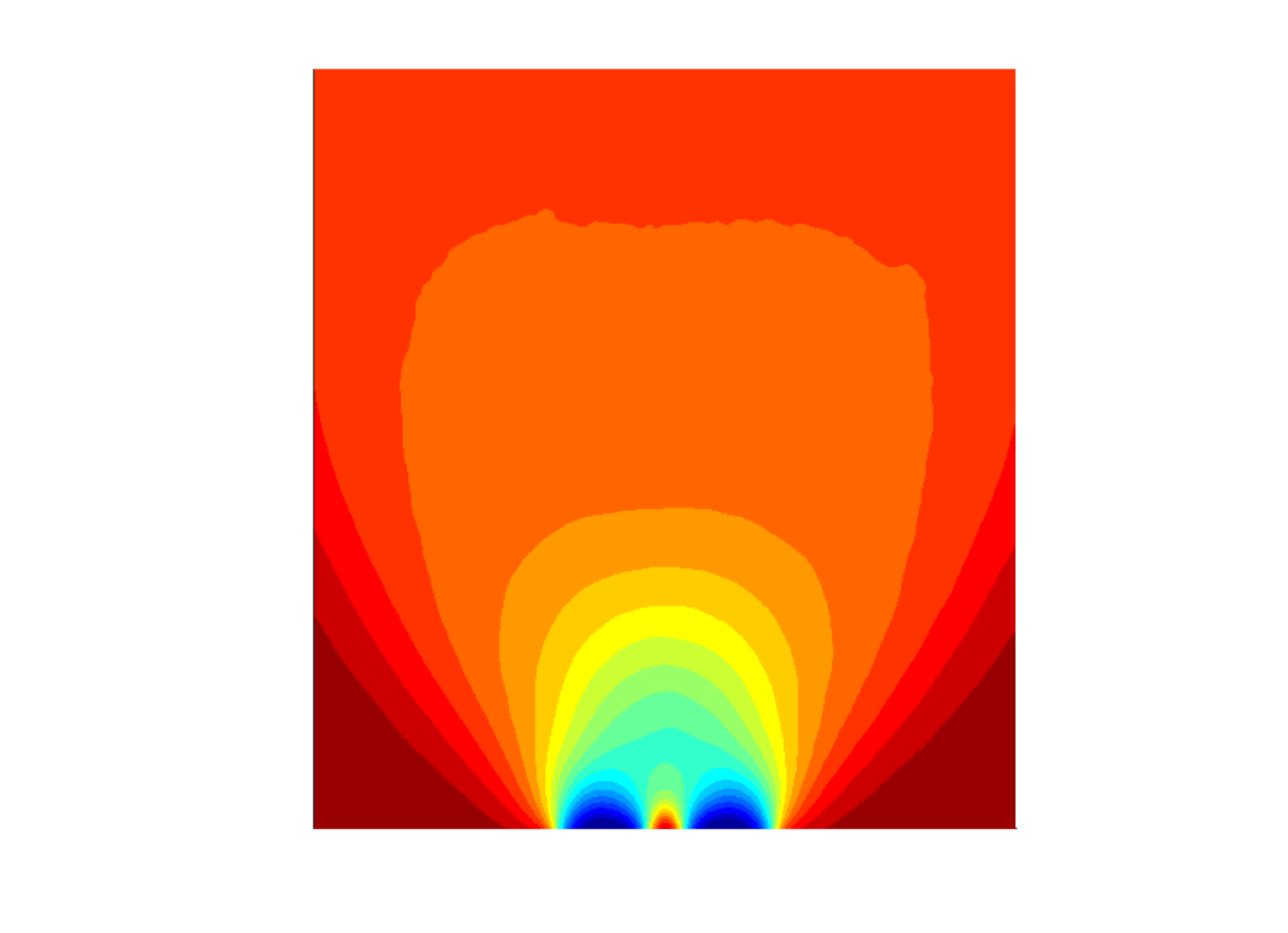}}
	\hspace{0.01cm}
	\subfloat[][]{\includegraphics[clip,trim=3.15cm 1.2cm 2.65cm 0.6cm,width=0.14\linewidth]{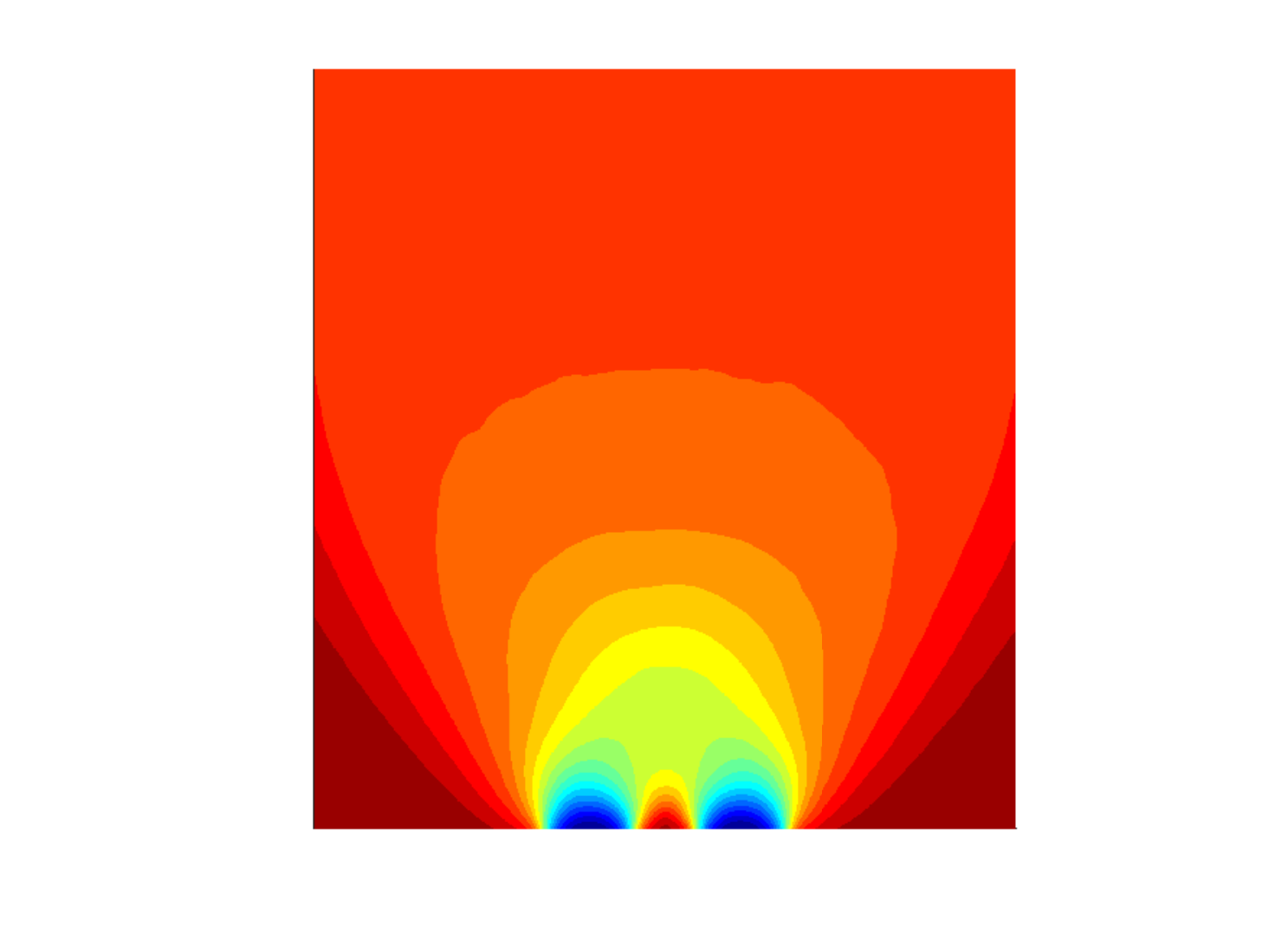}}
	\hspace{0.01cm}
	\subfloat[][]{\includegraphics[clip,trim=3.15cm 1.2cm 2.65cm 0.6cm,width=0.14\linewidth]{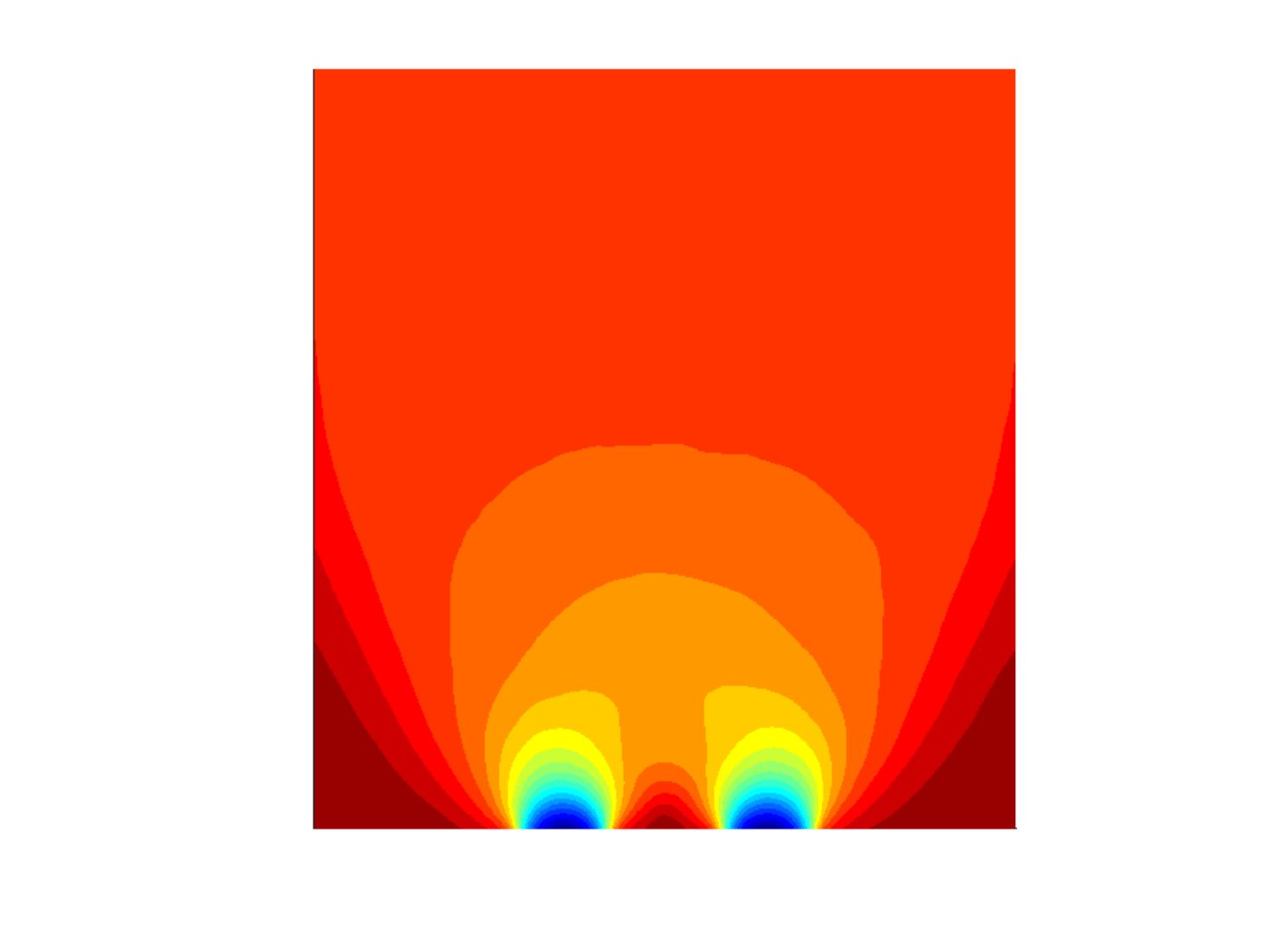}}
	\hspace{0.01cm} 
	\subfloat[][]{\includegraphics[clip,trim=3.15cm 1.2cm 2.65cm 0.6cm,width=0.14\linewidth]{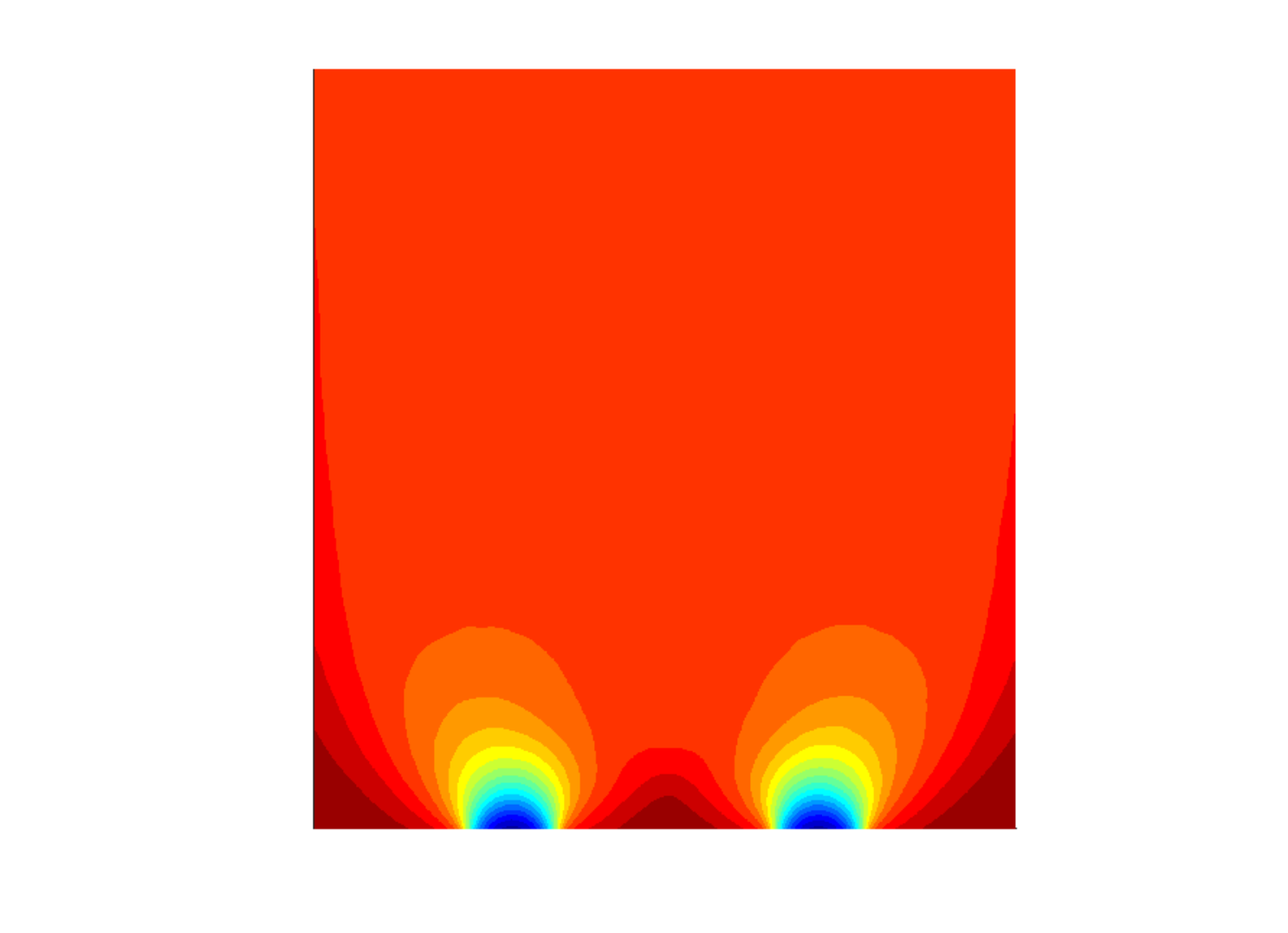}}
	\hspace{0.01cm}
	\includegraphics[width=0.054\linewidth]{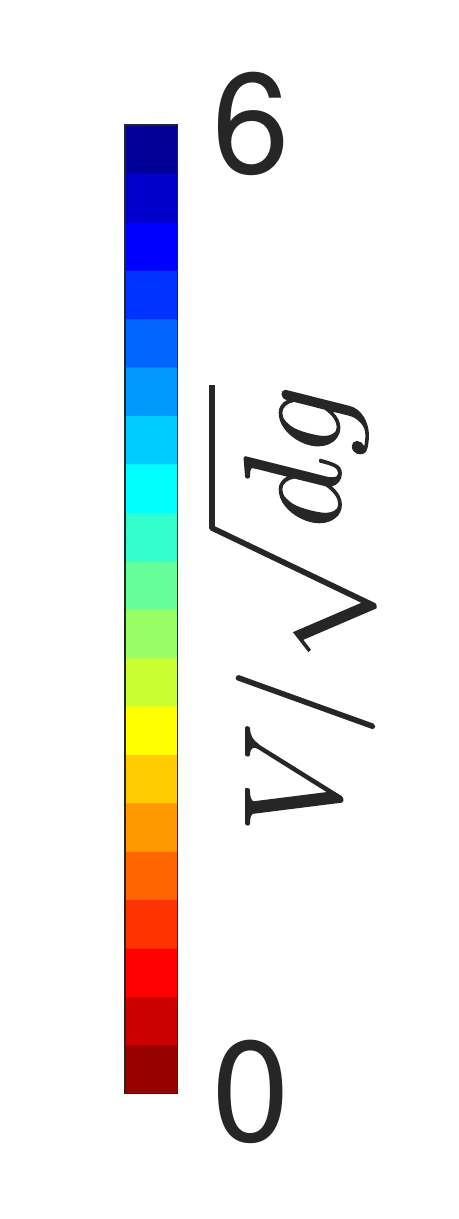}
	\caption{\label{fields:velm}Spatial distribution of velocity $V$ at different spacings $L/d$ = a) 0.0, b) 2.5, c) 5, d) 10, e) 20 and f) 40 between the orifices, each of width $W/d=20$, placed on the silo base. The fraction of dumbbells for all the cases is $X_{db}=0.5$. }
\end{figure}

\begin{figure}	
	\centering
	\subfloat[][]{\includegraphics[clip,trim=3.15cm 1.2cm 2.65cm 0.6cm,width=0.14\linewidth]{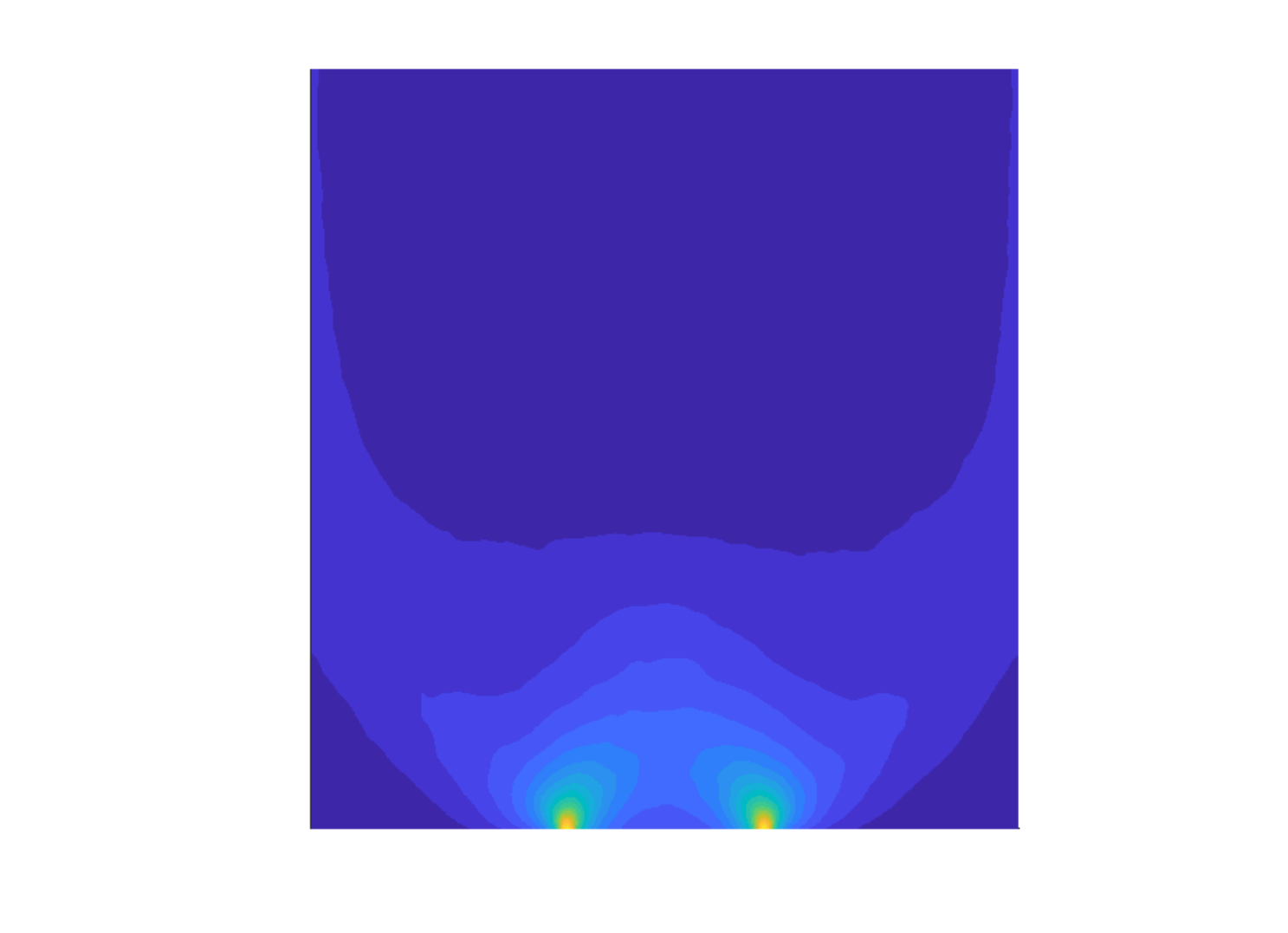}}
	\hspace{0.01cm}
	\subfloat[][]{\includegraphics[clip,trim=3.15cm 1.2cm 2.65cm 0.6cm,width=0.14\linewidth]{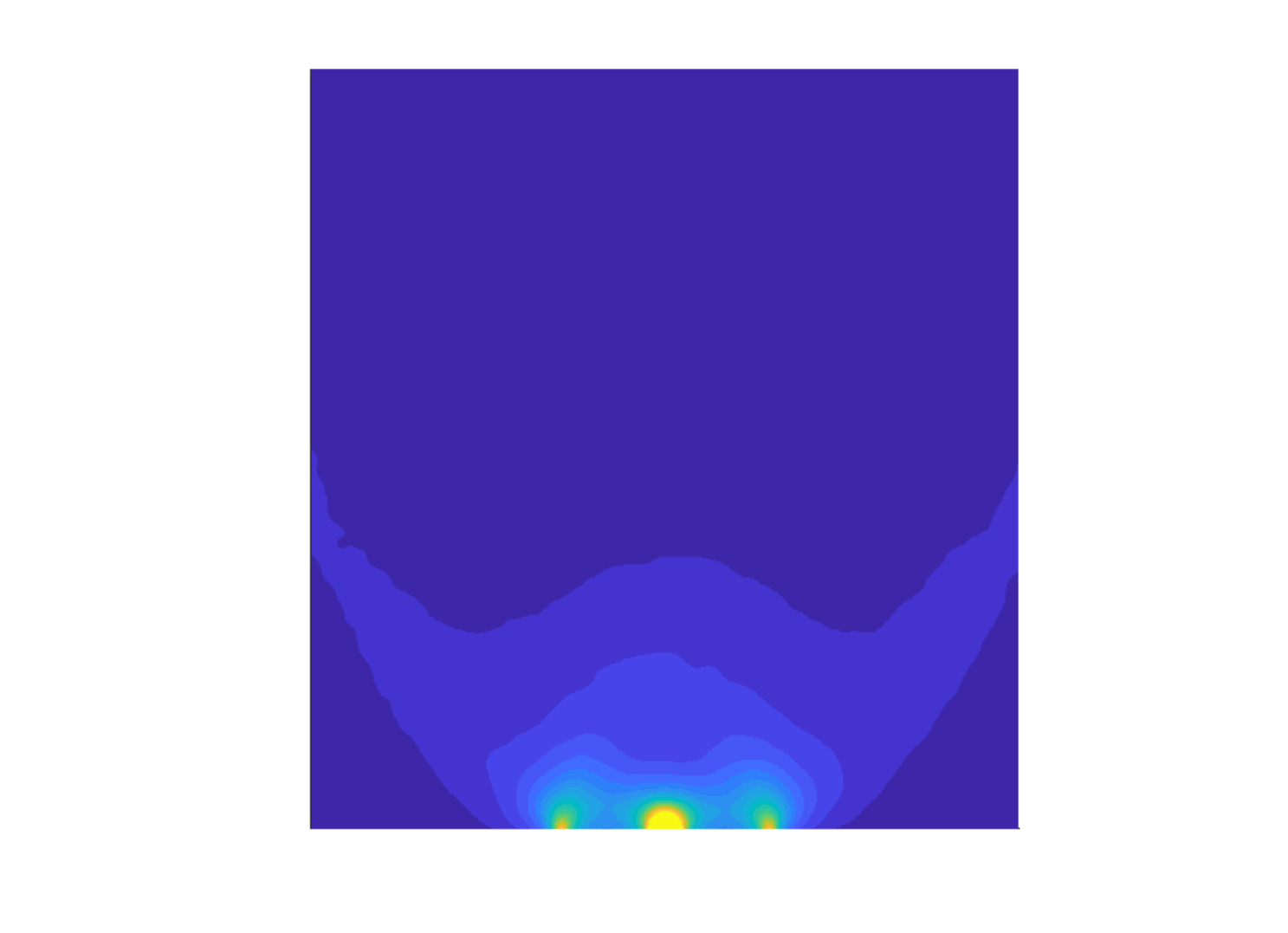}}
	\hspace{0.01cm}
	\subfloat[][]{\includegraphics[clip,trim=3.15cm 1.2cm 2.65cm 0.6cm,width=0.14\linewidth]{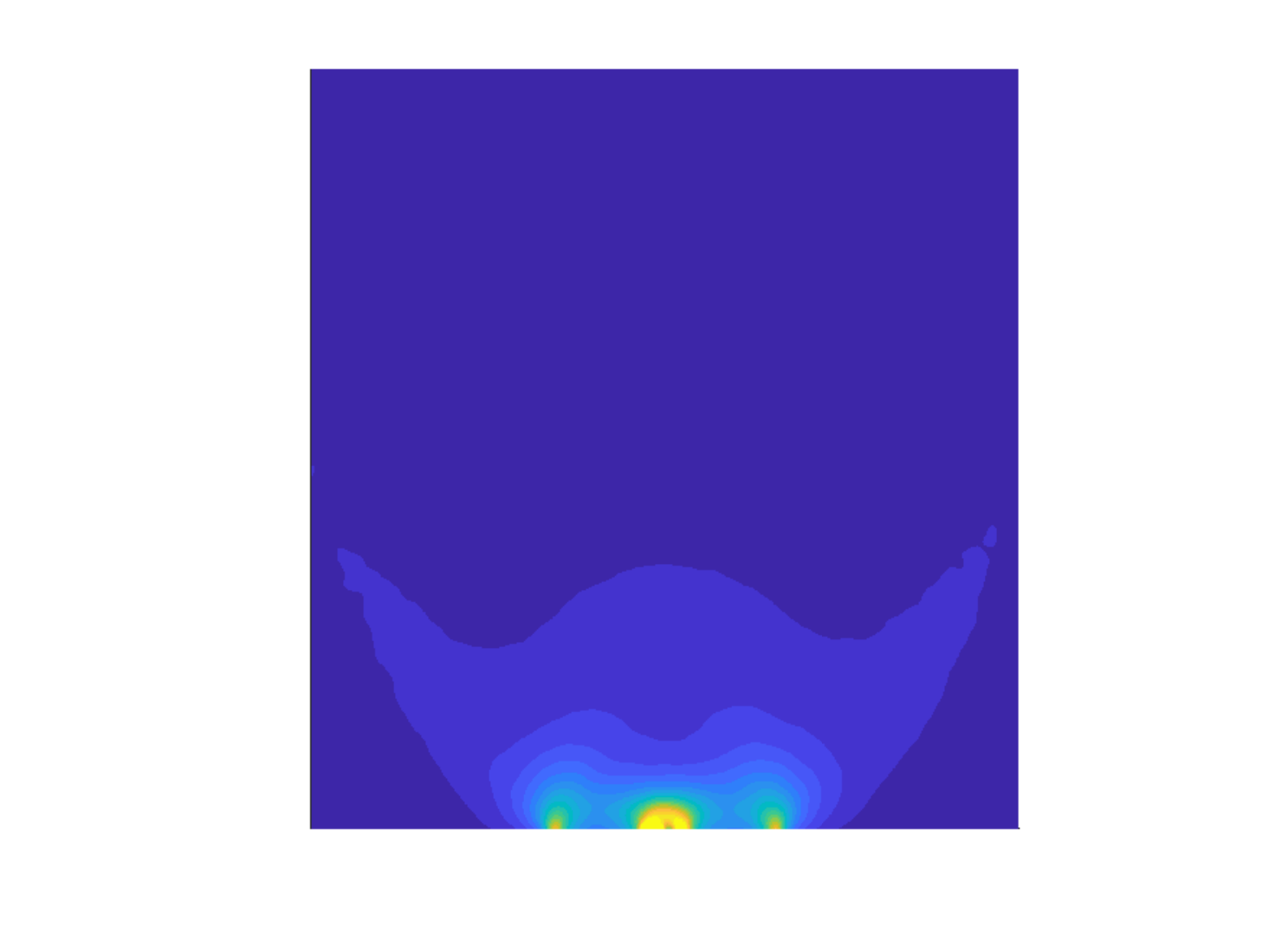}}
	\hspace{0.01cm}
	\subfloat[][]{\includegraphics[clip,trim=3.15cm 1.2cm 2.65cm 0.6cm,width=0.14\linewidth]{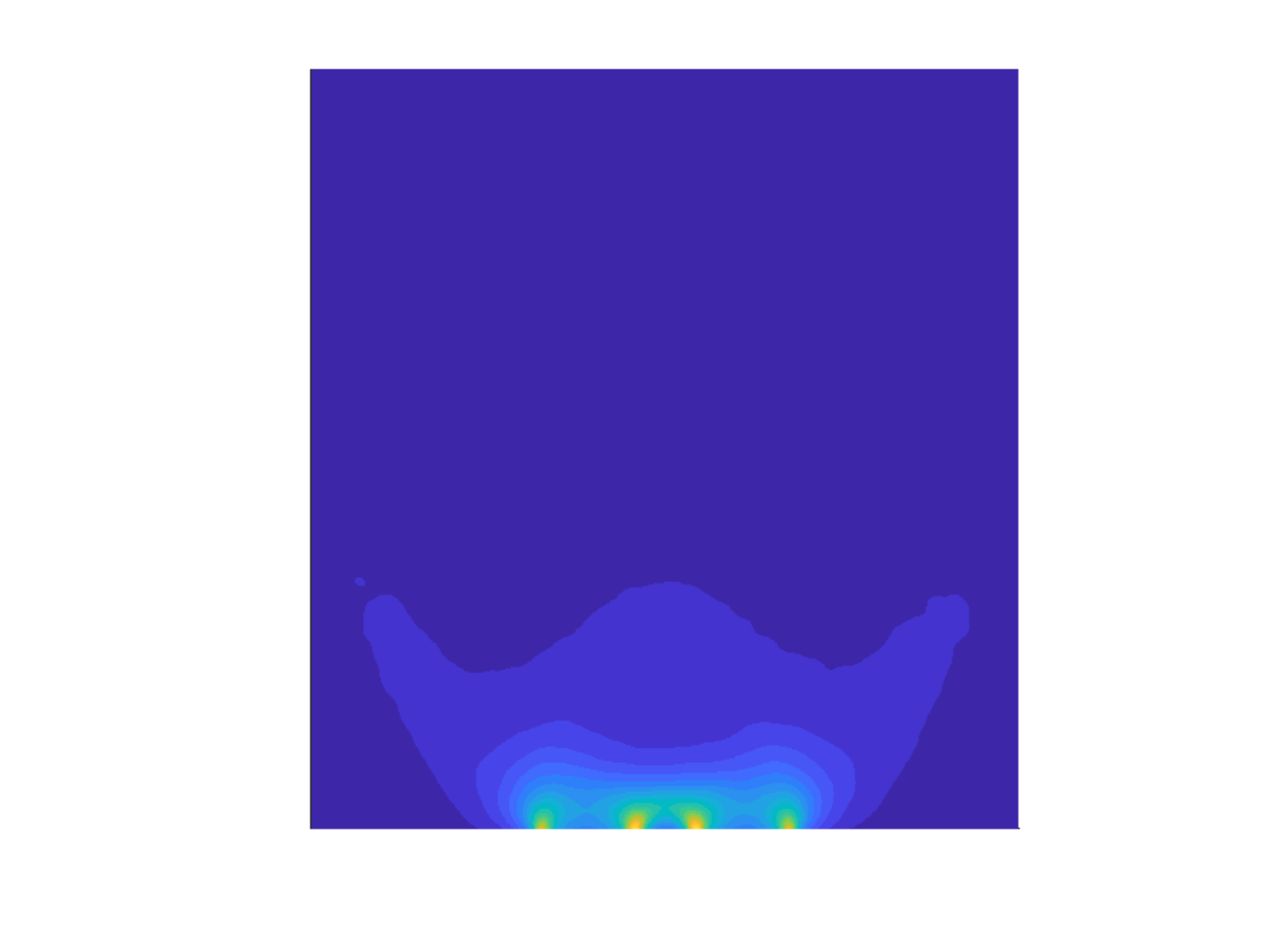}}
	\hspace{0.01cm}
	\subfloat[][]{\includegraphics[clip,trim=3.15cm 1.2cm 2.65cm 0.6cm,width=0.14\linewidth]{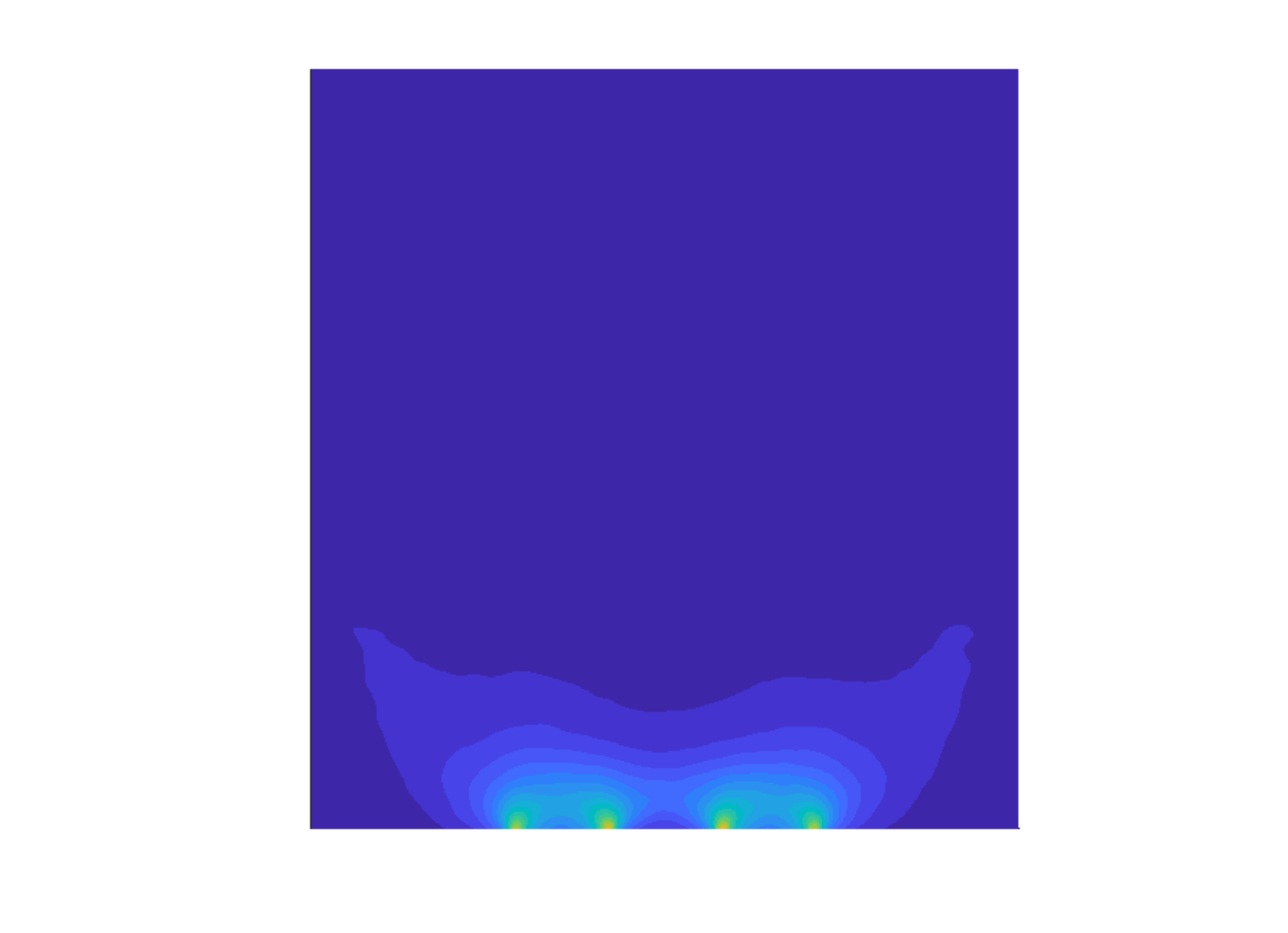}}
	\hspace{0.01cm} 
	\subfloat[][]{\includegraphics[clip,trim=3.15cm 1.2cm 2.65cm 0.6cm,width=0.14\linewidth]{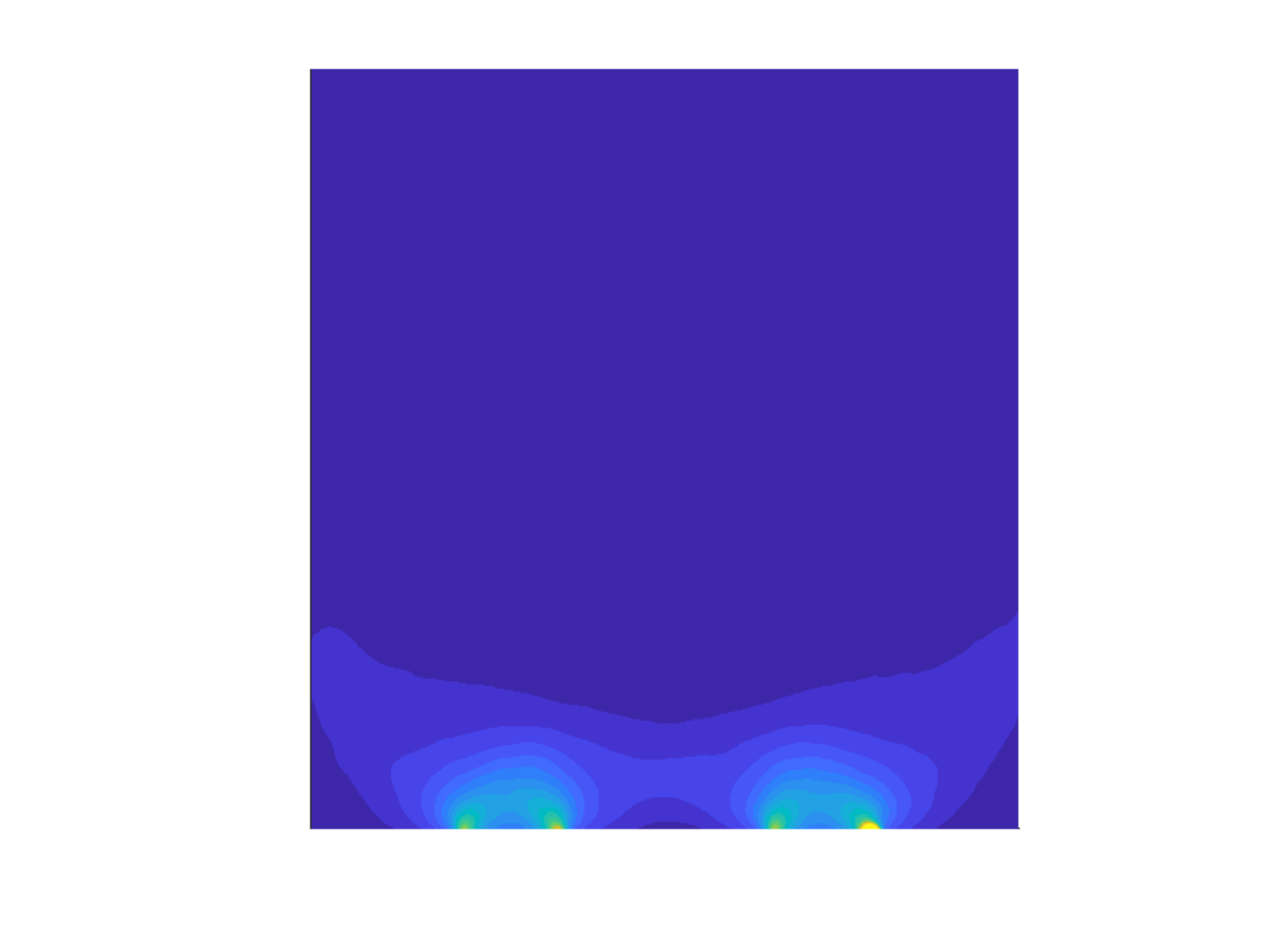}}
	\hspace{0.01cm}
	\includegraphics[width=0.054\linewidth]{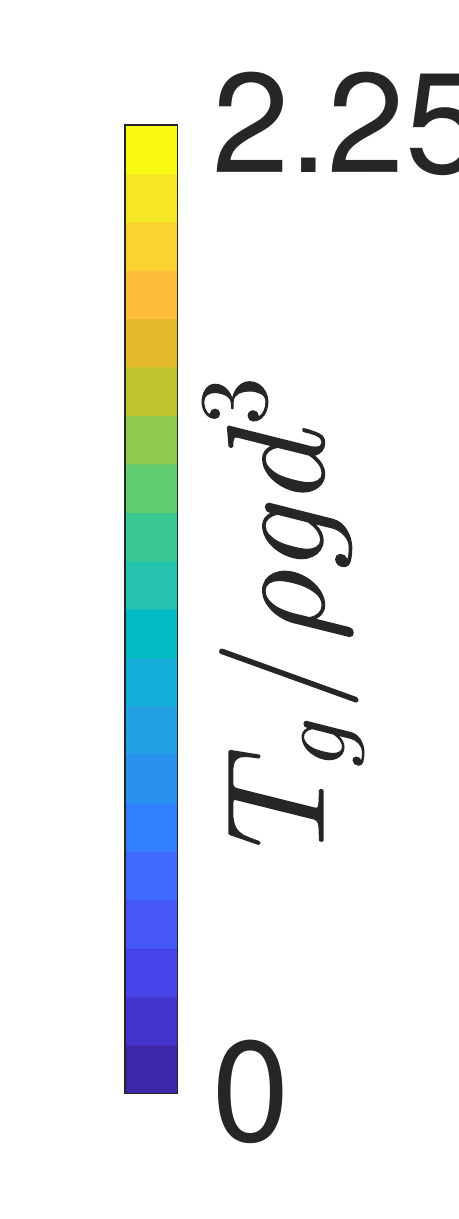}
	\caption{\label{fields:tempm}Spatial distribution of granular temperature $T_g$ at different spacings $L/d$ = a) 0.0, b) 2.5, c) 5, d) 10, e) 20 and f) 40 between the orifices, each of width $W/d=20$, placed on the silo base. The fraction of dumbbells for all the cases is $X_{db}=0.5$. }
\end{figure}

Velocity fields of a mixture of discs and dumbbells at various inter-orifice distances $L/d$ are displayed in figure \ref{fields:velm}. In the bulk, the velocities $V$ of the particles are almost constant and $V$ is found to decrease with an increase in $L/d$. The orifices are found to interact until $L/d=20$ and then for $L/d=40$, they cease to interact as they are wide apart. Qualitatively similar behaviour has been noticed in \citep{multiori} while spherical beads are discharging through two orifices placed on the base of a flat-bottomed quasi-2D hopper. The stagnant zone is found to expand between the orifices from $L/d\ge10$ forming an upward-pointed triangle due to the availability of a flat base. However, for $L/d<10$, the stagnant zone is almost negligible as the base is so small that hardly two or three particles can stay on the base which would be discharged from either of the orifices. As stagnant zone hinders the flow, consequently velocity is found to decrease with an increase in $L/d$ due to the expansion of the stagnant zone. Granular temperature $T_g$ is found to be almost negligible in the bulk(figure \ref{fields:tempm}) as the velocity is almost constant. However, $T_g$ decreases with an increase in the inter-orifice distance $L/d$ in the region above the orifice due to decrease in the collisions resulting from a decrease in the particle velocities as noticed in figure \ref{fields:velm}. 

\begin{figure}	
	\centering
	\subfloat[][]{\includegraphics[clip,trim=3.15cm 1.2cm 2.65cm 0.6cm,width=0.14\linewidth]{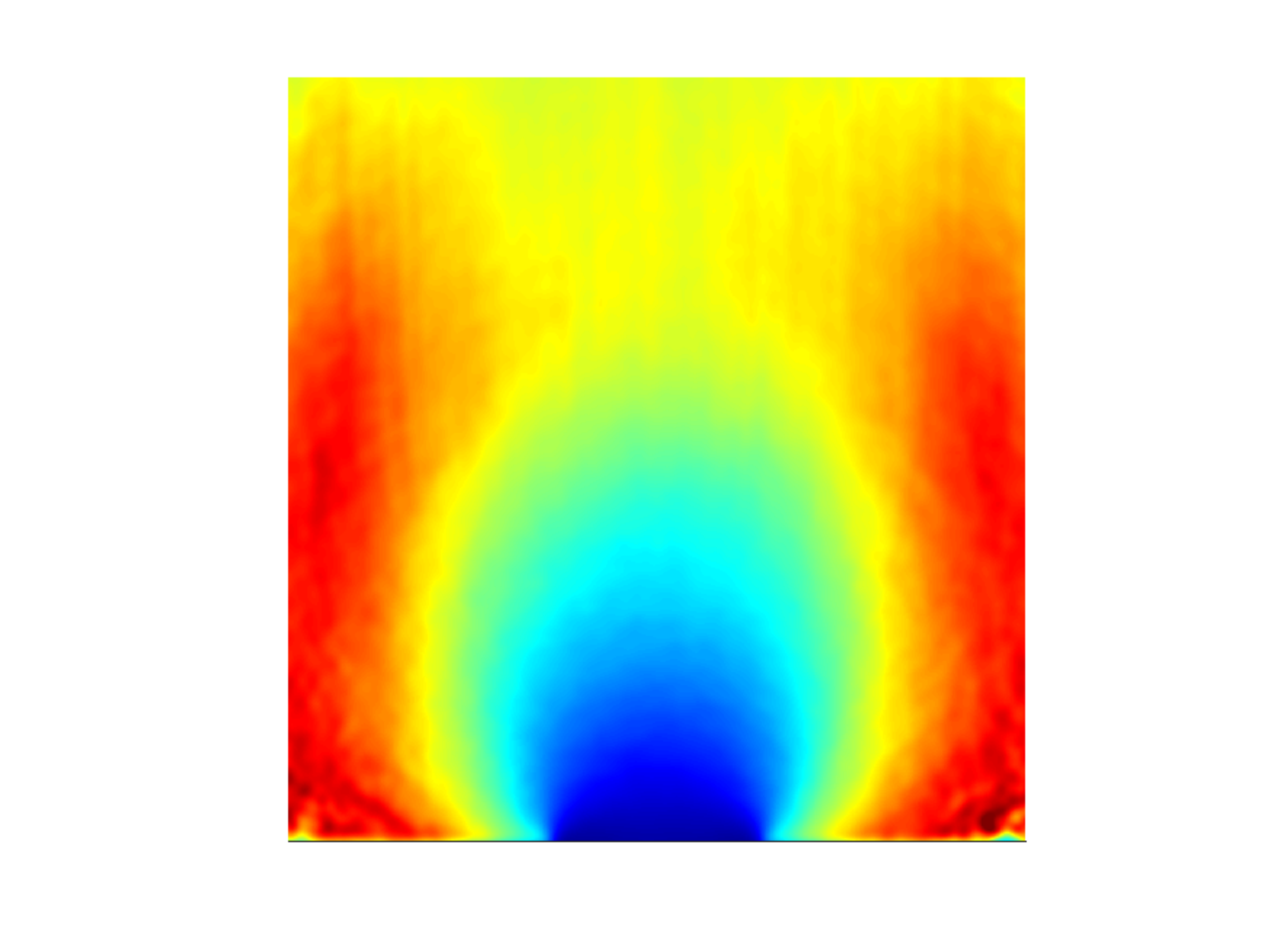}}
	\hspace{0.01cm}
	\subfloat[][]{\includegraphics[clip,trim=3.15cm 1.2cm 2.65cm 0.6cm,width=0.14\linewidth]{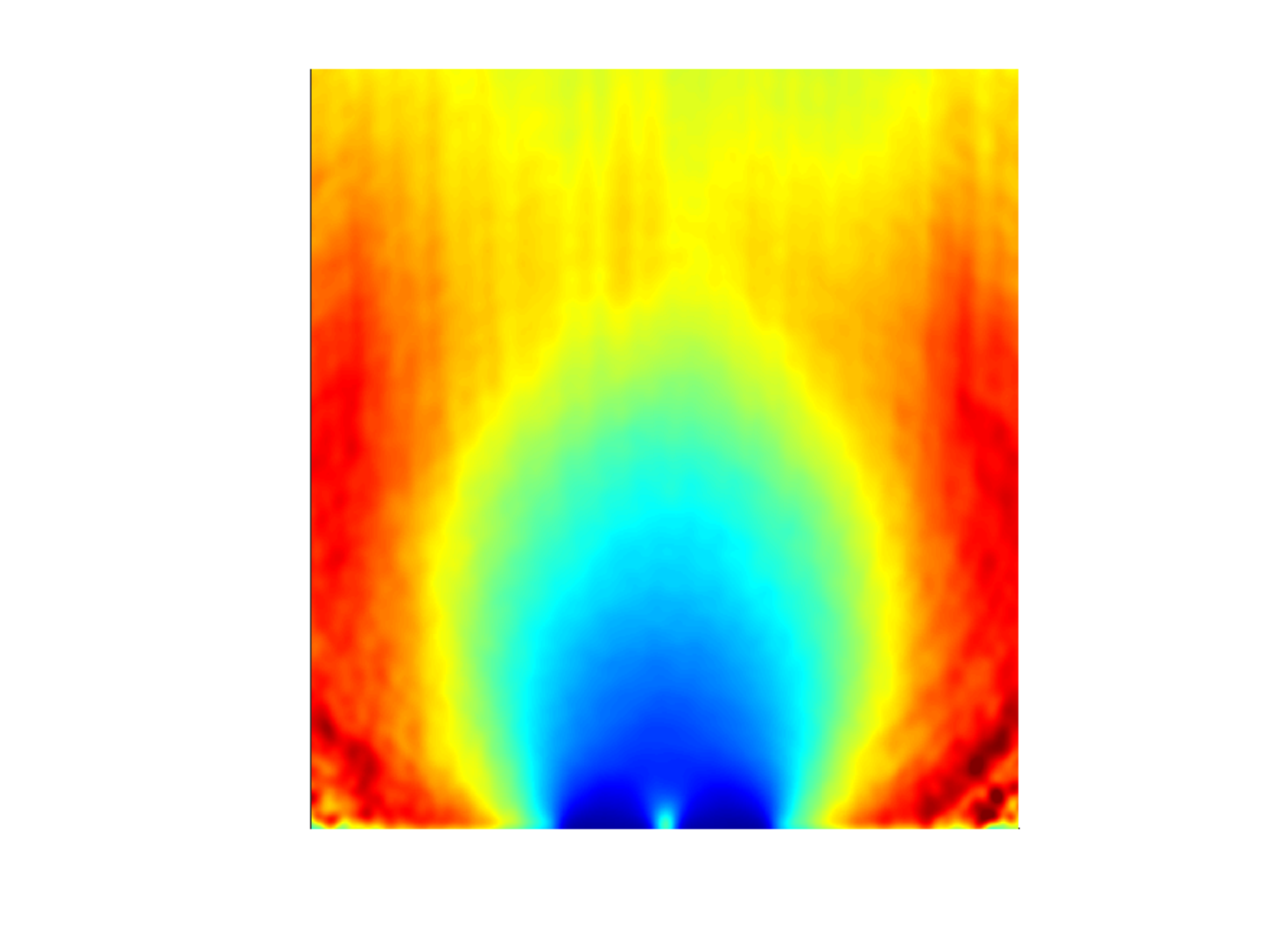}}
	\hspace{0.01cm}
	\subfloat[][]{\includegraphics[clip,trim=3.15cm 1.2cm 2.65cm 0.6cm,width=0.14\linewidth]{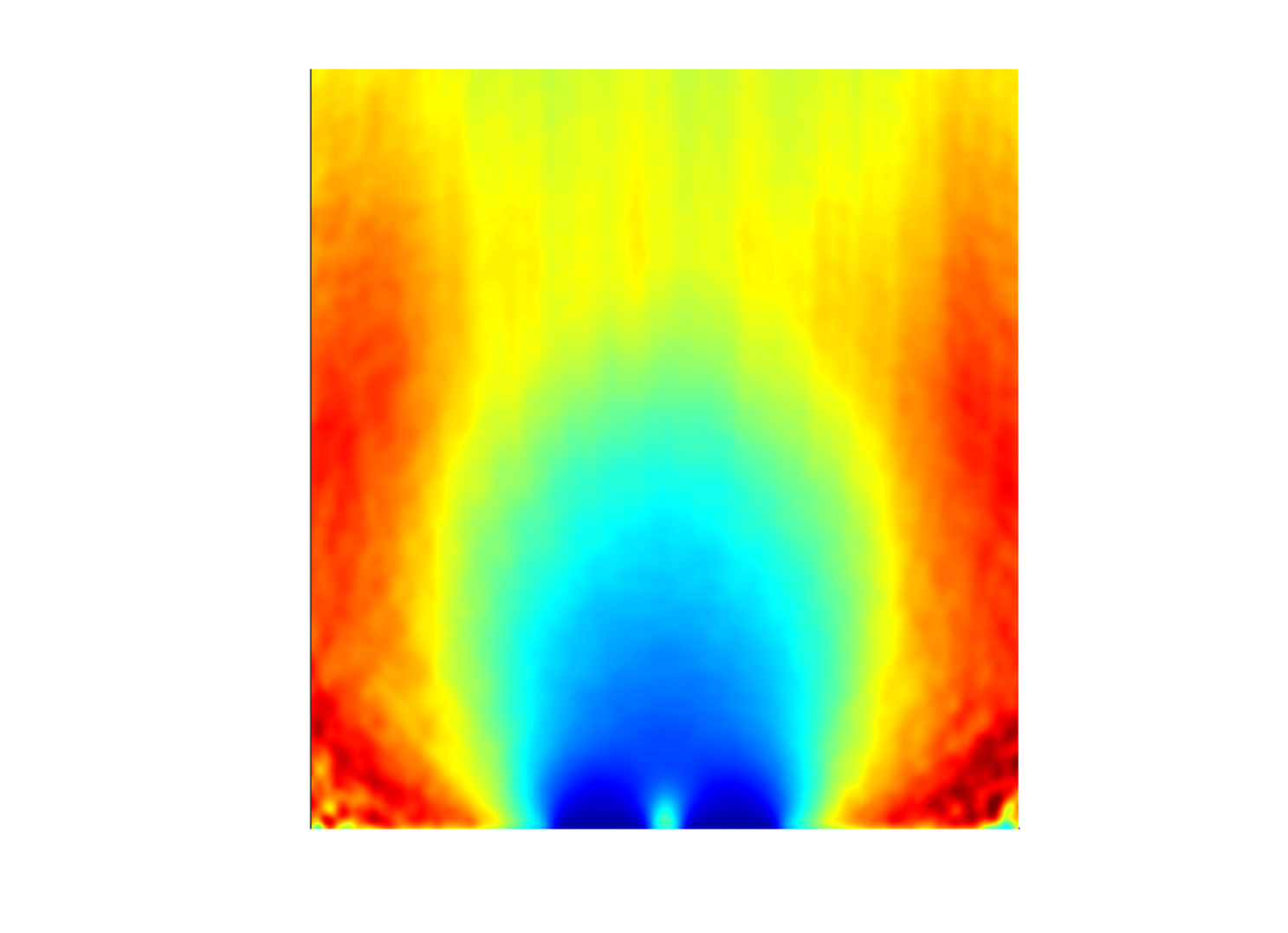}}
	\hspace{0.01cm}
	\subfloat[][]{\includegraphics[clip,trim=3.15cm 1.2cm 2.65cm 0.6cm,width=0.14\linewidth]{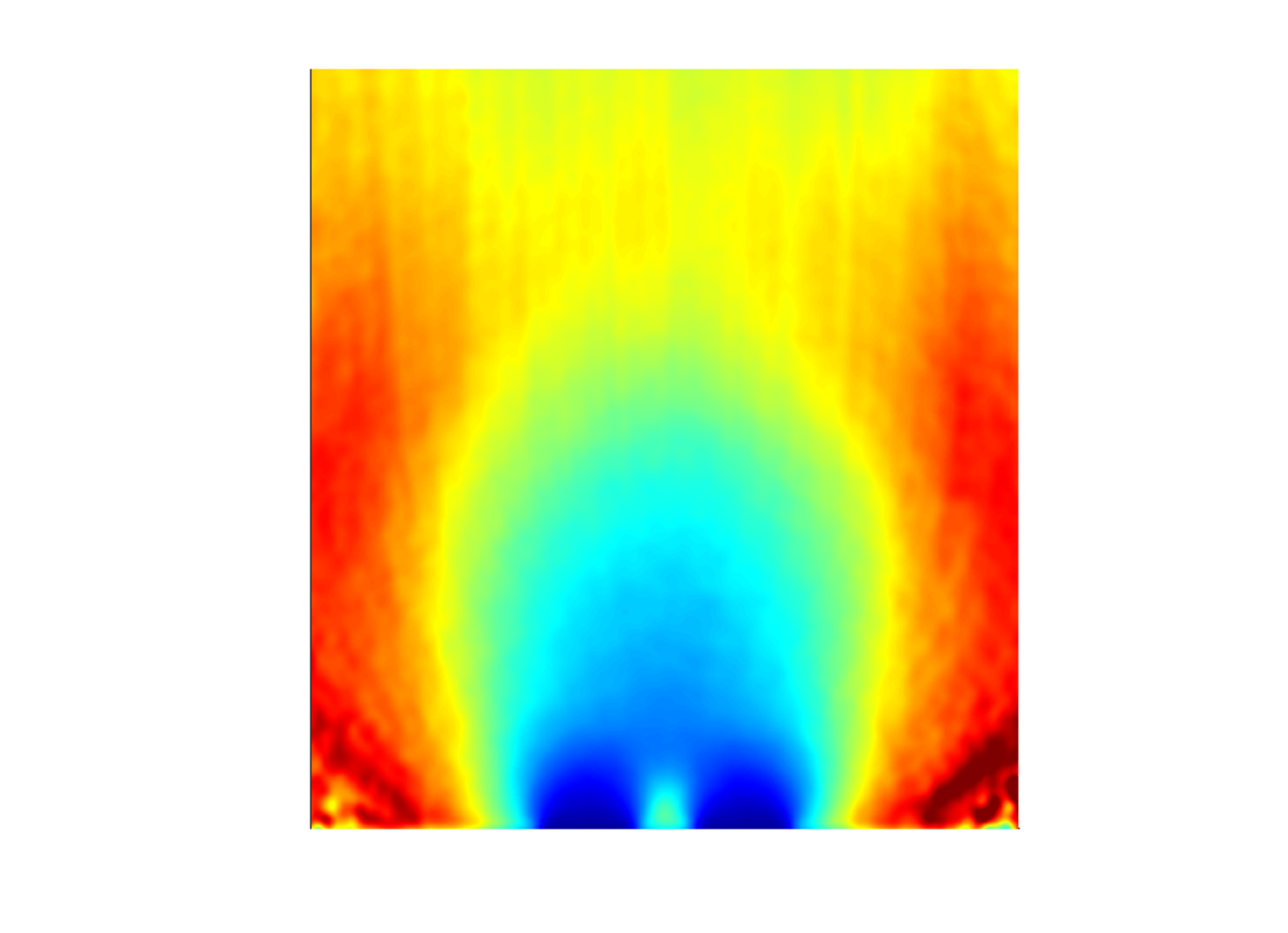}}
	\hspace{0.01cm}
	\subfloat[][]{\includegraphics[clip,trim=3.15cm 1.2cm 2.65cm 0.6cm,width=0.14\linewidth]{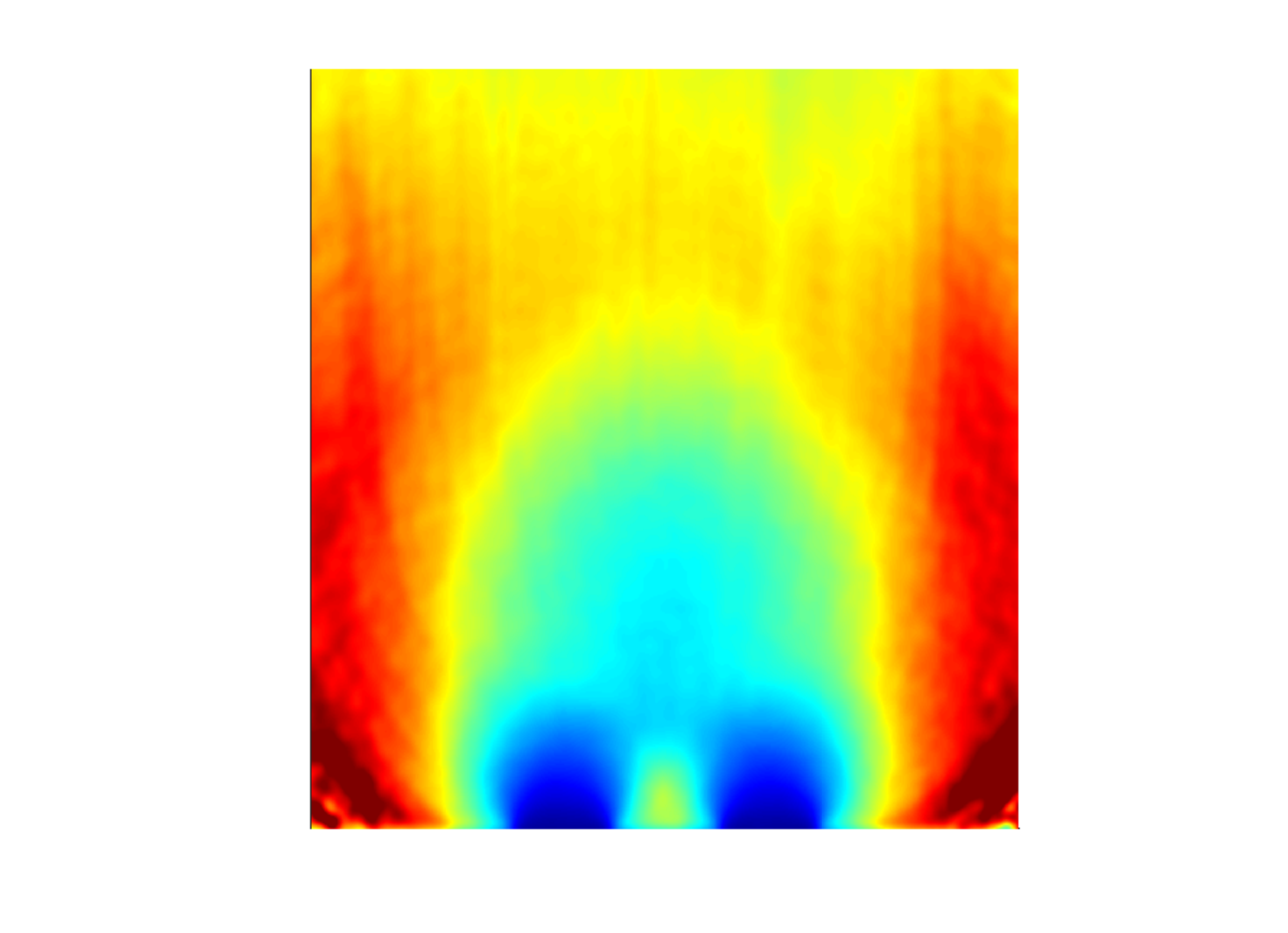}}
	\hspace{0.01cm} 
	\subfloat[][]{\includegraphics[clip,trim=3.15cm 1.2cm 2.65cm 0.6cm,width=0.14\linewidth]{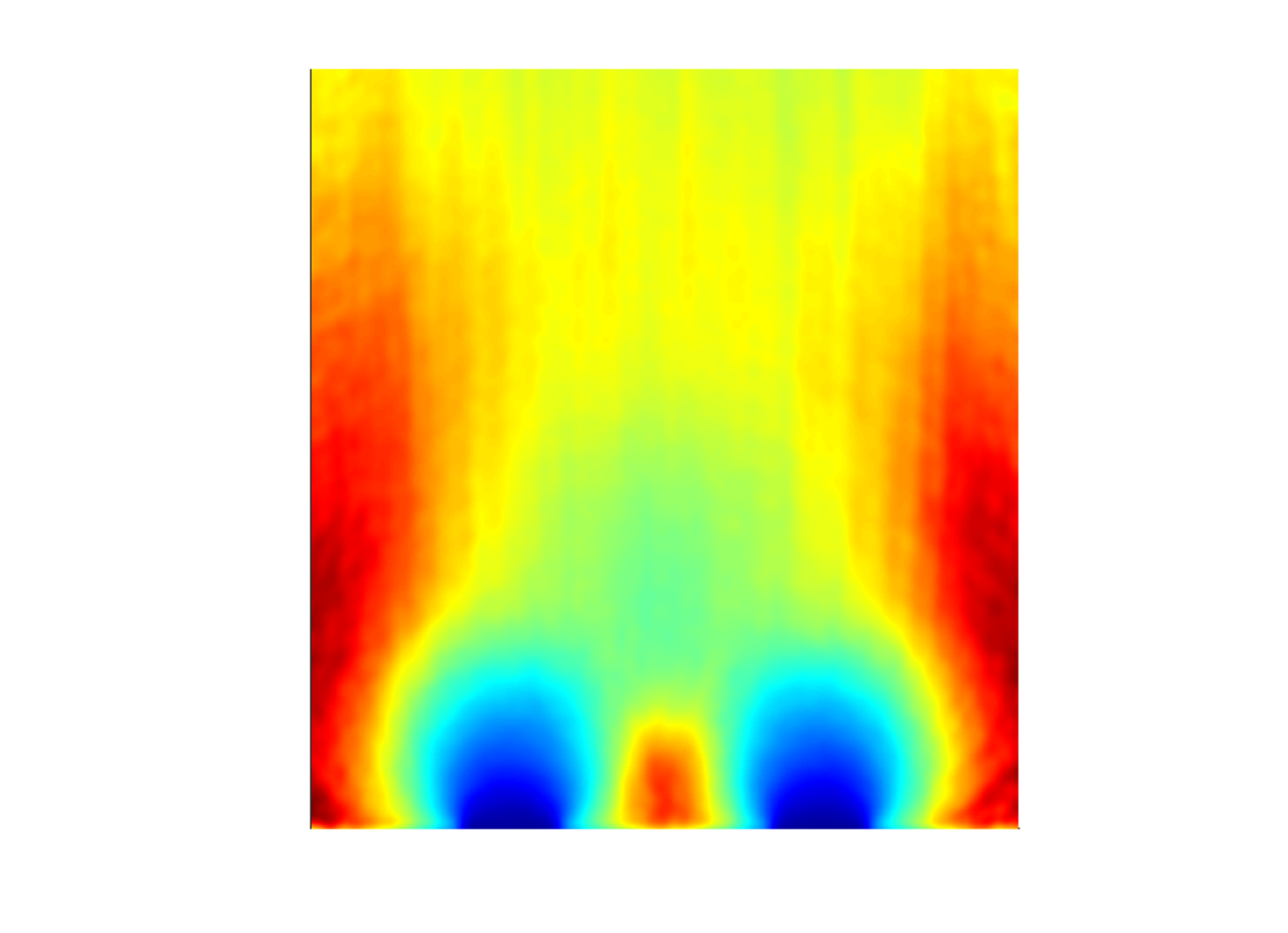}}
	\hspace{0.01cm}
	\includegraphics[width=0.054\linewidth]{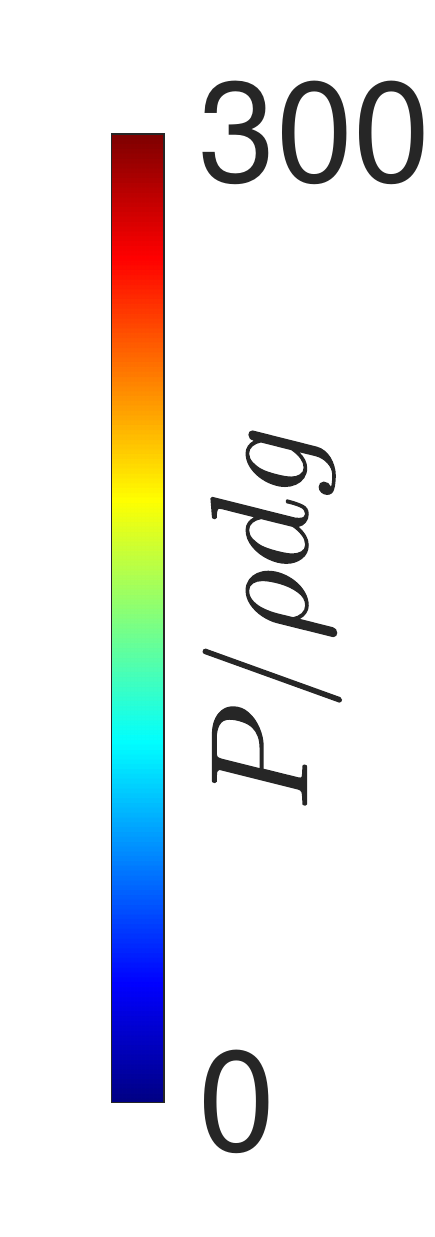}
	\caption{\label{fields:pressm}Spatial distribution of pressure $P$ at different spacings $L/d$ = a) 0.0, b) 2.5, c) 5, d) 10, e) 20 and f) 40 between the orifices, each of width $W/d=20$, placed on the silo base. The fraction of dumbbells for all the cases is $X_{db}=0.5$. }
\end{figure}

\begin{figure}	
	\centering
	\subfloat[][]{\includegraphics[clip,trim=3.15cm 1.2cm 2.65cm 0.6cm,width=0.14\linewidth]{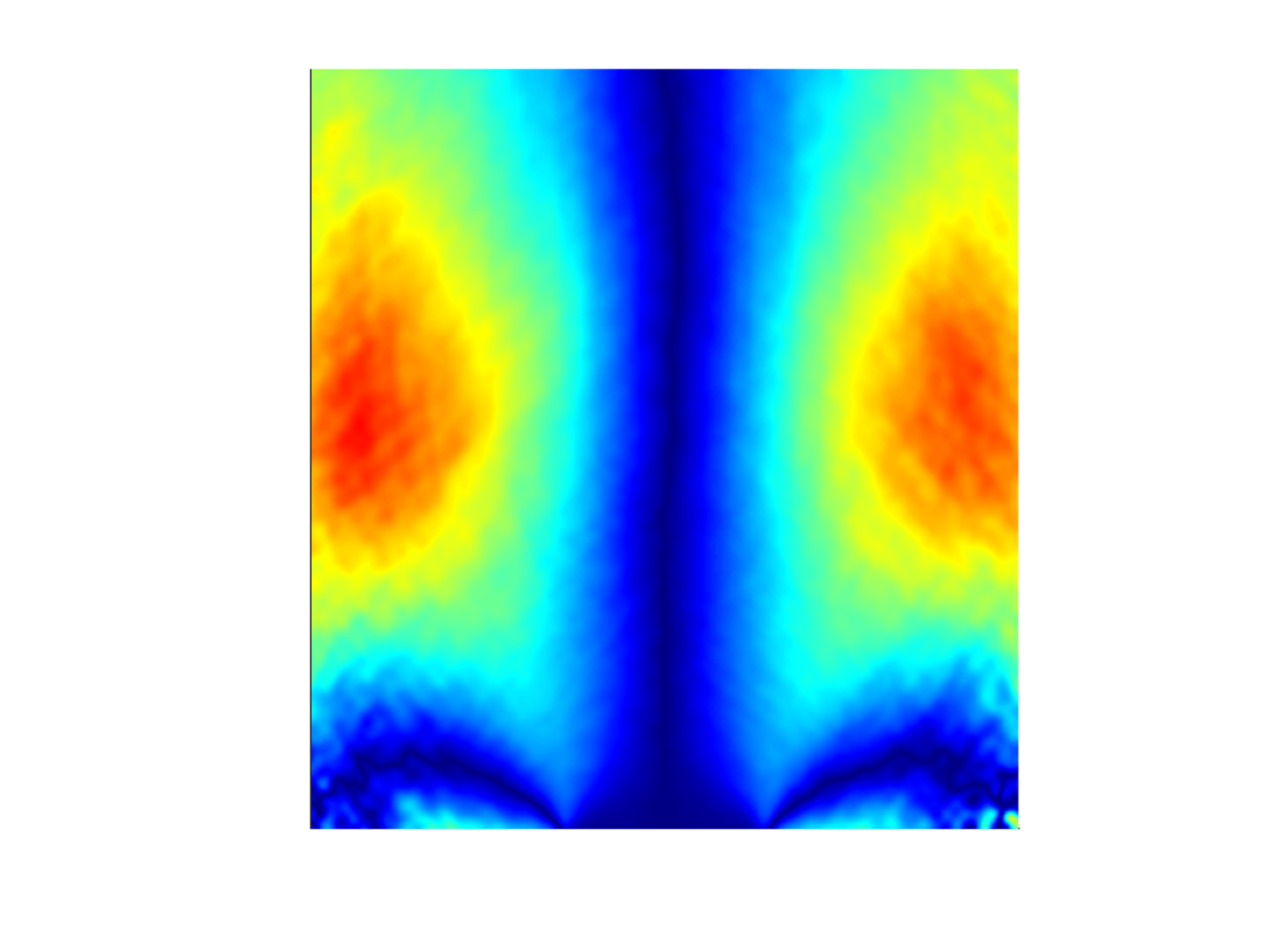}}
	\hspace{0.01cm}
	\subfloat[][]{\includegraphics[clip,trim=3.15cm 1.2cm 2.65cm 0.6cm,width=0.14\linewidth]{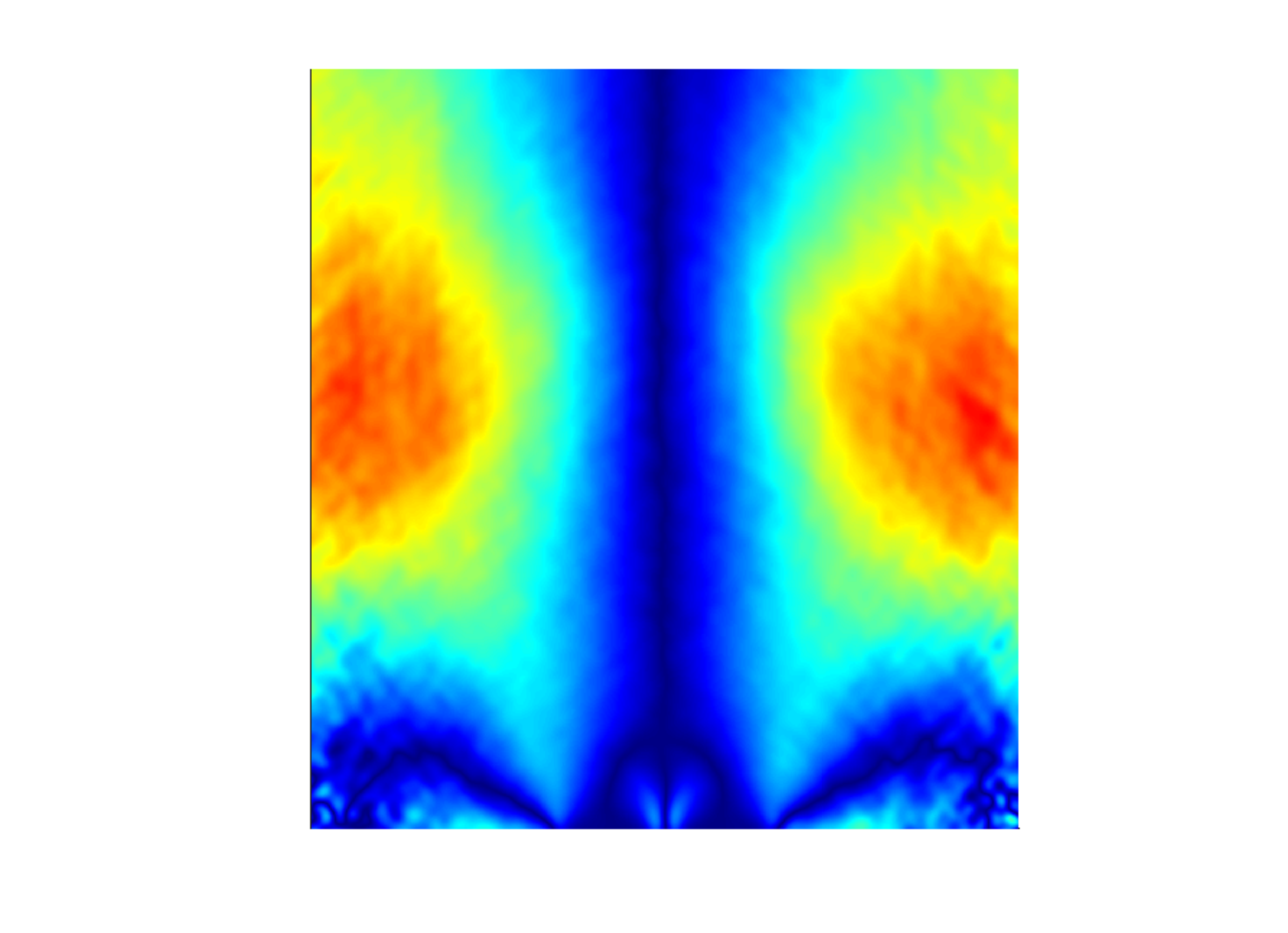}}
	\hspace{0.01cm}
	\subfloat[][]{\includegraphics[clip,trim=3.15cm 1.2cm 2.65cm 0.6cm,width=0.14\linewidth]{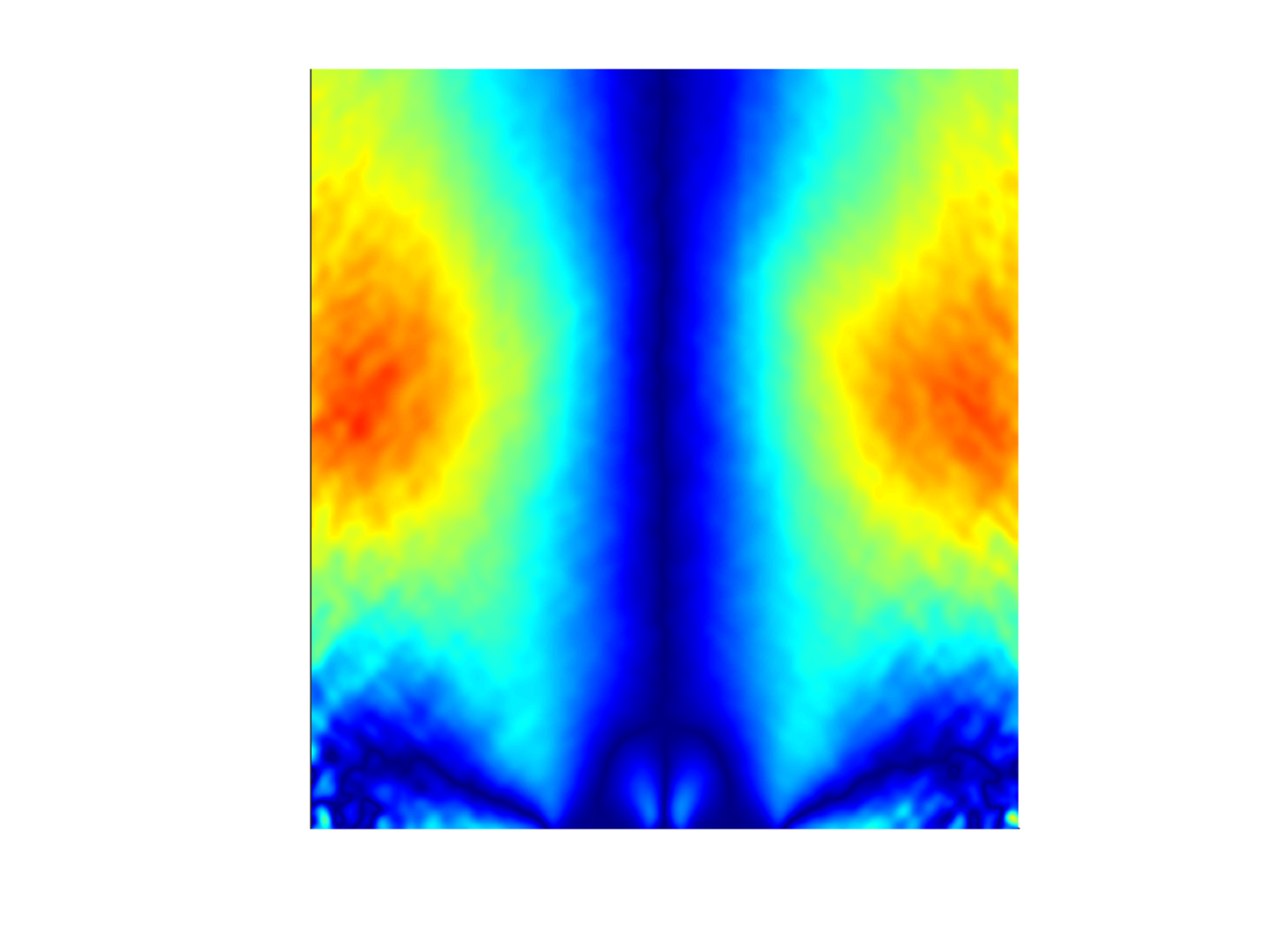}}
	\hspace{0.01cm}
	\subfloat[][]{\includegraphics[clip,trim=3.15cm 1.2cm 2.65cm 0.6cm,width=0.14\linewidth]{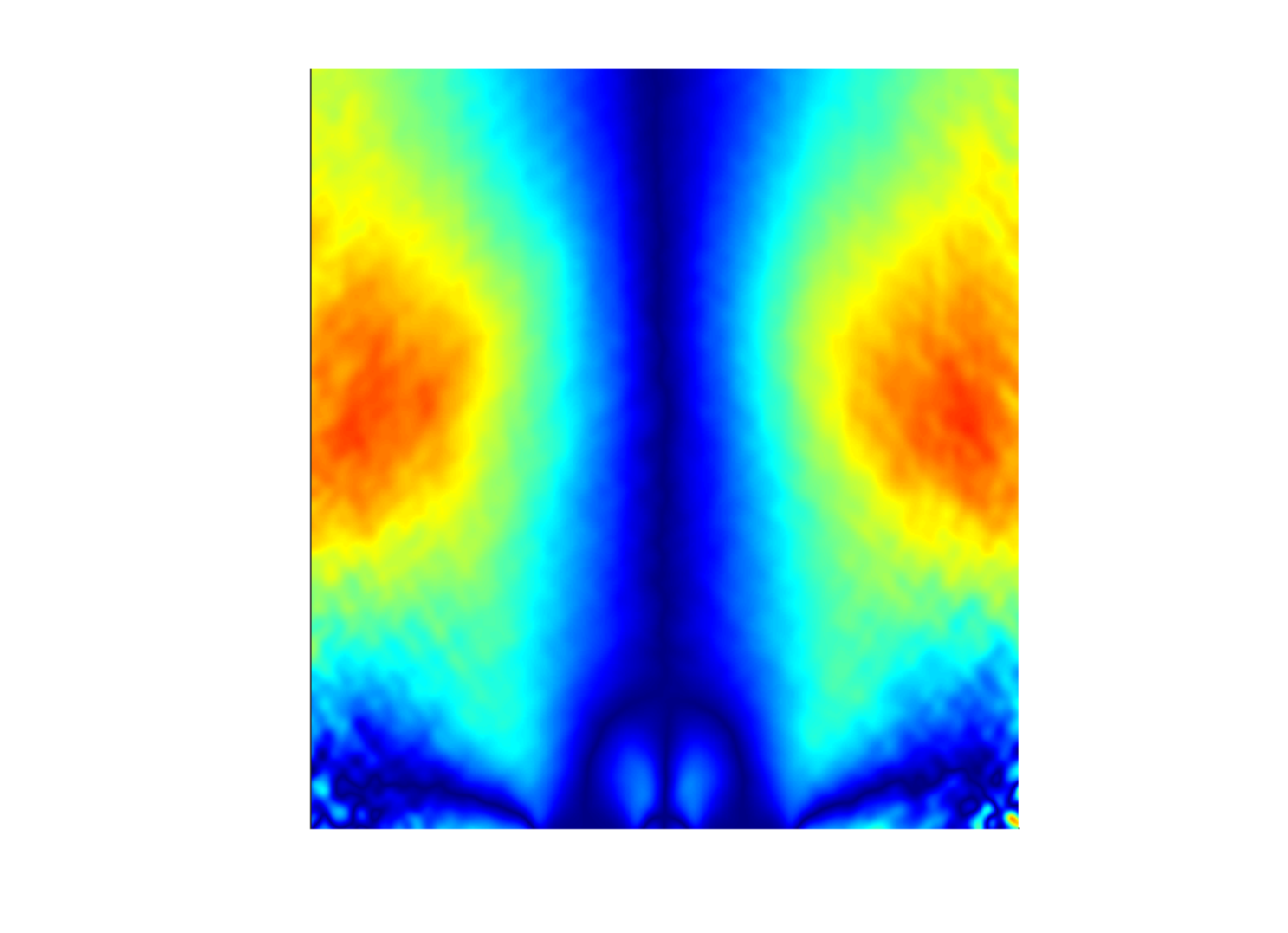}}
	\hspace{0.01cm}
	\subfloat[][]{\includegraphics[clip,trim=3.15cm 1.2cm 2.65cm 0.6cm,width=0.14\linewidth]{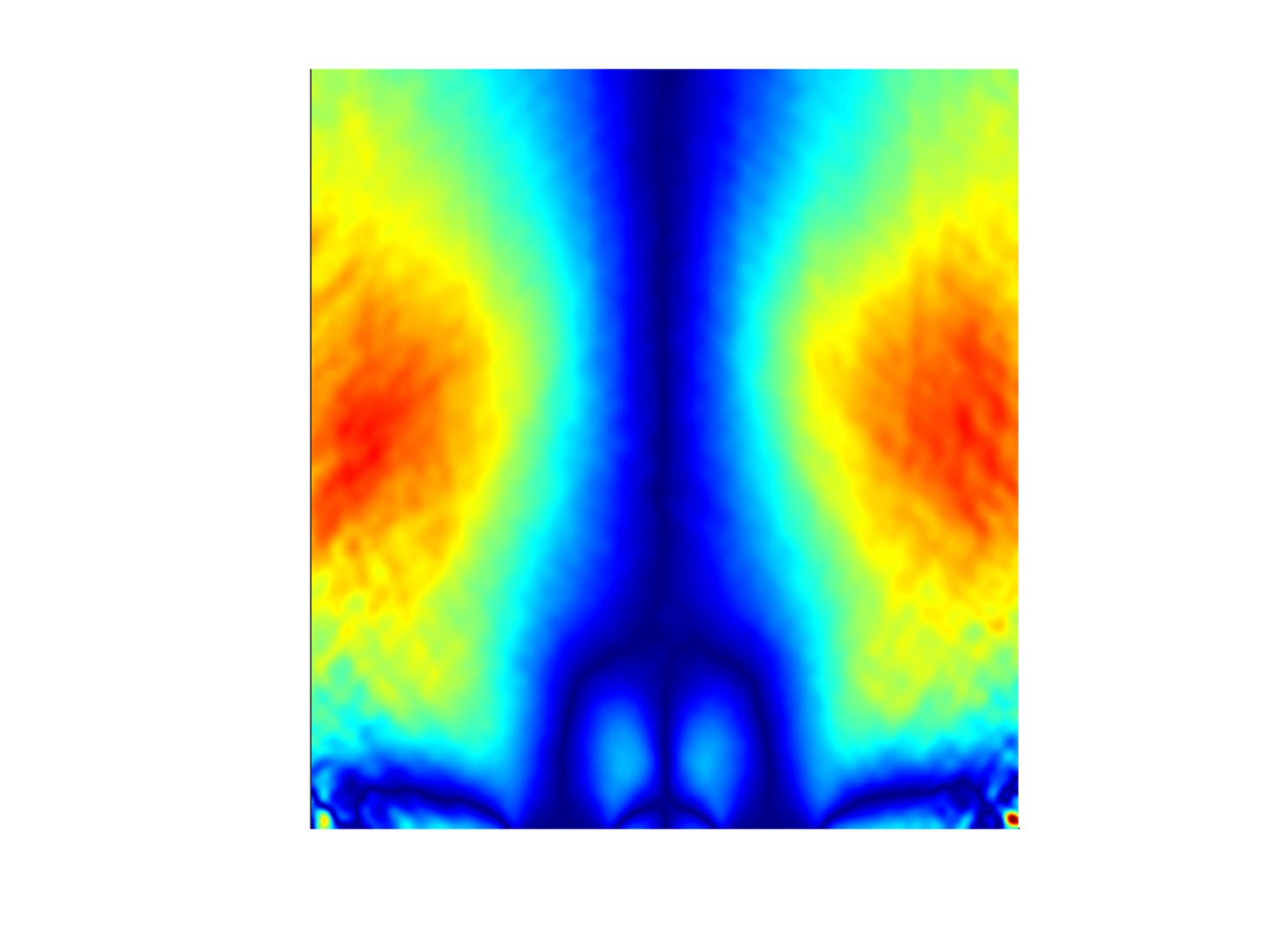}}
	\hspace{0.01cm} 
	\subfloat[][]{\includegraphics[clip,trim=3.15cm 1.2cm 2.65cm 0.6cm,width=0.14\linewidth]{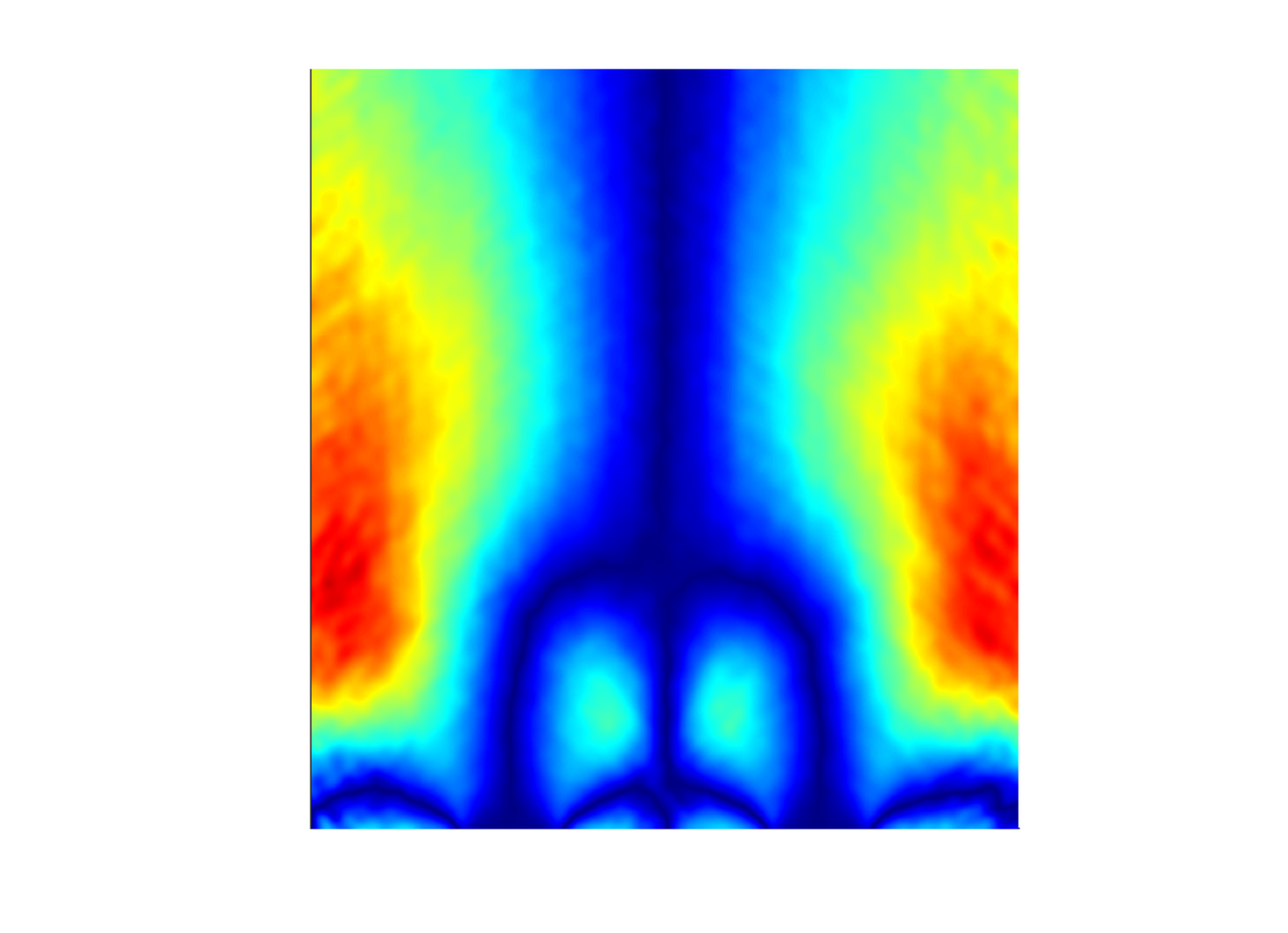}}
	\hspace{0.01cm}
	\includegraphics[width=0.054\linewidth]{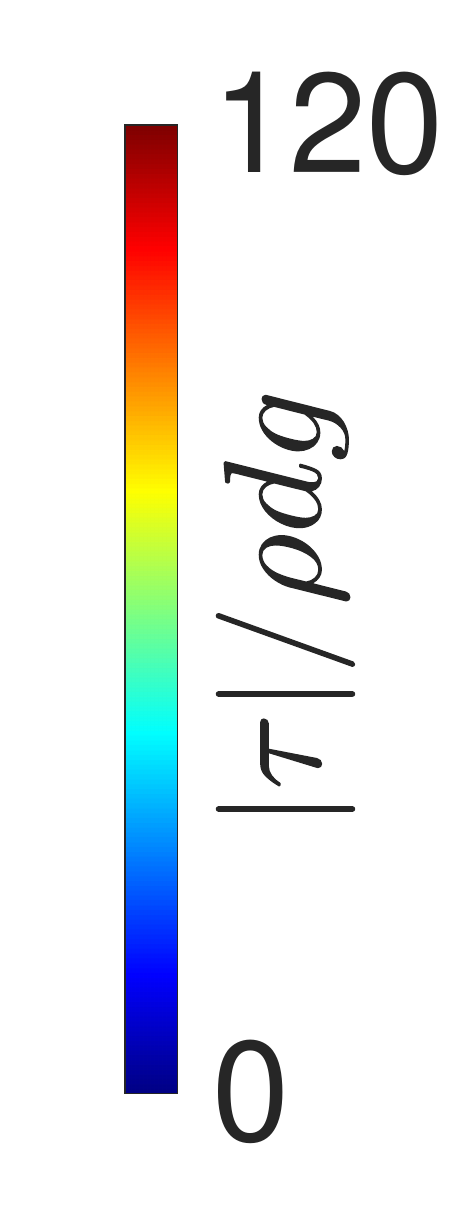}
	\caption{\label{fields:shearm}Spatial distribution of shear stress $|\tau|$ at different spacings $L/d$ = a) 0.0, b) 2.5, c) 5, d) 10, e) 20 and f) 40 between the orifices, each of width $W/d=20$, placed on the silo base. The fraction of dumbbells for all the cases is $X_{db}=0.5$. }
\end{figure}

Pressure fields are illustrated in the figure \ref{fields:pressm} where the orifices are found to interact until $L/d=10$ in the region above the orifice as if there is a single orifice. However, at $L/d=20$, a weak interaction is noticed and at $L/d=40$, it is completely absent as the orifices are wide apart. The pressure is more near the walls as compared to the bulk because the force chains are usually more stronger near the walls as they can be supported by the walls. The pressure is found to be least in the region above the orifices due to dilation as noticed in figure \ref{fields:pfm}. Moreover, due to the expansion of stagnant zone as observed in figure \ref{fields:velm}, pressure is found to increase in the region between the two orifices as the load from the particles flowing above is supported by the base wall between the two orifices. Shear stress seems to be almost independent of inter-orifice distance $L/d$ in the bulk as shown in the figure \ref{fields:shearm} except at very large $L/d$. The areas of deep blue with least $|\tau|$ in the bulk correspond to the flowing zone. Moreover, shear stress is noticed to be maximum near the walls, a behaviour reminiscent in the fluid flow.  The stagnant zone developed at the centre of the silo base at $L/d\ge10$ hinders the movement of particles discharging through each of the orifices thus resulting in a higher shear stress.  Figure \ref{fields:shearm} displays an increase in shear stress with an increase in the inter-orifice distance in the region between the orifices and the stagnant zone at the centre of the silo base due to an expansion of the stagnant zone as observed in the figure \ref{fields:velm}.

\section{Conclusion\label{sec:conclusion}}

In this work, we performed numerical simulations to study the dynamics of granular mixtures for two silo cases differing in orifice positioning. 

\subsection{Lateral orifice}
Here, we studied the effect of the fraction of dumbbells on the mixture of dumbbells and discs flowing through an orifice placed on the sidewall. Flow rate is found to decrease with an increase in the fraction of dumbbells. This can be due to an increase in the dynamic friction with an increase in $X_{db}$ resulting from the interlocking of the dumbbells. At any fraction of dumbbells, flow rate $Q$ is found to scale with $(W/d)^{1.5+X_{db}/2}$ where, $W/d$ is the width of the lateral orifice. This is a modified Beverloo's law which includes not only orifice width but also the fraction of dumbbells. Moreover, at any orifice width, the flow rate is observed to scale with $1-0.65 \times X_{db}$. The ordering of dumbbells is found to decrease with an increase in the fraction of dumbbells near the lateral orifice. This results in a decrease in the area fraction and consequently a decrease in the flow rate. Moreover, maximum velocity $V_{max}$ is found to decrease with an increase in the fraction of dumbbells in the region beside the orifice thus complimenting the flow rate trends. Area fraction in the region beside the lateral orifice is found to be more as compared to the region above the orifice placed on a silobase as the particles lying above the lateral orifice slides into the region beside the orifice. Self-similar profiles of horizontal and vertical velocities are observed in the region beside the orifice for a mixture of dumbbells and discs. Pressure is noticed to be more towards the left side wall and least in the region beside the orifice. Shear stress is maximum in the region close to the left side wall as it lies between the wall and the flowing zone. 

\subsection{Multiple orifices on the silo base}
Here, we analysed how the distance between the two orifices placed on the silo base influences the rheology of a mixture of dumbbells and discs. The flow rate $Q$ is found to be maixmum when the inter-orifice distance $L/d$ is zero for all mixture concentrations. It decreases gradually with an increase in the inter-orifice distance and gets saturated when the distance between the two orifices is very large. The flow rate scaling with the inter-orifice distance is also reported. Time-averaged velocity fields revealed an increase in the stagnant zone present in between the two orifices. The hindrance offered by this stagnant zone along with the one present beside the side walls might be the reason for a decrease in $Q$ with an increase in $L/d$. The two orifices were found to interact until $L/d=20$ and then they cease to interact for larger inter-orifice spacing. Inter-orifice distance has little effect on the area fraction. In the region above the orifices, shear stress is found to increase with an increase in the inter-orifice spacing due to an expansion of stagnant zone between the orifices.

\section*{Conflicts of interest}There are no conflicts to declare.

\section*{Electronic Supplementary Information} The ESI consists of flow fields of various parameters at a lateral orifice width $W/d=25$. For each parameter, five flow fields are demonstrated each corresponding to a different fraction of dumbbells $X_{db}$ ranging from 0.0 to 1.0.
\section*{Acknowledgements}
We thank Sonu Kumar for his valuable discussions regarding coarse graining methodology. Also, we would like to thank IITG for providing HPC faclity for performing our simulations.



\balance


\bibliography{main} 
\bibliographystyle{rsc} 

\end{document}